\documentclass[amssymb,useAMS,prd,aps,amsmath,onecolumn,superscriptaddress,nofootinbib,preprintnumbers]{revtex4}

\usepackage{amsmath}
\usepackage[utf8]{inputenc}
\usepackage[T1]{fontenc}
\usepackage{float}
\usepackage{multirow}
\usepackage{graphicx}
\usepackage[caption=false]{subfig}

\usepackage{array}
\usepackage{psfrag}
\usepackage{braket}
\usepackage[usenames,dvipsnames]{color}
\usepackage[colorlinks=true, linkcolor=BrickRed, citecolor=Blue, urlcolor=Blue, filecolor=Blue]{hyperref}
\usepackage{epsfig,amssymb,fancyhdr}   
\usepackage[normalem]{ulem}

\usepackage{slashed}
\usepackage{mathtools}
\usepackage[titletoc]{appendix}
\usepackage{datetime}
\usepackage{alltt} 
\usepackage{esvect}

\usepackage{epstopdf}

\usepackage{diagbox}

\newcommand{\beq}{\begin{equation}}
\newcommand{\eeq}{\end{equation}}
\newcommand{\ben}{\begin{eqnarray}}
\newcommand{\een}{\end{eqnarray}}
\newcommand{\bi}{\begin{itemize}}
\newcommand{\ei}{\end{itemize}}

\newcolumntype{P}[1]{>{\centering\arraybackslash\Large}p{#1}}

\newcolumntype{b}{>{\large}c}

\newcommand{\remove}[1]{}

\DeclarePairedDelimiter{\evdel}{\langle}{\rangle}
\newcommand{\ev}{\evdel} 
\DeclarePairedDelimiter\abs{\lvert}{\rvert}%

\def\refe@jnl#1{{#1}}
\def\aj{\refe@jnl{Astron.~J.}}
\def\araa{\refe@jnl{Annu.~Rev.~Astron.~Astrophys.}}
\def\apj{\refe@jnl{Astrophys.~J.}}
\def\apjl{\refe@jnl{Astrophys.~J.~Lett.}}
\def\aap{\refe@jnl{Astron.~Astrophys.}}
\def\mnras{\refe@jnl{Mon.~Not.~R.~Astron.~Soc.}}
\def\prd{\refe@jnl{Phys.~Rev.~D}}
\def\fcp{\refe@jnl{Fund.~Cos.~Phys.}}
\def\physrep{\refe@jnl{Phys.~Rep.}}
\def\physlett{\refe@jnl{Phys.~Lett.}}

\def\pe{{p_{e^-}}}

\def\pe1{{p_{e_1}}}

\begin{document}

\hfill {\tt IFIC/17-54}

\hfill {\tt IPPP/17/84}

\vspace{1cm}

\title{Dark matter-neutrino interactions through the lens of their cosmological implications}

\author{Andr\'es Olivares-Del Campo}
\email{andres.olivares@durham.ac.uk}
\affiliation{Institute for Particle Physics Phenomenology, Durham University, South Road, Durham, DH1 3LE, United Kingdom}
\author{C\'eline B\oe hm} 
\email{c.m.boehm@durham.ac.uk}
\affiliation{Institute for Particle Physics Phenomenology, Durham University, South Road, Durham, DH1 3LE, United Kingdom}
\affiliation{LAPTH, U. de Savoie, CNRS,  BP 110, 74941 Annecy-Le-Vieux, France}
\affiliation{Perimeter Institute, 31 Caroline St N., Waterloo Ontario, Canada N2L 2Y5}
\author{Sergio Palomares-Ruiz}
\email{sergiopr@ific.uv.es}
\affiliation{Instituto de F\'{\i}sica Corpuscular (IFIC), CSIC-Universitat de Val\`encia, Apartado de Correos 22085, E-46071 Val\`encia, Spain}
\author{Silvia Pascoli} 
\email{silvia.pascoli@durham.ac.uk}
\affiliation{Institute for Particle Physics Phenomenology, Durham University, South Road, Durham, DH1 3LE, United Kingdom}

\preprint{}
\begin{abstract}
Dark matter and neutrinos provide the two most compelling pieces of evidence for new physics beyond the Standard Model of Particle Physics but they are often treated as two different sectors. The aim of this paper is to determine whether there are viable particle physics frameworks in which dark matter can be coupled to active neutrinos. We use a simplified model approach to determine all possible renormalizable scenarios where there is such a coupling, and study their astrophysical and cosmological signatures. We find that dark matter-neutrino interactions have an impact on structure formation and lead to indirect detection signatures when the coupling between dark matter and neutrinos is sufficiently large. This can be used to exclude a large fraction of the parameter space. In most cases, dark matter masses up to a few MeV and mediator masses up to a few GeV are ruled out. The exclusion region can be further extended when dark matter is coupled to a spin-1 mediator or when the dark matter particle and the mediator are degenerate in mass if the mediator is a spin-0 or spin-$1/2$ particle.
\end{abstract}
\maketitle

\section{Introduction}

Collisionless dark matter (DM) has been the main paradigm for the last four decades. However, at the very least, DM needs to have interactions to be produced in the early Universe. Interactions with neutrinos are the most intriguing of all because they involve particles which provide the only evidence of new physics beyond the Standard Model (SM) so far. It is therefore natural to ask whether DM and neutrino properties are related and  whether these two species can interact directly. 

Non-vanishing DM-neutrino (DM-$\nu$) interactions have several important cosmological and astrophysical consequences. They can explain the observed DM relic density if DM has been thermally produced and annihilations into neutrinos are the dominant channel. They can also lead to DM indirect detection signatures if DM annihilates or decays into neutrinos in the galaxy~\cite{Yuksel:2007ac, PalomaresRuiz:2007eu, PalomaresRuiz:2007ry, Kile:2009nn, Covi:2009xn, Primulando:2017kxf} or via cosmic neutrino signals~\cite{Beacom:2006tt, Moline:2014xua, Moline:2016fdo}. They can further produce dips in the diffuse supernova neutrino background~\cite{Farzan:2014gza} and lead to possible anisotropies in the angular distribution of high-energy neutrinos caused by their scattering off DM particles~\cite{Arguelles:2017atb}. DM-$\nu$ interactions can also erase primordial DM fluctuations, and eventually lead to a suppression of large scale structures (LSS) in the Universe~\cite{Boehm:2003xr} and of the number of satellites in Milky Way-like galaxies~\cite{Boehm:2014vja, Schewtschenko:2014fca, Schewtschenko:2015rno}, potentially solving the ``too-big-to-fail'' problem of cold DM. This effect is referred to as the collisional damping~\cite{Boehm:2000gq, Boehm:2001hm, Boehm:2004th}. Finally, DM-$\nu$ interactions can be at the origin of neutrino masses in radiative models~\cite{Ma:2006km, Boehm:2006mi, Farzan:2010mr, Farzan:2010wh}.

Here, we consider all possible renormalizable direct interactions between active neutrinos and DM. We do not attempt to build a complete ultra-violet (UV) model but adopt instead a phenomenological approach in which we consider the cosmological and experimental consequences of such interactions. It is not trivial to build a full theory and the latter will often involve a more extended sector than just neutrinos and DM. A first difficulty arises from the fact that left-handed neutrinos belong to $SU(2)$ doublets. Therefore, naively, one expects large charged lepton-DM couplings in the presence of interactions with neutrinos\footnote{We refer to Ref.~\cite{Lindner:2010rr} for a systematic study of the gauge invariant combinations that lead to sizable DM annihilation into neutrinos, while considering experimental constraints for the DM coupling to charged leptons.}. To avoid this issue, one could however invoke some mixing between neutrinos and heavy neutral fermions so that DM-$\nu$ interactions effectively arise only below the electroweak symmetry breaking scale (see, e.g., Ref.~\cite{Farzan:2016wym}). Secondly, in models which also aim to generate neutrino masses, the introduction of a Dirac mass term can break the symmetry needed to ensure DM stability, by mixing its neutral fermions with neutrinos. Nevertheless, models in which neutrino masses are generated radiatively can overcome this issue by preserving a $Z_2$ or larger symmetry~\cite{Ma:2006km}.

In what follows, we restrict the extensions of the SM to models with one additional DM particle and one mediator. We adopt the same simplified approach as in Ref.~\cite{Boehm:2003hm}, where a generic light spin-1 gauge boson and/or a heavy neutral (scalar or vector fermion) were proposed as mediators of the DM interactions, but we extend it to account for other mediators. We consider all possible cases involving fermions and bosons, in an effort to get a systematic assessment of the allowed parameter space. Given that we focus uniquely on a DM-$\nu$ interaction term, we disregard constraints from colliders or beam dump experiments, since these bounds explicitly assume a coupling to either charged leptons or quarks through mixing or through additional interactions. We note that such constraints would be relevant in an UV-complete model though.

In Section~\ref{sec:cosmosig}, we start by summarizing the constraints that apply to scenarios with DM-$\nu$ interactions. There is a total of twelve relevant scenarios, which are presented in Section~\ref{sec:scenarios}. In Section~\ref{sec:results}, we determine the allowed parameter space for each of these scenarios and conclude in Section~\ref{sec:conclusion}.

\section{Cosmological and observational signatures}
\label{sec:cosmosig}
DM-$\nu$ interactions induce a variety of signatures. In this work, we mostly focus on the collisional damping effect and on indirect detection signatures stemming from DM annihilations into neutrinos in the Milky Way halo. For reference, we also mention the constraints obtained in the case of thermal DM, if one assumes the same DM-$\nu$ interaction to be responsible for the observed DM abundance.

\subsection{Relic density} 

Assuming equal number densities for DM particles and antiparticles, the typical value for the thermal average of the annihilation cross section (times the relative velocity) that is needed to explain the observed abundance is about $\ev{\sigma v_{\rm{r}}} \simeq 3 \times 10^{-26} \, \rm{cm^3}/\rm{s}$ for a constant cross section\footnote{If the cross section is $v^2$- or $v^4$-dependent, a value of $\ev{\sigma v_{\rm{r}}} \simeq 6 \times  10^{-26} \, \rm{cm^3}/\rm{s}$ or $\ev{\sigma v_{\rm{r}}} \simeq 9 \times 10^{-26} \ \rm{cm^3}/\rm{s}$ is then required at freeze-out, respectively~\cite{Giacchino:2013bta}.}. For the current analysis, we assume the DM particle and the mediator to be the only beyond the SM particles, but in a complete model, other DM annihilation channels or DM production mechanisms may be at work. As a result, the DM annihilation cross section into neutrinos may not contribute significantly to the relic density calculations. 

In what follows, we use this value to rapidly assess whether a candidate is over- or under-abundant. A more careful approach would require to solve the Boltzmann equation. However, this would need to be done on a case-by-case basis, since DM-$\nu$ interactions could change the value of the cosmological parameters and, in particular, $\Omega_{\rm{DM}} h^2$~\cite{Wilkinson:2014ksa,DiValentino:2017oaw}. Such a detailed analysis is beyond the scope of this paper, but we have checked that, for representative mass benchmark points, we could explain $\Omega_{\rm{DM}} h^2=0.11$ with an accuracy of a few percent.

\subsection{CMB and structure formation \label{sec:colldamp}}
DM-$\nu$ interactions can leave a visible imprint in the matter and light distribution across the sky. Even though they are expected to happen in the early Universe, such interactions can  alter the physics of the cosmic microwave background (CMB) and structure formation~\cite{Boehm:2000gq, Boehm:2004th, Mangano:2006mp}. 

There are two main effects. Firstly, DM is no longer collisionless; as the DM particles scatter off neutrinos, they diffuse out and wash out the smallest primordial fluctuations. This collisional damping effect translates into a suppressed (oscillating) matter power spectrum, which can mimic a warm DM spectrum~\cite{Boehm:2001hm}. Secondly, neutrinos stay collisional for much longer than in the standard case. This reduces their free-streaming length and increases their ability to cluster and form large-scale structures.

By confronting the CMB and LSS predictions to observations, one can get an upper bound on the strength of DM-$\nu$ interactions. Using Planck's angular matter power spectra, one obtains that the DM-$\nu$ elastic scattering cross section cannot exceed $\sigma_{\rm{el}} < 6 \, \times \, 10^{-34} \ \left(\frac{m_{\rm{DM}}}{\rm{MeV}}\right) \ \rm{cm^2}$~\cite{Escudero:2015yka, DiValentino:2017oaw}. This limit is based on physical processes that took place in the linear regime and is therefore fairly robust. Nevertheless, it would be a bit stronger with extremely precise polarised data. An alternative is to require the matter distribution in the early Universe to be compatible with Lyman-$\alpha$ observations. This means that the damping can only happen at small scales, which translates into a constraint on the elastic scattering cross section of~\cite{Wilkinson:2014ksa} 
\begin{equation}
\sigma_{\rm{el}} < 10^{-36} \ \left(\frac{m_{\rm{DM}}}{\rm{MeV}}\right) \ \rm{cm^2} ~, 
\label{constant}
\end{equation}
for a constant elastic cross section, and
\begin{equation}
 \sigma_{\rm{el}}< 10^{-48} \ \left(\frac{m_{\rm{DM}}}{\rm{MeV}}\right) \ \left(\frac{T_\nu}{T_0}\right)^2 \  \rm{cm^2} ~, \label{temperature}  
\end{equation}
for a temperature-dependent cross section, where $T_0 = 2.35 \times 10^{-4}$~eV is the photon temperature today. While there are uncertainties regarding the use of Lyman-$\alpha$ emitters to constrain the matter power spectrum, similar limits have been derived using the number of satellite companions of the Milky Way~\cite{Boehm:2014vja, Bertoni:2014mva, Schewtschenko:2014fca, Schewtschenko:2015rno}. Such limits are conservative and could become much stronger with a better understanding of the role of baryons in galaxy formation, since astrophysical feedback processes may also reduce the number of satellites (see, e.g., Ref.~\cite{Fattahi:2016nld}).

\subsection{Neutrino reheating bounds}
\label{sec:reheating}

DM annihilations into neutrinos after the neutrino decoupling from electrons, i.e., at $T \lesssim T_{\rm dec} \sim 2.3$ MeV~\cite{Enqvist:1991gx}, can reheat the neutrino sector and lead to visible signatures. The subsequent change in the neutrino energy density is equivalent to increasing the number of relativistic degrees of freedom, $N_{\rm{eff}}$, in the early Universe, according to
\begin{equation}
\rho_{\nu} \equiv \rho_\gamma \left[1 + \frac{7}{8} \, \left(\frac{T_\nu}{T_\gamma}\right)^{4/3} \, N_{\rm{eff}} \right] ~,
\end{equation}
where $\rho_\gamma$ is the energy density of photons. However $N_{\rm{eff}}$ cannot be arbitrarily large as this would impact the formation of light elements at the time of big bang nucleosynthesis (BBN)~\cite{Kolb:1986nf, Serpico:2004nm, Ho:2012ug, Berezhiani:2012ru, Nollett:2013pwa, Nollett:2014lwa} and the CMB angular power spectrum at decoupling~\cite{Smith:2011es, Hamann:2011ge, Archidiacono:2011gq, Hamann:2011hu, Nollett:2011aa, Boehm:2012gr, Steigman:2013yua, Archidiacono:2013lva, Boehm:2013jpa, DiValentino:2013qma, DiValentino:2016ikp}. This condition eventually rules out DM candidates much lighter than a few MeVs~\cite{Boehm:2013jpa, Wilkinson:2016gsy}.

The derivation of the precise value of the DM mass bound assumes DM was in thermal equilibrium with neutrinos. Nevertheless, even in the case of non-thermal DM, a limit on $N_{\rm{eff}}$ could be set, if DM annihilates (or decays) into neutrinos after BBN and before decoupling.

\subsection{Signatures in neutrino detectors}\label{SKSec}

\begin{figure*}[t]
\begin{center}
\includegraphics[width=0.95\textwidth]{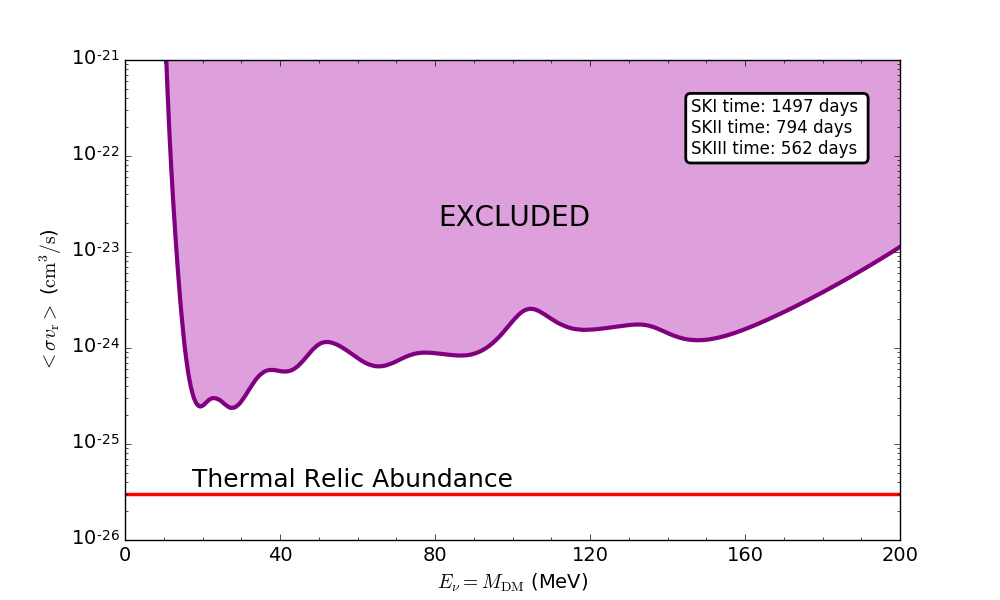}
\end{center}
\caption{Limits at the 90\% C.L. on the DM annihilation cross section as a function of the DM mass. We have considered annihilations in the Milky Way, with $R_0 = 8.5$ kpc, $\rho_0 = 0.3$ GeV $\text{cm}^{-3}$. The solid line represents the combined analysis with data from the three different SK phases. The thick red line corresponds to the value that is needed to explain the observed abundance in thermal DM scenarios, i.e., $\langle \sigma v_r \rangle = 3\times 10^{-26}\; \rm{cm}^3/\rm{s}$.} 
\label{fig:SK}
\end{figure*}

DM annihilations in high density regions like the Milky Way may lead to a detectable monochromatic flux of neutrinos (and antineutrinos) in neutrino detectors~\cite{PalomaresRuiz:2007eu, Frankiewicz:2015zma}. This occurs because each neutrino produced from DM annihilations in the Milky Way has an energy equal to the DM mass. Assuming for simplicity that DM annihilates into the three neutrino flavours with the same probability (hence, the factor of 3 in the following equation), the differential neutrino and antineutrino flux per flavour produced (and at Earth\footnote{Note that, because of the very long propagation distances, $\Delta m^2 \, L/E \gg 1$, an incoherent mixture of mass eigenstates arrives at Earth. This implies that whatever the flavor fractions at the source are, the $\nu_e$ fraction at the detector would be within $[4/7, \, 5/2]/3$, and our results apply to all cases within a factor of 2 or so.}) by DM annihilations in the Milky Way halo is given by~\cite{PalomaresRuiz:2007eu}
\begin{equation} \label{eq:flux}
\frac{d\phi}{dE_\nu} = \frac{\ev{\sigma v_{\rm r}}}{2} \, \mathcal{J}_{\rm avg} \, \frac{R_0 \, \rho^2_0}{m^2_{\rm DM}} \, \frac{1}{3} \, \delta(E_\nu - m_{\rm DM}) \equiv \Gamma(\ev{\sigma v_{\rm r}}, m_{\rm DM}) \, \delta(E_\nu - m_{\rm DM}) ~, 
\end{equation}
where $R_0 = 8.5$~kpc is the Sun's distance to the galactic center and $\rho_0 = 0.3 \, \rm{GeV}/\rm{cm^{3}}$ is the DM density at that position. The factor of 2 accounts for the situation where the DM particles are their own antiparticles and must therefore be omitted in scenarios where this is not the case. In this expression, $\mathcal{J}_{\rm{avg}}$ is the integral of the square of the DM density in the halo along the line of sight averaged over the Milky Way halo and can be evaluated as~\cite{Yuksel:2007ac}
\begin{equation}
\mathcal{J}_{\rm avg} = \frac{1}{2 \, R_0 \, \rho^2_0} \int^1_{-1} \int^{l_{\rm{max}}}_0\rho^2(r) \, dl \, d(\cos{\psi}) \ .
\end{equation}
In this equation, $r = \sqrt{R^2_0 - 2 \, l \, R_0 \, \cos{\psi} + l^2}$, with $\psi$ being the angle between the galactic centre and the line of sight and the upper limit of integration $l_{\rm max} = \sqrt{R^2_{\rm halo} - \sin{\psi}^2 R^2_0}$ a function of the radius of the halo $R_{\rm halo}$. This quantity can be estimated numerically using different DM halo profiles. It is equal to $\mathcal{J}_{\rm avg}=3$ and $\mathcal{J}_{\rm avg} = 8$, assuming the NFW and Moore profiles, respectively~\cite{Navarro:1995iw, Kravtsov:1997dp, Moore:1999gc}. In what follows, we use an intermediate value of $\mathcal{J}_{\rm avg} = 5$ (see Ref.~\cite{Yuksel:2007ac}).

Here we improve and update the analysis performed\footnote{Note that the limit is slightly worse than the analysis in Ref.~\cite{PalomaresRuiz:2007eu}. Moreover, during the writing of this manuscript, Ref.~\cite{Primulando:2017kxf} performed a similar analysis and obtained bounds an order of magnitude better than the ones presented in this paper.} in Refs.~\cite{PalomaresRuiz:2007eu, PalomaresRuiz:2007ry}. In these references, it was found that neutrino experiments with a low-energy threshold such as Super-Kamiokande (SK) can be used to place limits on DM-$\nu$ interactions for DM candidates with masses in the MeV range. In what follows, we combine SK data from the diffuse supernova neutrino searches, using the energy resolutions, thresholds and efficiencies from SK-I, II, III phases~\cite{Bays:2011si, Hosaka:2005um, Cravens:2008aa, Abe:2010hy}. Given that the maximum observed positron energy is 88 MeV, we simulate the expected neutrino signal from DM annihilations and the relevant backgrounds for neutrino energies between 10-130 MeV, for each phase. We include both the interactions of antineutrinos with free protons (inverse beta decay) and the interactions of neutrinos and antineutrinos with bound nucleons, as done in Refs.~\cite{PalomaresRuiz:2007eu, PalomaresRuiz:2007ry}. At these energies, the two main sources of background\footnote{We are not including other two subleading sources of background, such as neutral current (NC) elastic events and low-energy muons or pions misidentified as electrons/positrons. Our results are not significantly affected by this.} are the atmospheric $\nu_e$ (and $\bar\nu_e$) flux and the Michel positrons (and electrons) from the decays at rest of invisible muons (i.e., muons below detection threshold produced by atmospheric neutrinos), which we compute as in Refs.~\cite{PalomaresRuiz:2007eu, PalomaresRuiz:2007ry}, with the updated energy resolutions and efficiencies.

We define one $\chi^2_i$ for each SK phase, as defined in Refs.~\cite{PalomaresRuiz:2007eu, PalomaresRuiz:2007ry}, which depends on the rate of DM events ($\alpha$) and the rate of events for each of the two sources of background we consider. We then combine these  $\chi^2_i$ into a single quantity, $\chi^2 = \chi^2_{\rm I} + \chi^2_{\rm II} + \chi^2_{\rm III}$, as done in Ref.~\cite{Bernal:2012qh}. The total $\chi^2$ is minimized with respect to the rate of events of the two background components, resulting in a function of the event rate from DM annihilations, i.e., $\chi^2(\alpha)$. The 90\% confidence level (C.L.) limit on the DM event rate, $\alpha_{90}$, is obtained from
\begin{equation}
\frac{\int_0^{\alpha_{90}} \chi^2(\alpha) \, d\alpha}{\int_0^{\infty} \chi^2(\alpha) \, d\alpha} = 0.9 ~. 
\label{eq:alpha90}
\end{equation}
We show in Fig.~\ref{fig:SK} the 90\% C.L. limit on the annihilation cross section, which is obtained by solving~\cite{Bernal:2012qh}
\begin{equation}
\Gamma(\ev{\sigma v_{\rm r}}, m_{\rm DM}) \, A_{\rm{tot}} = \alpha_{90} \hspace{3mm} , \,  
A^{\rm{tot}} \equiv \frac{\sum_i A_i \, t_i}{\sum_i t_i} ~, 
\end{equation}
where $A_i$ ($i = \{\rm I, II, III\}$) is the number of events for a monochromatic neutrino flux, $\delta(E_\nu - m_{\rm DM})$, for each SK phase at the detector, and $t_i$ are the corresponding data-taking times.

We further constrain the parameter space by superimposing the results from the analysis carried out by the SK collaboration for GeV neutrinos produced at the galactic centre~\cite{Frankiewicz:2015zma}. This allows us to derive bounds on the DM annihilation cross section for DM masses between 1--950~GeV. The DM mass range between 130~MeV and 1~GeV is constrained using the upper bound which was derived in Ref.~\cite{Beacom:2006tt} by comparing the diffuse neutrino flux from all halos in the universe with the atmospheric neutrinos measured at neutrino detectors such as SK, Fr\'ejus and Amanda. Finally, for the low DM mass region between 2--17~MeV, we use the measured antineutrino flux at Borexino~\cite{Bellini:2010gn} and convert this into a conservative upper bound on the annihilation cross section. The last two bounds could be improved by one or two orders of magnitude with a more detailed analysis. Nevertheless, as we will see, these constraints allow us to exclude regions of the parameter space where collisional damping could be relevant.

Note that all these constraints are derived for a constant annihilation cross section. Limits on velocity-dependent cross sections would require to consider the velocity distribution of the DM particles within the halo, e.g., in a similar way as done in Ref.~\cite{Diamanti:2013bia}. However, we have checked that one could derive  reasonably accurate limits in the velocity-dependent case by simply rescaling the constant cross section limits by the appropriate power of the DM velocity in the halo ($v_{\rm{halo}} = 10^{-3}$~c). 

It is important to also notice that these constraints assume a sizable annihilation rate in the Milky Way halo and, therefore, no asymmetry in the DM and anti DM number densities (should the DM and its antiparticle be different)~\cite{Barr:1990ca, Barr:1991qn, Kaplan:1991ah}.

\section{Scenarios} 
\label{sec:scenarios}

With these constraints in mind, we can perform a systematic study of the viability of DM-$\nu$ interactions. Assuming that DM only interacts with left-handed active neutrinos, we can establish the list of all possible renormalizable scenarios (i.e., the combination of DM and mediator particles) consistent with Lorentz invariance\footnote{Scenarios with DM candidates of any spin interacting with a triplet scalar mediator have a very similar phenomenology to those cases with a spin-1 mediator. Thus, we disregard them for simplicity.}.

In what follows, we refer to the DM candidate as $\chi$. The mediator is referred to as $Z'$ if it is a spin-1 particle, $\phi$ if it is a spin-0 particle and $N$ if it is a spin-1/2 particle. The expressions for the effective Lagrangian and for the approximate elastic scattering and annihilation cross sections are summarized in Table~\ref{table2}. When the DM candidate is a spin-1 particle, we consider a real vector candidate, since the only difference with the case of complex vector DM coupled to a Dirac (Majorana) mediator is a factor of $\frac{1}{4}$ ($\frac{5}{12}$) in the annihilation cross section\footnote{This is due to the fact that a real vector DM with a fermion mediator annihilates via a $t$- and a $u$- channel, while a complex vector DM proceeds via a $t$- channel only.}. To perform our calculations, we have assumed that

\begin{itemize}
 \setlength\itemsep{0em}
\item There are only left-handed neutrinos in the final and/or initial states. For simplicity, we do not differentiate between neutrino species.
\item Neutrino masses can be neglected. Nevertheless, the neutrino mass generation mechanism and the particular nature (Dirac vs Majorana) of neutrinos would impose further constraints on the parameter space.
\item The elastic scattering cross section could be safely averaged over the range $\cos \theta \in [-0.95,0.95]$, to avoid the co-linear divergence.
\item The DM-$\nu$ coupling is equal to 1 ($g=1$). This means that we are probing the regime where DM is strongly coupled to neutrinos. Limits can be rescaled accordingly when $g\ll 1$ and will be discussed in the text.
\item For the calculations of the elastic scattering we have assumed that neutrinos follow a Fermi-Dirac distribution with temperature $T_\nu$ and consequently, $\ev{E^2_\nu}\simeq1 2.9 \, T^2_\nu$ and $\ev{E^4_\nu}\simeq 396 \, T^4_\nu$.
\end{itemize}

We have checked that all the scenarios considered in this letter predict a late kinetic decoupling (in agreement with the way the collisional damping bound was derived) and that the elastic cross section calculations are  valid at low energies (see Appendix \ref{App:el} for the full expressions and their approximations at low and high energies). The elastic scattering cross section depends on the neutrino temperature $T_\nu$. The latter differs from the photon temperature after the standard neutrino decoupling; however the difference can be neglected.  We note in addition that the DM-$\nu$ interactions may modify the neutrino temperature by reheating the neutrino sector due to DM annihilations, as discussed in Section~\ref{sec:reheating}. However the difference between the neutrino and photon temperatures is bound to be small, owing to the $N_\mathrm{eff}$ constraint. Hence we have approximated the neutrino temperature to $ T_{\nu} = T_{\gamma}$ throughout this work.


    

\renewcommand{\arraystretch}{2.3}
\begin{table*}[!ht]
\resizebox{\textwidth}{!}{%
\begin{tabular}{ | c | b | P{6.3cm} | P{4cm} |}
\hline
\hspace{0.1cm} \textbf{Scenario} \hspace{0.1cm} & \hspace{0.1cm} \textbf{Lagrangian $\mathbf{(\mathcal{L}_{\rm int}})$} \hspace{0.1cm} & \hspace{0.1cm} $\mathbf{\sigma v_{\rm r}}$  \hspace{0.1cm}& \hspace{0.1cm} $\mathbf{\sigma_{\rm el}}$ \hspace{0.1cm} \\ \hline
\begin{tabular}{@{}c@{}}
Complex DM \\ Dirac Mediator \end{tabular} &\multirow{7}{*}{ $\;\; -\,g\,\chi \, \overline{N_{\rm R}} \, \nu_{\rm{L}}   \ +\ \rm{h.c.} \;\;$} & $\;\;\frac{g^4}{12 \, \pi}\frac{m^2_{\rm DM}}{(m^2_{\rm DM} \, + \, m^2_{\rm N})^2} \, v^2_{\rm CM} \;\; $ & $\;\; \frac{g^4}{32 \, \pi} \frac{m_{\rm DM}^2 \, y^2}{(m_{\rm N}^2 \, - \, m_{\rm DM}^2)^2} \;\;$\\ \cline{1-1}\cline{3-4} 
\begin{tabular}{@{}c@{}} Real DM \\ Dirac Mediator\end{tabular} & &$ \;\; \frac{4 \, g^4}{15 \, \pi}\frac{m^6_{\rm DM}}{ (m^2_{\rm DM} \, + \, m^2_{\rm N})^4} \, v^4_{\rm CM} \; \;$ &$\;\; \frac{g^4}{8 \, \pi}\frac{\, m_{\rm DM}^6 \, y^4}{(m_{\rm N}^2 \, - \, m_{\rm DM}^2)^4} \;\;$ \\ \cline{1-1}\cline{3-4}
\begin{tabular}{@{}c@{}} Complex DM \\ $\;$ Majorana Mediator \end{tabular}&  &$\;\;\frac{g^4}{16 \, \pi}\frac{m^2_{\rm N}}{(m^2_{\rm DM} \, + \, m^2_{\rm N})^2} \;\; $  & $\;\; \frac{g^4}{32 \, \pi} \frac{m_{\rm DM}^2 \, y^2}{(m_{\rm N}^2 \, - \, m_{\rm DM}^2)^2} \;\;$\\ \cline{1-1}\cline{3-4}
\begin{tabular}{@{}c@{}} Real DM \\ Majorana Mediator \end{tabular} &  & $\;\;\frac{g^4}{4 \, \pi}\frac{m^2_{\rm N}}{(m^2_{\rm DM} \, + \, m^2_{\rm N})^2} \;\; $ & $\;\; \frac{g^4}{8 \, \pi}\frac{m_{\rm DM}^6 \, y^4}{(m_{\rm N}^2 \, - \, m_{\rm DM}^2)^4} \;\;$\\ \hline
\begin{tabular}{@{}c@{}}  Dirac DM \\ Scalar Mediator \end{tabular}& \multirow{3}{*}{$\;\; -g \overline{\chi_{\rm R}} \nu_{\rm{L}} \phi \;\; +\;\; \rm{h.c.}\;\;$} & $\;\;\frac{g^4}{32 \, \pi}\frac{m^2_{\rm DM}}{(m^2_{\rm DM} \, + \, m^2_\phi)^2} \;\;$ & $\;\; \frac{g^4}{32 \, \pi}\frac{m^2_{\rm DM} \, y^2}{(m^2_{\rm DM} \, - \, m^2_\phi)^2}\;\;$\\ \cline{1-1}\cline{3-4}
\begin{tabular}{@{}c@{}}  Majorana DM \\ Scalar Mediator \end{tabular}&  & $\;\;\frac{g^4}{12 \, \pi}\frac{m^2_{\rm DM}}{(m^2_{\rm DM} \, + \, m^2_\phi)^2} \, v^2_{\rm CM} \;\;$  & $\;\;\frac{g^4}{16 \, \pi} \, \frac{m^2_{\rm DM} \, y^2}{(m^2_{\rm DM} \, - \, m^2_\phi)^2} \;\;$\\ \hline
\begin{tabular}{@{}c@{}}  Vector DM \\ Dirac Mediator \end{tabular}& \multirow{3}{*}{$\;\;-\, g\overline{N_{\rm L}}\gamma^\mu \chi_\mu \nu_{\rm{L}} \ +\ \rm{h.c.}\;\;$} & $\;\;\frac{2 \, g^4}{9 \, \pi}\frac{m^2_{\rm DM}}{ (m^2_{\rm DM} \, + \, m^2_{\rm N})^2} \;\;$ & \multirow{3}{*}{$\frac{g^4}{4 \, \pi}\frac{m^2_{\rm DM} \, y^2}{(m^2_{\rm DM} \, - \, m^2_{\rm N})^2}$}\\ \cline{1-1}\cline{3-3}
\begin{tabular}{@{}c@{}}  Vector DM \\ Majorana Mediator \end{tabular}&  & $\;\;\frac{g^4}{6 \, \pi}\frac{m^2_{\rm N}}{(m^2_{\rm DM} \, + \, m^2_{\rm N})^2}\;\;$  & \\ \hline
\begin{tabular}{@{}c@{}} Complex DM \\ Vector mediator \end{tabular} & \begin{tabular}{@{}c@{}}$\;\;-\,g_\chi Z'^\mu( (\partial_\mu \chi)\chi^\dagger-(\partial_\mu \chi)^\dagger\chi) \,\;\;$\\$\;\;-\, g_\nu\overline{\nu_{L}}\gamma^\mu Z'_\mu \nu_{\rm{L}}\;\;$\end{tabular} & $\;\;\frac{g_\chi^2 \, g_\nu^2}{3 \, \pi} \frac{m^2_{\rm DM}}{(4 \, m_{\rm DM}^2 \, - \, m_{\rm{Z}'}^2)^2} \, v^2_{\rm CM} \;\; $ & $\;\;\frac{g_\chi^2 \, g_\nu^2}{8 \, \pi}\frac{m^2_{\rm DM} \, y^2}{ m^4_{\rm{Z}'}}\;\;$\\\hline
\begin{tabular}{@{}c@{}}  Dirac DM \\ Vector Mediator \end{tabular}& \begin{tabular}{@{}c@{}}$\;\;-\,g_{\chi_{\rm{L}}} \overline{\chi_{\rm{L}}}\gamma^\mu Z'_\mu \chi_{\rm{L}} \,-\,g_{\chi_{\rm{R}}} \overline{\chi_{\rm{R}}}\gamma^\mu Z'_\mu \chi_{\rm{R}} \,\;\;$\\  $\;\;\,- \,g_\nu\overline{\nu_{L}}\gamma^\mu Z'_\mu \nu_{\rm{L}}\;\;$\end{tabular} & $\;\;\frac{g_\chi^2 \, g^2_\nu}{2 \, \pi}\frac{m^2_{\rm DM}}{(4 \, m^2_{\rm DM} \, - \, m^2_{\rm{Z}'})^2}  \;\;$ & $\,\,\frac{g_\chi^2 \, g^2_\nu}{8 \, \pi}\frac{m^2_{\rm DM} \, y^2}{m^4_{\rm{Z}'}}\;\;$\\ \hline
\begin{tabular}{@{}c@{}}  Majorana DM \\ Vector Mediator \end{tabular}& \begin{tabular}{@{}c@{}}$\;\;-\,\frac{g_{\chi}}{2} \bar{\chi}\gamma^\mu Z'_\mu\gamma^5 \chi  \,\;\;$\\  $\;\;\,- \,g_\nu\overline{\nu_{L}}\gamma^\mu Z'_\mu \nu_{\rm{L}}\;\;$\end{tabular} & $\;\;\frac{g_\chi^2 \, g^2_\nu}{12 \, \pi}\frac{m^2_{\rm DM}}{(4 \, m^2_{\rm DM} \, -\, m^2_{\rm{Z}'})^2} \, v^2_{\rm CM} \;\;$ & $\,\,\frac{3 \, g_\chi^2 \, g^2_\nu}{32 \, \pi}\frac{m^2_{\rm DM} \, y^2}{m^4_{\rm{Z}'}} \;\;$\\ \hline
\begin{tabular}{@{}c@{}}  Vector DM \\ Vector Mediator \end{tabular}& \begin{tabular}{@{}c@{}}$\;\;-\,g_{\chi} \frac{1}{2}\chi^{ \mu}\partial_\mu \chi^\nu Z'_\nu  \ +\ \rm{h.c.}  \,\;\;$\\  $\;\;\,- \,g_\nu\overline{\nu_{L}}\gamma^\mu Z'_\mu \nu_{\rm{L}}\;\;$\end{tabular} & $\;\;\frac{g_\chi^2 \, g^2_\nu}{\pi}\frac{m^2_{\rm DM}}{(4 \, m^2_{\rm DM} \, - \, m^2_{\rm{Z}'})^2}  \, v^2_{\rm CM} \;\;$ & $\,\,\frac{g_\chi^2 \, g^2_\nu}{8 \, \pi}\frac{m^2_{\rm DM} \, y^2}{m^4_{\rm{Z}'}}\;\;$\\ \hline
\end{tabular}}
\caption{This table presents the relevant terms in the Lagrangian, the  approximate expressions for the annihilation cross section and the low-energy limit of the elastic scattering for all possible scenarios that involve DM-$\nu$ interactions (12 in total). Only the leading terms in $v_{\rm{CM}}$ and $y = (s - m_{\rm {DM}}^2)/m_{\rm DM}^2 \simeq 2 \, E_{\nu}/m_{\rm DM}$ (with $s$ the usual Mandelstam variable) are presented for the thermally averaged annihilation cross section and the elastic scattering cross section, respectively. We refer the reader to Appendix~\ref{App:el} for the full expressions of the elastic scattering cross sections.}
\label{table2}
\end{table*}

\section{Results} 
\label{sec:results}

We now discuss the main features of the scenarios tabulated in Tab.~\ref{table2} and the  constraints associated with them. We will discuss the scenarios with spin-1 mediators separately from the scenarios with spin-0 and spin-1/2 mediators as they lead to  different phenomenology.

\subsection{Scenarios with scalar or fermion mediators}

\subsubsection{General considerations}

Eight out of the twelve scenarios tabulated in Tab.~\ref{table2} involve spin-0 and spin-1/2 mediators. Many share common properties, so we will articulate the discussion accordingly. In all of these eight scenarios, a left-handed neutrino couples directly to the DM candidate and the mediator must be heavier than the DM candidate to prevent DM from decaying. This stability condition excludes half the parameter space of the ($m_{\rm{DM}}$, $m_{\rm{mediator}}$) plane, as shown in Fig.~\ref{Fig2Paper2}.

In all these eight configurations, the DM annihilation cross section never involves an $s$- channel and is therefore never resonantly enhanced\footnote{In the case of a triplet scalar mediator, the annihilation cross section proceeds via a $s$- channel, but we have not considered it in the eight scenarios above.}. Furthermore, in most cases, we observe that the annihilation cross section is dominated by a velocity-independent term, except for complex scalar or Majorana DM for which it is $v^2$-suppressed and for real DM candidates, as it is $v^4$-suppressed. As expected, a velocity-suppressed cross section weakens the indirect detection constraint (since $v_{\rm r} \sim 10^{-3}~c$ in the halo), which in turns opens up the parameter space, as shown explicitly in Fig.~\ref{Fig2Paper2} (right) (see Section.~\ref{SKSec} for details).

The elastic scattering cross section associated with these scenarios depends on the square of the neutrino energy ($E_{\nu}^2$). The only exception occurs for real DM candidates in which case the elastic scattering cross section varies as $E_{\nu}^{4}$ (see Table~\ref{table2}). We note also that, for a very strong mass degeneracy ($m_{\rm mediator} - m_{\rm DM} \ll E_\nu$), the denominator of the propagator depends solely on the transferred momentum, which is similar to the neutrino energy. Consequently, the elastic cross section no longer depends on the neutrino energy~\cite{Boehm:2003hm} and can be considerably enhanced. This is shown as the region along the diagonal in Fig.~\ref{Fig2Paper2}. 

All (elastic scattering and annihilation) cross sections depend on both the DM and mediator masses, as well as the coupling $g$. One can therefore constrain both the DM and mediator masses using the collisional damping and indirect detection constraints for a fixed value of the coupling $g$, which we take to be unity in the figures for definiteness.  

\cleardoublepage

\subsubsection{Fermion DM and scalar mediator} 

Most of the scenarios listed in this section predict a similar phenomenology. For illustration purposes, we shall focus on fermion DM particles coupled to a scalar mediator. However, the discussion below can be easily extended to other scenarios.

The corresponding Lagrangian is given by
\begin{equation}  
\label{eq:LANEQ}
\mathcal{L}_{\rm{int}} \supset-\,g\,\phi \, \overline{\chi_{\rm R}} \, \nu_{\rm{L}}    \ +\ \rm{h.c.} ~,
\end{equation}
where $\chi$ is the DM and can be either a Dirac or Majorana particle. Since the neutrino is a member of an $SU(2)$ doublet, one can consider two minimal extensions of the SM which include such a coupling. First, $\chi_{\rm{R}}$ can be promoted to a $SU(2)$ doublet like in supersymmetric models~\cite{Jungman:1995df} or supersymmetry-inspired models~\cite{Boehm:2003hm}. This would constrain the DM mass to be heavier than few GeVs or even few TeVs in the presence of co-annihilations~\cite{Boehm:1999bj, Bagnaschi:2015eha}. Second, we can assume $\chi_{\rm{R}}$ to be a singlet and the scalar $\phi$ a $SU(2)$ doublet like in inert doublet models~\cite{Goudelis:2013uca}. This would also imply that the DM necessarily interacts with charged leptons, a scenario which is strongly constrained by cosmological observations, astrophysics and particle physics experiments~\cite{Kopp:2014tsa}. Therefore, such interactions would need to be suppressed, for instance by a
very heavy charged mediator~\cite{Boehm:2003hm, Vasquez:2009kq}.

In order to consider masses below the 100~GeV scale for the DM and the mediator, both fields need to be singlets. The required coupling in Eq.~(\ref{eq:LANEQ}) can then be generated via mixing with extra scalar or fermion doublets. If the mixing occurs via an extra fermion doublet $R$, the strongest constraints arise from lepton flavour violating processes at one loop and from measurements of the anomalous magnetic moments of the electron and the muon~\cite{Lindner:2010rr, Boehm:2003hm, Boehm:2007na}. On the other hand, if one introduces another scalar doublet, $\eta$, that mixes with the scalar DM singlet, $\phi$, there are tight, though model-dependent, constraints in the effective DM-$\nu$ coupling from the requirement that $2\rightarrow 2$ scalar processes must be  unitary~\cite{Akeroyd:2000wc}.

\begin{figure}[t]
  \centering   \subfloat{{\includegraphics[width=0.45\linewidth, height=7cm]{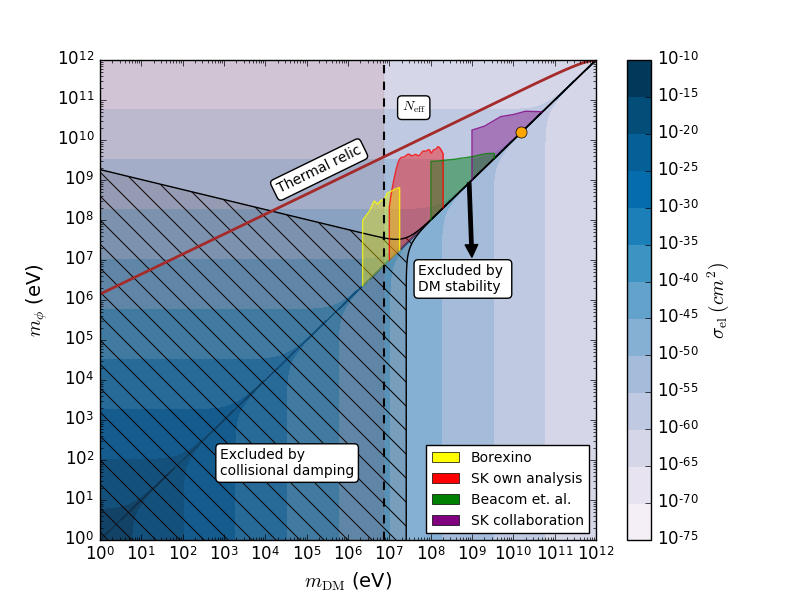} }}%
    \qquad   \subfloat{{\includegraphics[width=0.45\linewidth, height=7cm]{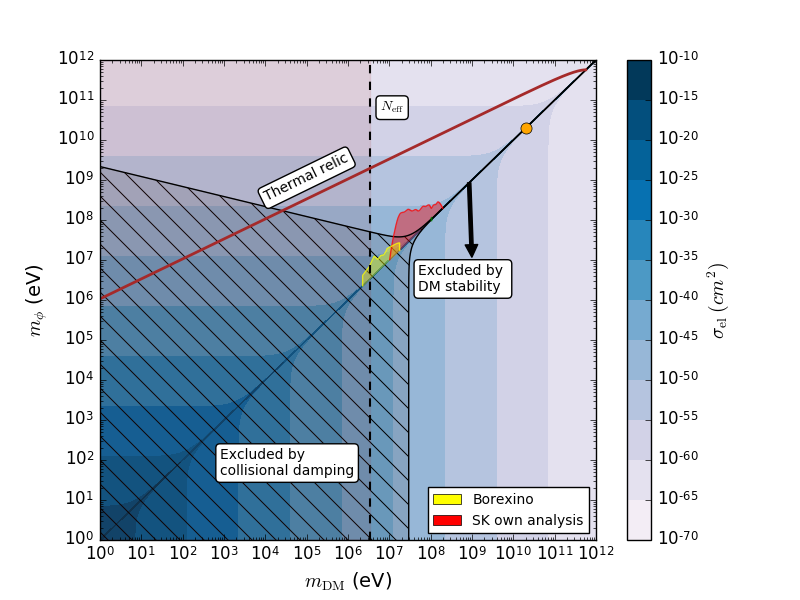} }}%
\caption{Elastic scattering of Dirac DM (left) and Majorana DM (right) coupled to a scalar mediator in the $m_{\rm{\phi}}-m_{\rm{DM}}$ plane for $g=1$. Different regions are constrained by: the collisional damping limit (dashed region and black line along the diagonal up to the orange dot), a conservative bound from the antineutrino flux at Borexino~\cite{Bellini:2010gn} (in yellow), our analysis at SuperKamiokande (SK) described in Section~\ref{SKSec} (in red), the analysis done in Ref.~\cite{Yuksel:2007ac} using results from SK, Fr\'ejus and Amanda (in green), and the analysis done by the SK collaboration for GeV neutrinos produced at the galactic centre~\cite{Frankiewicz:2015zma} (in purple). The parameters that give rise to the right relic abundance (brown line) are shown as a reference. The dashed line refers to the DM mass upper bound derived from $N_{\rm{eff}}$ in~\cite{Boehm:2013jpa, Wilkinson:2016gsy} as discussed in Section \ref{sec:reheating}. \label{Fig2Paper2}} 
\end{figure}

\paragraph{\textbf{Annihilation cross section.}}

Dirac particles annihilate via a constant cross section while the cross section is $v^2$-dependent for Majorana particles. Nevertheless, both models can explain the observed DM abundance if the value of their  annihilation cross section is of the order of $\ev{\sigma v_{\rm{r}}} \simeq 3 \times  10^{-26} \ \rm{cm^3}/\rm{s}$ and $\ev{\sigma v_{\rm{r}}} \simeq 6 \times  10^{-26} \ \rm{cm^3}/\rm{s}$, respectively, represented by the brown lines in Fig.~\ref{Fig2Paper2}. For the parameters below that line, the annihilation cross section is larger than the thermal value. Hence, $\chi$'s cannot constitute all the DM unless one invokes a different production mechanism, such as the decay of an unstable heavy particle (see Ref.~\cite{Baer:2014eja} for a recent review of non-thermal DM production mechanisms) or a regeneration mechanism~\cite{Williams:2012pz}. In contrast, configurations above the brown line over-predict the DM abundance and require, e.g., additional annihilation channels to explain the observed abundance.

\begin{figure}[t]
  \centering
    \subfloat{{\includegraphics[width=0.45\linewidth, height=7cm]{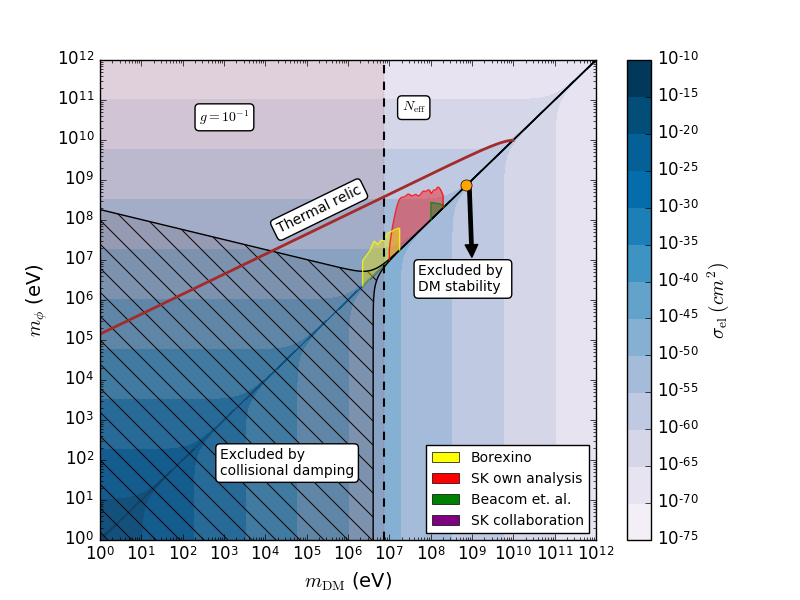} }}%
    \qquad
 \subfloat{{\includegraphics[width=0.45\linewidth, height=7cm]{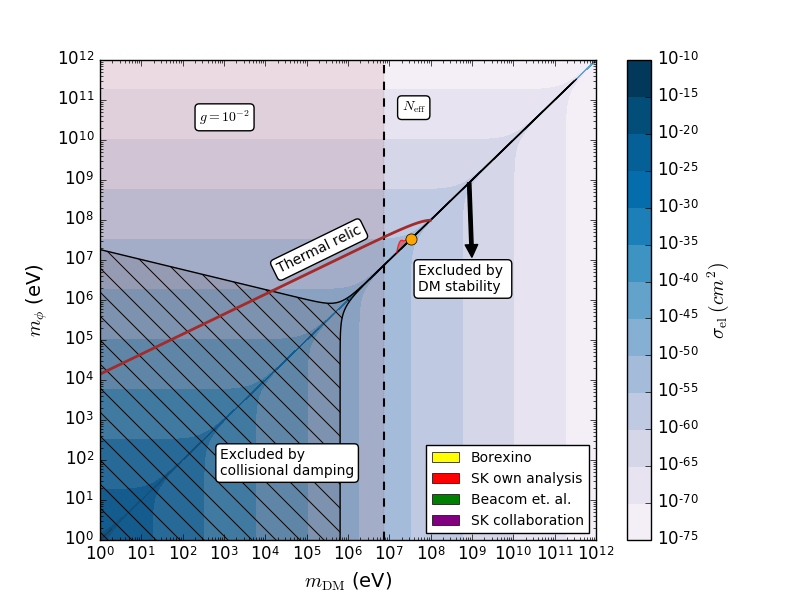} }}%
\caption{Same as Fig.~\ref{Fig2Paper2} but for $g=10^{-2}$ (left) and for $g=10^{-1}$ (right).\label{g1}}    
\end{figure}

We note that, in these fermion DM scenarios, the mediator needs to be light if the DM is weakly coupled to neutrinos, i.e., $g\ll 1$. Furthermore, the DM cannot be too heavy. Indeed the stability condition, $m_{\rm{DM}} < m_\phi$, leads to an upper limit on the DM mass of $m_{\rm{DM}} < 10^4 \ (100)$~ MeV for $g = 10^{-1} \ (10^{-2})$ (see Fig.~\ref{g1}). If we further impose the limit on $N_{\rm{eff}}$ and require thermal DM, then any coupling smaller than $g\ll 10^{-2}$ is ruled out.

As expected from the velocity-dependence of the cross section, indirect detection searches are much more sensitive to Dirac DM candidates than Majorana DM particles  (see Fig.~\ref{Fig2Paper2}). Dirac DM candidates strongly coupled to neutrinos ($g \sim 1$ and $\ev{\sigma v_{\rm r}} > 10^{-26} \ \rm{cm^3/s}$) are excluded  by a combination of low-energy neutrino detectors (such as Borexino) and high-energy experiments, including SK, even when their mass is up to $\sim 100$~GeV. As such, limits from future neutrino detectors combined with those from CMB and gamma-ray observations \cite{Boehm:2002yz,Chen:2003gz, Padmanabhan:2005es, Mapelli:2006ej, Zhang:2006fr, Ripamonti:2006gq, Chuzhoy:2007fg, Finkbeiner:2008gw, Natarajan:2008pk, Natarajan:2009bm, Galli:2009zc, Slatyer:2009yq, Cirelli:2009bb, Kanzaki:2009hf, Chluba:2009uv, Valdes:2009cq, Natarajan:2010dc, Galli:2011rz, Hutsi:2011vx, Evoli:2012zz, Giesen:2012rp, Evoli:2012qh, Slatyer:2012yq, Frey:2013wh, Cline:2013fm, Weniger:2013hja, Lopez-Honorez:2013lcm, Diamanti:2013bia, Madhavacheril:2013cna, Ade:2015xua, Slatyer:2015jla, Slatyer:2015kla, Poulin:2015pna, Kawasaki:2015peu, Liu:2016cnk, Oldengott:2016yjc, Slatyer:2016qyl} could rule out the entire thermal DM region below $\sim 100$~GeV. Note that the exclusion limit in our plot assumes that, somehow, the candidate's relic density matches the observed DM abundance, despite the large annihilation cross section into neutrinos and thus,  would require a regeneration mechanism.

The bounds derived above become significantly weaker when the value of the coupling $g$ becomes smaller (see Fig.~\ref{g1}). In fact there is no observable signal at SK (and at future neutrino detectors) when $g$ becomes smaller than $g = 10^{-1} \ (10^{-2})$ if the DM mass is a few GeV (MeV). Note however, that, since the annihilation cross section controls both the thermal relic density and the indirect detection constraints, it is always possible to test thermal DM candidates in future neutrino experiments as long as $g \geq 10^{-2}$.

\paragraph{\textbf{Elastic scattering cross section.}}

The elastic scattering cross section is similar for Majorana and Dirac DM. It reads
\begin{equation}
\sigma_{\rm{el}} \simeq 1.1 \, (2.2) \, \times 10^{-41} \  g^4 \ \left(\frac{T_\nu}{T_0} \right)^2 \ \left(\frac{m_\phi}{\rm{MeV}}\right)^{-4} \ \left(1 - \left(\frac{m_{\rm DM}}{m_\phi} \right)^2 \right)^{-2} \, \rm{cm^2,} 
\label{el1b}
\end{equation}
for Dirac (Majorana) DM candidates. The difference stems from the additional $s$- channel diagram in the Majorana case. When the DM and mediator masses are degenerated, the elastic scattering cross section in the low-energy regime becomes 
\begin{equation}
\sigma_{\rm{el}} = 1 \, (2) \times g^4 \ \frac{1}{32 \ \pi \ m^2_{\rm DM}} \simeq 4 \ (8) \times 10^{-24} \ g^4 \ \left(\frac{m_{\rm DM}}{\rm{MeV}} \right)^{-2} \, \rm{cm^2,} 
\label{el1bdeg}
\end{equation}
for Dirac (Majorana) DM candidates. Therefore, the collisional damping constraint can only exclude  masses below $\sim \mathcal{O}(10)$~GeV. In general, collisional damping bounds  require rather large values of the elastic scattering cross section, i.e., light mediators, ($m_\phi \in [{\cal{O}}(10), {\cal{O}}(10^3)]$~MeV), and light DM particles (with a mass in the sub--10~MeV range), or degenerate values of the DM and mediator masses between $m_{\rm{DM}}\sim [10, 10^4]$~MeV (for $g=1$), to enhance the elastic scattering cross section. Given the $N_{\rm{eff}}$ bound on the DM mass and the Borexino constraints, the first possibility (light DM and light mediators) is mostly excluded for any value of the coupling. The second option (degenerate masses) is ruled out by the other indirect detection searches for a large coupling ($g=1$). 

The exclusion region for fermion DM candidates weakly coupled to neutrinos (i.e., $g \ll 1$) is shown in Fig.~\ref{g1}. As one can see, the regions excluded by indirect detection searches and the collisional damping mechanism become smaller. As a result, Dirac DM candidates heavier than a few GeVs are now allowed.

One can obtain an expression for the elastic scattering cross section that is independent of the coupling $g$ by combining the elastic scattering and annihilation cross sections when $m_\phi \gg m_{\rm{DM}}$. The latter reads
\begin{equation}
\sigma_{\rm{el}} \simeq  2.6 \, (19) \times 10^{-54} \left(\frac{T_\nu}{T_0} \right)^2  \ \left(\frac{m_{\rm DM}}{\rm{MeV}}\right)^{-2}  \  \left( \frac{\ev{\sigma v_{\rm r}}}{3 \times 10^{-26} \, \rm{cm^3/s}}\right)\ \rm{cm^2} ~, \label{elCOMBINED}
\end{equation}
for Dirac (Majorana) DM, assuming $v_{\rm CM}\simeq 1/3$ at freeze-out. This expression can be used to set a lower bound on the DM mass using the collisional damping constraint in Eq.~(\ref{temperature}) and requiring DM annihilations into neutrinos to explain the observed DM abundance. As can be readily seen from Eq.~(\ref{elCOMBINED}), thermal candidates must be heavier than $m_{\rm{DM}} > 14$~keV (or $m_{\rm{DM}} > 34$~keV). These constraints are not as stringent as the limits from $N_{\rm{eff}}$ described in Section~\ref{sec:reheating}, which impose $m_{\rm{DM}} > 10$~MeV for Dirac DM and $m_{\rm{DM}} > 3.5$~MeV for Majorana DM~\cite{Boehm:2013jpa, Wilkinson:2016gsy}. However, unlike the $N_{\rm{eff}}$ constraint, the collisional damping bound remains valid in the case of asymmetric DM candidates, and also enables to constrain the mediator mass. Furthermore, it is worth recalling that we have used a conservative limit from collisional damping, which may improve with a better knowledge of the matter distribution in the early Universe and an improved understanding of the role of baryonic physics in galaxy formation.

Finally, the lower limit on the DM mass that we have found by combining the annihilation and scattering cross sections should remain the same when $g \ll 1$, because both the annihilation and elastic scattering cross sections scale in the same way with respect to the coupling $g$.

Other scenarios where the mediator is either a fermion or a scalar show a similar behaviour to the one discussed here, if no lepton number violating (LNV) process occurs. Specifically, scenarios in which a vector DM candidate is coupled to a Dirac mediator would have a very similar behaviour to the case of a Dirac DM particle coupled to a scalar mediator. Indeed, the leading term in the annihilation cross section is velocity-independent and features the same DM and mediator mass dependence. The scenario with a complex scalar DM coupled to a Dirac mediator is analogous to the Majorana DM case, and given the $v^2$ dependence of the annihilation cross section, indirect DM searches are less sensitive to it.  The only case which is somewhat different is the real scalar DM and Dirac mediator scenario, since the annihilation cross section scales as $v^4$. Therefore, finding evidence for this scenario using indirect detection searches would be very challenging. In this case, unless $m_{\rm DM}\sim m_{\rm N}$, the elastic cross section is severely suppressed, as it varies as $E_\nu^4$. Therefore, for this scenario we only expect sizable collisional damping when the mediator and DM masses are similar and in the $\sim [4,10^4]$~MeV range, i.e., above the $N_{\rm{eff}}$ bound (see Appendix~\ref{FullResults} for the relevant results of the remaining scenarios).

\subsubsection{Scalar DM and Majorana mediator} \label{sec:LNV}

When the mediator is a Majorana particle, LNV processes are allowed and change the phenomenology significantly. In fact, LNV processes may constitute the dominant annihilation channels. This is the case for example when the DM is a spin-0 or a spin-1 particle that interacts with active neutrinos through the exchange of a Majorana fermion $N_{\rm{R}}$.  We will focus here on the spin-0 DM scenario for concreteness, but similar conclusions apply to a spin-1 DM candidate. The term in the Lagrangian describing this interaction corresponds to
\begin{equation}
\mathcal{L}_{\rm{int}} \supset-\,g\,\chi \, \overline{N_{\rm{R}}} \, \nu_{\rm{L}} \, + \, \rm{h.c.} ~,   
\label{lag1}
\end{equation}
and applies regardless the DM candidate, $\chi$, is a real or complex scalar. Note that the same interaction term can also lead to neutrino masses at loop level~\cite{Boehm:2006mi, Farzan:2010mr, Farzan:2014aca, Arhrib:2015dez}. 

\paragraph{\textbf{Annihilation cross section.}}

The two dominant annihilation channels are $\chi\chi^{\ast} \rightarrow \nu \nu$ and $\chi \chi^{\ast} \rightarrow \bar{\nu} \bar{\nu}$, which violate lepton number by two units. We ignore annihilations into $\chi \chi^{\ast} \rightarrow \nu \bar{\nu}$, even though they also take place in LNV scenarios because the associated cross section is $v^2$-suppressed. This scenario provides a natural implementation of thermal light DM candidates while keeping the mass of the mediator very heavy. The annihilation cross section is proportional to 
\begin{equation} \ev{\sigma v_{\rm r}} \propto g^4 \, \frac{m_{\rm N}^2}{\left(m_{\rm DM}^2 + m_{\rm N}^2 \right)^2} \propto \frac{g^4}{m_{\rm N}^2} ~,
\end{equation} 
when $m_{\rm{N}} \gg m_{\rm{DM}}$.  The relic density does not constrain the DM mass, but only the mediator mass and the coupling $g$, satisfying the relation 
\begin{equation}
m_{\rm{N}} \simeq \mathcal{O}(1) \, g^{2} \, \left(\frac{\ev{\sigma v_{\rm r}}}{3 \times 10^{-26} \, \rm{cm^3/s}}\right)^{- \frac{1}{2}} \, \rm{TeV.}
\label{RD_mnTeV}
\end{equation}

Hence, the DM candidate could be light while the mediator could be very heavy, i.e., with a mass of a few TeVs for $g=1$. Since the leading term in the annihilation cross section is velocity-independent, we expect a copious production of neutrinos in the galactic halo. As a result, indirect detection searches set significant constraints and exclude a large fraction of the parameter space for DM particles with a mass in between $[2,10^4]$~MeV.

\paragraph{\textbf{Elastic scattering cross section.}}

The elastic scattering is mostly controlled by the mediator mass and $E_0$, and reads
\begin{equation}
\sigma_{\rm{el}} \simeq  1.2 \times 10^{-41} \  g^4 \ \left(\frac{T_\nu}{T_0} \right)^2 \ \left(\frac{m_{\rm{N}}}{\rm{MeV}}\right)^{-4}  \ \left( 1 - \left(\frac{m_{\rm{DM}}}{m_{\rm{N}}} \right)^2 \right)^{-2} \ \rm{cm^2} ~, 
\label{ELFINAL}
\end{equation}
assuming $m_{\rm{N}} \gg m_{\rm{DM}}$. As for the previous scenario, this cross section can be significantly enhanced if both the mediator and DM masses are degenerate.

Observable collisional damping effects  require either very light DM particles, $m_{\rm DM} < 2$~MeV or degenerate DM and mediator masses with values below $m_{\rm DM} < 10$~GeV. This is however excluded by the $N_{\rm{eff}}$ bound and indirect DM searches, respectively. It should be pointed out also that, for these configurations, the annihilation cross section exceeds the canonical value $\ev{\sigma v_{\rm{r}}} \simeq 3 \times 10^{-26} \, \rm{cm^3}/\rm{s}$ when $g=1$. Therefore, one may have to invoke a regeneration mechanism to explain the observed DM abundance in these scenarios. Nevertheless, this may be ruled out ultimately by indirect detection searches~\cite{Williams:2012pz}. Furthermore, if the DM candidate is a real scalar, the elastic scattering cross section scales as $E^4_\nu$ and it is therefore very suppressed. Consequently, there is no room for significant collisional damping in this case (see Appendix~\ref{FullResults}).

\begin{figure}[t]
  \centering    \subfloat{{\includegraphics[width=0.45\linewidth, height=7cm]{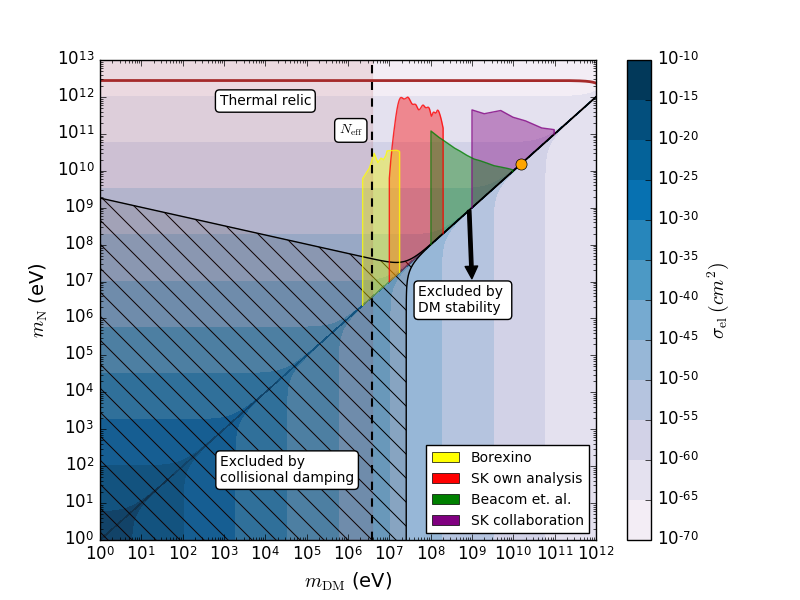} }}%
    \qquad
  \subfloat{{\includegraphics[width=0.45\linewidth, height=7cm]{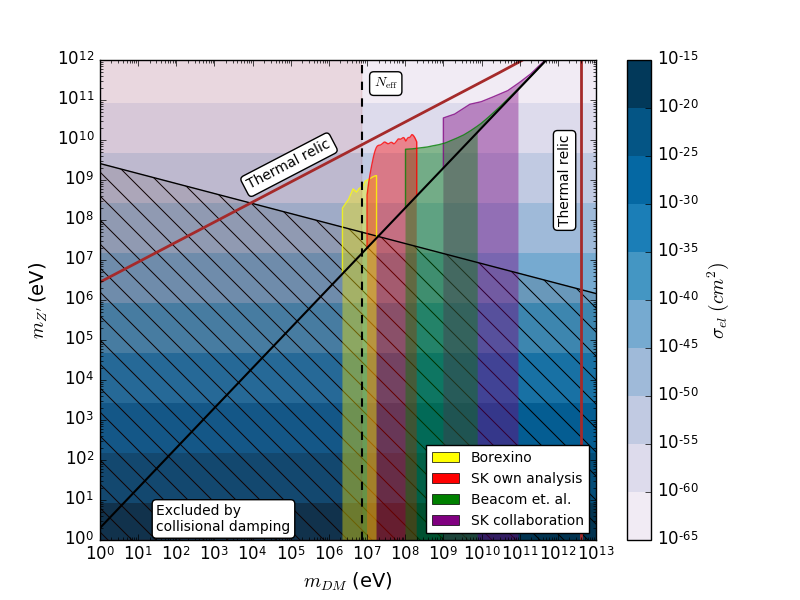} }}%
\caption{Same as Fig.~\ref{Fig2Paper2} but for complex DM with a Majorana mediator (left) and for Dirac DM with a vector mediator (right). \label{Fig5}}    
\end{figure}

\subsection{Scenarios with a vector mediator}

We now discuss scenarios where the mediator is a spin-1 particle. There are four possible Lagrangians to describe DM-$\nu$ interactions in presence of such a mediator, which are shown in Table~\ref{table2}. For concreteness, let us focus on scenarios with a spin-0 or spin-1/2 (Dirac) DM candidate. The associated Lagrangians read
\begin{equation}
\mathcal{L}_{\rm{int}} \supset \,\begin{cases}
- \,g_\nu\overline{\nu_{\rm L}}\gamma^\mu Z'_\mu \nu_{\rm{L}}\, - g_\chi Z'^\mu\big{(} (\partial_\mu \chi)\chi^\dagger-(\partial_\mu \chi)^\dagger\chi \big{)} \,, \\[3ex]
- \,g_\nu\overline{\nu_{\rm L}}\gamma^\mu Z'_\mu \nu_{\rm{L}}\, -  g_{\chi_{\rm{L},\rm{R}}} \overline{\chi_{\rm{L},\rm{R}}}\gamma^\mu Z'_\mu \chi_{\rm{L},\rm{R}} ~.
\end{cases}
\label{eq:Lvecmed}
\end{equation}

In both cases, the first term represents the coupling of the spin-1 particle to active neutrinos while the second term represents the spin-1 coupling to the DM particle. The top (bottom) line corresponds to a vector coupling to a complex scalar or Dirac DM candidate,  respectively\footnote{If DM is a Majorana particle, the coupling is chiral and $g_{\chi_{\rm{L}}} = -g_{\chi_{\rm{R}}}$, so that $g_{\rm{vector}} = \frac{g_{\rm{R}} + g_{\rm{L}}}{2} = 0$, while $g_{\rm{axial}} = \frac{g_{\rm{R}} - g_{\rm{L}}}{2} = 2 \, g_{\rm{R}}$.}. This type of interaction was initially introduced in Refs.~\cite{Boehm:2001hm, Boehm:2003hm} as an attempt to build viable models of sub-GeV DM candidates and illustrates the new collisional damping effects described in Refs.~\cite{Boehm:2001hm, Boehm:2004th}. As such, these model building efforts provided an exception to the Hut~\cite{Hut:1977zn} and Lee-Weinberg~\cite{Lee:1977ua} calculations which forbid light thermal DM candidates. 

More recently, models where the DM is coupled to a light spin-1 mediator have been proposed in the context of self-interacting DM~\cite{Tulin:2013teo} and models with both self-interactions and DM-$\nu$ interactions (all mediated by a spin-1 boson) have also been considered in Refs.~\cite{Aarssen:2012fx, Miranda:2013wla, Bringmann:2016din}. In these references, collisional damping -- that stems from early DM interactions -- is neglected and the thermal-production assumption has been relaxed.

The phenomenology of these scenarios with spin-1 mediators is different from that associated with spin-0 and spin-1/2 mediators. Firstly, the absence of a direct coupling between the DM candidate and neutrinos ensures the stability of the DM candidate. Mediators lighter than the DM are allowed (unlike for the spin-0 and spin-1/2 mediator cases). Secondly, since DM can be heavier than the mediator, DM particles can also annihilate into two spin-1 particles. This process may actually be the dominant annihilation channel, depending on the exact value of the couplings.

\paragraph{\textbf{Annihilation cross section.}}

For concreteness, we shall consider Dirac DM candidates coupled to vector boson mediators, i.e., $g_{\chi_{\rm{L}}} = g_{\chi_{\rm{R}}}$. When the mediator is heavier than the DM particles, the only possible annihilation channel is a neutrino/antineutrino pair. The associated cross section is given by
\begin{equation}
\ev{\sigma v_{\rm r}} = \frac{g_\chi^2 \, g^2_\nu}{2 \, \pi}\frac{m^2_{\rm DM}}{(4 \, m^2_{\rm DM} - m^2_{\rm{Z}'})^2} \simeq  \frac{g_\chi^2 \, g^2_\nu}{2 \, \pi}\frac{m^2_{\rm DM}}{m_{\rm{Z}'}^4} ~,
\end{equation}
which can  become resonant since it proceeds via an $s$- channel diagram. We do not illustrate the impact of the resonance on the parameter space but a smaller value of the coupling would be required to explain the observed DM abundance\footnote{In fact, using the Breit-Wigner form of the propagator, a DM mass of the order of $\mathcal{O}(100)$~TeV would be needed to explain the observed DM abundance for $g_\chi = 1$.}.

When the mediator is lighter than the DM particles, annihilations can be both i) into neutrino/antineutrino pairs, with a cross section of the order of
\begin{equation}
\ev{\sigma v_{\rm r}}_{\nu \bar{\nu}} \simeq \frac{g_\chi^2 \, g^2_\nu}{32 \, \pi}\frac{1}{m_{\rm DM}^2} ~, 
\label{nunuann}
\end{equation}
and ii) into two vector bosons, with a cross section of the order of 
\begin{equation}
\ev{\sigma v_{\rm r}}_{Z' \, Z'} = \frac{g^4_\chi}{8 \, \pi \, m^2_{\rm DM}} \, \sqrt{1 - \frac{m^2_{\rm{Z}'}}{m^2_{\rm DM}}} \simeq \frac{g^4_\chi}{8 \, \pi} \ \frac{1}{m^2_{\rm DM}} ~, 
\label{ZZann}
\end{equation}
which does not have a resonant structure since the $Z'$ are produced via a $t$- channel diagram.

Both final states eventually contribute to the relic density calculations. However one may dominate over the other one, depending on the relative strength of $g_\chi$ and $ g_\nu$ (hence the two thermal lines in Fig.~\ref{Fig5}). One expects a lower limit on the $Z'$ mass if $Z'$'s are copiously produced by DM annihilations and decay into neutrinos after the standard neutrino decoupling, as this would lead to an increase in $N_{\rm{eff}}$. To avoid such a limit, one can invoke additional $Z'$ decay channels to suppress the branching fraction into neutrinos\footnote{A dedicated analysis would be required to obtain a precise bound on the $Z'$ mass in an UV-complete model.}.

Here, we only consider the $\chi \bar{\chi} \rightarrow \nu \bar{\nu}$ channel in order to derive the constraints from indirect detection searches. For mediators lighter than the DM mass, the DM annihilation into two vector bosons could also yield a signal in neutrino detectors if the produced $Z'$'s decay into neutrino/antineutrino pairs. This signal would however generate a box-shaped energy spectrum that depends on the $Z'$ branching ratio into neutrinos and on the $m_{\rm{DM}}/m_{Z'}$ ratio~\cite{Ibarra:2012dw}. For simplicity, we do not consider this case.

Note that the Sommerfeld enhancement of the annihilation cross section when $m_{Z'} \ll m_{\rm{DM}}$ at the time of freeze-out is expected to be small and consequently, we disregard this effect\footnote{We expect order $\mathcal{O}(1)$ corrections to our relic density results~\cite{Feng:2010zp, Hannestad:2010zt}.}. However, the Sommerfeld effect might be important at late-times and it might increase the neutrino production in the galactic halo, in particular for $v^2$-dependent cross sections and DM masses above 1~TeV~\cite{Hisano:2003ec, Das:2016ced}. Nevertheless, our analysis focuses on smaller DM masses and we do not consider the effects of such enhancements, although this might rule out tiny DM mass regions between the 1--100 TeV regime, depending on the particular scenario~\cite{ElAisati:2017ppn}.

The scalar and Majorana DM case is similar except that the annihilation cross section is $v^2$-dominated and therefore suppressed with respect to the Dirac DM case. Yet, despite the $v^2$-dependent suppression, indirect detection searches rule out the parameter space where large collisional damping effects would be expected (see Appendix~\ref{FullResults}).

\paragraph{\textbf{Elastic scattering cross section.}}

The elastic scattering cross section for spin-1 mediator scenarios is independent of the DM mass. It reads
\begin{equation}
\sigma_\text{el} \simeq 4.4 \times  10^{-41} \, g_\nu^2 \, g_\chi^2 \, \left(\frac{T_\nu}{T_0} \right)^2 \,  \left(\frac{m_{\rm{Z}'}}{\rm{MeV}} \right)^{-4} \text{cm}^2 ~, 
\label{FermionVectorEl}
\end{equation}
which proceeds via a $t$- channel diagram and is proportional to $E_\nu^2$, in contrast to previous scenarios in which the cross section is energy-independent in the regime of degenerate DM and mediator masses. Moreover, in contrast to scenarios with a spin-0 and a spin-$1/2$ mediator, the $m_{Z'} < m_{\rm DM}$ region could give rise to measurable collisional damping for $m_{\rm DM} >$ few MeV. However, for constant annihilation cross sections, indirect detection constraints imply that only DM masses above $\gtrsim 100$ GeV and mediators between $[1, 10]$~MeV would produce sizable collisional damping for $g_\nu = g_\chi =1$. This is alleviated if the DM annihilation cross section is velocity-dependent (for Majorana, scalar and vector DM candidates). In such cases, collisional damping could be important for $m_{\rm DM} \sim [1, 10]$~MeV and $m_{\rm DM} \gtrsim 100$~MeV with $m_{Z'} \sim [1, 100]$~MeV. We disregard the indirect detection constraints from the $Z'$ decay into a neutrino/antineutrino pair since they are model-dependent. Moreover, for $g_\chi \sim g_\nu \ll 1 $ indirect detection constraints weaken, allowing for sizable collisional damping for $m_{\rm DM} \sim [0.4, 1]$~GeV and $\mathcal{O}(\mathrm{few})$~MeV mediators masses for $g = 10^{-1}$, while $m_{\rm DM} \gtrsim 100$~MeV and sub-MeV mediators are required for $g = 10^{-2}$ .

For a thermal DM candidate and $m_{\rm{DM}} \ll m_{Z'}$, 
\begin{equation}
\sigma_{\rm el} \simeq 7.7 \times 10^{-55} \ \left(\frac{T_\nu}{T_0} \right)^2 \ \left(\frac{m_{\rm DM}}{\rm{MeV}}\right)^{-2} \ \left(\frac{\ev{\sigma v_{\rm r}}}{3 \times 10^{-26} \  \rm{cm^3/s}}\right) \, \rm{cm^2} ~,
\label{el5COMBINED}
\end{equation}
so that, using the collisional damping and relic density constraints, we obtain a lower limit on the DM mass independent of the DM coupling to the mediator and neutrinos ($g_\chi$ and $g_\nu$, respectively). More specifically, we find $m_{\rm{DM}} \ge 9.2$~keV. This lower bound is again less constraining than the one derived by the change in $N_{\rm{eff}}$, which in turn, also excludes observable collisional damping for light DM candidates and $m_{Z'} \sim [10, 10^3]$~MeV. Nevertheless, indirect constraints could still constrain a large region of the parameter space when $g_\nu \ll 1$ if one considers the annihilation channel into a pair of $Z'$ for a strongly coupled dark sector ($g_\chi \simeq 1$).  

Finally, if the DM candidate is heavier than the mediator, to produce the correct DM relic density assuming only DM-$\nu$ interactions, Eq.~(\ref{eq:Lvecmed}), requires $m_{\rm DM} \simeq 4 \, g^2_\chi \left( 1 + \frac{1}{4}\left( \frac{g_\nu}{g_\chi}\right)^2\right)^{\frac{1}{2}}$~TeV. Therefore, in the $m_{\rm{DM}}\gg m_{Z'}$ limit and for a thermal DM candidate, the elastic cross section is
\begin{equation}
\sigma_{\rm el} \simeq 1.2 \times 10^{-47} \ g^2_\nu \left( 1 + \frac{1}{4}\left( \frac{g_\nu}{g_\chi}\right)^2\right)^{-\frac{1}{2}} \ \left(\frac{T_\nu}{T_0} \right)^2 \ \left(\frac{m_{\rm DM}}{\rm{MeV}}\right)\ \left(\frac{m_{\rm Z'}}{\rm{MeV}}\right)^{-4} \ \left(\frac{\ev{\sigma v_{\rm r}}}{3 \times 10^{-26} \  \rm{cm^3/s}}\right)^{\frac{1}{2}} \, \rm{cm^2} ~,
\end{equation}
which, when compared to the collisional damping constraint, Eq.~(\ref{temperature}), sets a lower bound in the mediator mass of $m_{Z'}\ge 2 \, g^\frac{1}{2}_\nu\left( 1 + \frac{1}{4}\left( \frac{g_\nu}{g_\chi}\right)^2\right)^{-\frac{1}{8}}$ MeV.

\section{Conclusions} 
\label{sec:conclusion}

In this paper, we have investigated the viability of scenarios in which DM is coupled to active neutrinos, by evaluating their cosmological effects (collisional damping and relic density) and their implications for indirect DM searches with neutrino detectors.

Using a simplified model approach and considering only renormalisable terms in the Lagrangian, we have identified twelve different scenarios. Many of these share some common properties and can be grouped according to the nature of the particle that mediates the interactions. For all these scenarios we have computed the elastic scattering and DM annihilation cross sections. The full expressions are given in the Appendix~\ref{App:el} while their dominant terms are given in Table~\ref{table2}. We have not explicitly assumed thermal DM. However, we do show the DM and mediator masses that lead to a thermal annihilation cross section of $\ev{\sigma v_{\text{r}}}\simeq 3 \times 10^{-26}\rm{cm}^3/\rm{s}$ (or $\ev{\sigma v_{\text{r}}}\simeq 6 \times 10^{-26}\rm{cm}^3/\rm{s}$ for $v^2$-dependent cross sections). For each of these scenarios, we constrain the parameter space by imposing the stability of the DM candidate and also that i) the DM-$\nu$ interactions are compatible with small scales Lyman-$\alpha$ forest data, ii) there are no anomalous neutrino signals at Borexino and SK experiments and finally iii) DM annihilations into neutrinos do not significantly change the CMB angular power spectrum.

We find that, generically, for scalar and fermion mediators that are much heavier than the DM particle, the annihilation cross section is either constant or velocity-dominated and scales as the square of the DM mass (except if LNV annihilation channels dominate, which occurs when the mediator is a Majorana particle). While the velocity dependence is not particularly important at the time of the DM freeze-out, it is crucial for annihilations in the Milky Way, as it significantly suppresses the neutrino signal. Therefore, only scalar DM-Majorana mediator, Dirac DM-scalar mediator and vector DM-Dirac mediator scenarios have strong indirect detection constraints. These bounds are not far from the values required for a thermal DM candidate. Hence, future neutrino experiments have the potential to significantly improve these constraints and exclude a large fraction of the thermal DM parameter space (assuming the DM annihilates into SM particles). This conclusion remains valid as long as $g \gg 10^{-2}$, since the relic density line and the indirect constraints are both proportional to $g^4$.

The elastic cross section typically scales as the neutrino temperature squared and can be resonantly enhanced if the DM and mediator masses are nearly degenerate. Observable collisional damping requires very large values of this cross section, which implies sub-MeV DM masses or the quasi-degenerate DM-mediator mass regime. The first possibility is, however, excluded by constraints from $N_{\rm{eff}}$, using CMB data~\cite{Boehm:2013jpa, Wilkinson:2016gsy}. The second case is viable, but only for velocity-dependent annihilation cross sections (scalar DM-Dirac mediator and Majorana DM-scalar mediator), so that the indirect detection bounds are weak and leave significant portions of the parameter space unconstrained, in particular, DM masses in the $\sim [100, 10^4]$~MeV range for $g=1$. The real scalar DM-Dirac mediator case is an exception, as the annihilation and the elastic scattering cross sections are suppressed by $v^4$ and $E^4_\nu$ terms respectively, which in turn might produce observable collisional damping if the mediator and DM masses are degenerate and in the $\sim [4, 10^4]$~MeV range.

If the mediator is a vector, the phenomenology is different because it can be lighter than the DM particle and moreover, when the DM and mediator particles are degenerate in mass, the annihilation cross section can be resonantly enhanced. If the DM particle is heavier than the vector mediator, the annihilation channel into two vectors is open and could dominate, depending on the parameters. The indirect DM searches apply to both mass regimes: if DM is heavier than the mediator, the constraints are similar to those obtained for the previous cases, i.e., strong and close to the thermal relic line for Dirac DM and significantly weaker for complex, Majorana and vector DM, due to the velocity dependence of the cross section. In the opposite case, i.e., DM lighter than the mediator, DM can annihilate both into neutrinos and into $Z'$, which can subsequently decay into neutrinos. Depending on the relative strength of the couplings $g_\nu$ and $g_\chi$, either of the two channels can dominate and lead to significant constraints on the parameter space. The values of the elastic cross section needed for collisional damping and to solve the missing satellite and the too-big-to-fail problems can be achieved even for heavy DM, if the mediator mass is in the $\sim [1-10]$~MeV range. 

In summary, we find that DM-$\nu$ interactions can have a strong impact on the early and present Universe and that the complementarity between cosmological and astrophysical constraints can test large areas of the allowed parameter space. These bounds  should be taken into account when considering a particular UV-complete model that generates interactions between DM and neutrinos and would be particularly relevant for models that generate neutrino masses while providing a DM candidate.

\section*{Acknowledgements}
AO would like to thank Kristian Moffat, Matheus Hostert and Alexis Plascencia for useful discussions. AO and SP are supported by the  European  Research  Council  under  ERC  Grant ``NuMass''  (FP7-IDEAS-ERC  ERC-CG  617143).  CB would like to thank the Perimeter Institute for hospitality.  This research was supported in part by Perimeter Institute for Theoretical Physics. Research at Perimeter Institute is supported by the Government of Canada through Industry Canada and by the Province of Ontario through the Ministry of Economic Development and Innovation. SPR is supported by a Ram\'on y Cajal contract, by the Spanish MINECO under grants FPA2014-54459-P and SEV-2014-0398, and by the Generalitat Valenciana under grant PROMETEOII/2014/049. SP acknowledges partial support from the Wolfson Foundation and the Royal Society and also acknowledges SISSA for support and hospitality during part of this work. The authors are also partially supported by the European Union's Horizon 2020 research and innovation program under the Marie Sk\l odowska-Curie grant agreements No. 690575 (RISE InvisiblesPlus) and 674896 (ITN Invisibles). SPR is also partially supported by the Portuguese FCT through the CFTP-FCT Unit 777 (PEst-OE/FIS/UI0777/2013).

\clearpage

\bibliography{Ref.bib}

\begin{thebibliography}{132}
\expandafter\ifx\csname natexlab\endcsname\relax\def\natexlab#1{#1}\fi
\expandafter\ifx\csname bibnamefont\endcsname\relax
  \def\bibnamefont#1{#1}\fi
\expandafter\ifx\csname bibfnamefont\endcsname\relax
  \def\bibfnamefont#1{#1}\fi
\expandafter\ifx\csname citenamefont\endcsname\relax
  \def\citenamefont#1{#1}\fi
\expandafter\ifx\csname url\endcsname\relax
  \def\url#1{\texttt{#1}}\fi
\expandafter\ifx\csname urlprefix\endcsname\relax\def\urlprefix{URL }\fi
\providecommand{\bibinfo}[2]{#2}
\providecommand{\eprint}[2][]{\url{#2}}

\bibitem[{\citenamefont{Y{\"u}ksel et~al.}(2007)\citenamefont{Y{\"u}ksel,
  Horiuchi, Beacom, and Ando}}]{Yuksel:2007ac}
\bibinfo{author}{\bibfnamefont{H.}~\bibnamefont{Y{\"u}ksel}},
  \bibinfo{author}{\bibfnamefont{S.}~\bibnamefont{Horiuchi}},
  \bibinfo{author}{\bibfnamefont{J.~F.} \bibnamefont{Beacom}},
  \bibnamefont{and} \bibinfo{author}{\bibfnamefont{S.}~\bibnamefont{Ando}},
  \bibinfo{journal}{Phys. Rev.} \textbf{\bibinfo{volume}{D76}},
  \bibinfo{pages}{123506} (\bibinfo{year}{2007}), \eprint{0707.0196}.

\bibitem[{\citenamefont{Palomares-Ruiz and
  Pascoli}(2008)}]{PalomaresRuiz:2007eu}
\bibinfo{author}{\bibfnamefont{S.}~\bibnamefont{Palomares-Ruiz}}
  \bibnamefont{and} \bibinfo{author}{\bibfnamefont{S.}~\bibnamefont{Pascoli}},
  \bibinfo{journal}{Phys. Rev.} \textbf{\bibinfo{volume}{D77}},
  \bibinfo{pages}{025025} (\bibinfo{year}{2008}), \eprint{0710.5420}.

\bibitem[{\citenamefont{Palomares-Ruiz}(2008)}]{PalomaresRuiz:2007ry}
\bibinfo{author}{\bibfnamefont{S.}~\bibnamefont{Palomares-Ruiz}},
  \bibinfo{journal}{Phys. Lett.} \textbf{\bibinfo{volume}{B665}},
  \bibinfo{pages}{50} (\bibinfo{year}{2008}), \eprint{0712.1937}.

\bibitem[{\citenamefont{Kile and Soni}(2009)}]{Kile:2009nn}
\bibinfo{author}{\bibfnamefont{J.}~\bibnamefont{Kile}} \bibnamefont{and}
  \bibinfo{author}{\bibfnamefont{A.}~\bibnamefont{Soni}},
  \bibinfo{journal}{Phys. Rev.} \textbf{\bibinfo{volume}{D80}},
  \bibinfo{pages}{115017} (\bibinfo{year}{2009}), \eprint{0908.3892}.

\bibitem[{\citenamefont{Covi et~al.}(2010)\citenamefont{Covi, Grefe, Ibarra,
  and Tran}}]{Covi:2009xn}
\bibinfo{author}{\bibfnamefont{L.}~\bibnamefont{Covi}},
  \bibinfo{author}{\bibfnamefont{M.}~\bibnamefont{Grefe}},
  \bibinfo{author}{\bibfnamefont{A.}~\bibnamefont{Ibarra}}, \bibnamefont{and}
  \bibinfo{author}{\bibfnamefont{D.}~\bibnamefont{Tran}},
  \bibinfo{journal}{JCAP} \textbf{\bibinfo{volume}{1004}}, \bibinfo{pages}{017}
  (\bibinfo{year}{2010}), \eprint{0912.3521}.

\bibitem[{\citenamefont{Primulando and Uttayarat}(2017)}]{Primulando:2017kxf}
\bibinfo{author}{\bibfnamefont{R.}~\bibnamefont{Primulando}} \bibnamefont{and}
  \bibinfo{author}{\bibfnamefont{P.}~\bibnamefont{Uttayarat}}
  (\bibinfo{year}{2017}), \eprint{1710.08567}.

\bibitem[{\citenamefont{Beacom et~al.}(2007)\citenamefont{Beacom, Bell, and
  Mack}}]{Beacom:2006tt}
\bibinfo{author}{\bibfnamefont{J.~F.} \bibnamefont{Beacom}},
  \bibinfo{author}{\bibfnamefont{N.~F.} \bibnamefont{Bell}}, \bibnamefont{and}
  \bibinfo{author}{\bibfnamefont{G.~D.} \bibnamefont{Mack}},
  \bibinfo{journal}{Phys. Rev. Lett.} \textbf{\bibinfo{volume}{99}},
  \bibinfo{pages}{231301} (\bibinfo{year}{2007}), \eprint{astro-ph/0608090}.

\bibitem[{\citenamefont{Molin{\'e} et~al.}(2015)\citenamefont{Molin{\'e},
  Ibarra, and Palomares-Ruiz}}]{Moline:2014xua}
\bibinfo{author}{\bibfnamefont{A.}~\bibnamefont{Molin{\'e}}},
  \bibinfo{author}{\bibfnamefont{A.}~\bibnamefont{Ibarra}}, \bibnamefont{and}
  \bibinfo{author}{\bibfnamefont{S.}~\bibnamefont{Palomares-Ruiz}},
  \bibinfo{journal}{JCAP} \textbf{\bibinfo{volume}{1506}}, \bibinfo{pages}{005}
  (\bibinfo{year}{2015}), \eprint{1412.4308}.

\bibitem[{\citenamefont{Molin{\'e} et~al.}(2016)\citenamefont{Molin{\'e},
  Schewtschenko, Palomares-Ruiz, B{\oe}hm, and Baugh}}]{Moline:2016fdo}
\bibinfo{author}{\bibfnamefont{A.}~\bibnamefont{Molin{\'e}}},
  \bibinfo{author}{\bibfnamefont{J.~A.} \bibnamefont{Schewtschenko}},
  \bibinfo{author}{\bibfnamefont{S.}~\bibnamefont{Palomares-Ruiz}},
  \bibinfo{author}{\bibfnamefont{C.}~\bibnamefont{B{\oe}hm}}, \bibnamefont{and}
  \bibinfo{author}{\bibfnamefont{C.~M.} \bibnamefont{Baugh}},
  \bibinfo{journal}{JCAP} \textbf{\bibinfo{volume}{1608}}, \bibinfo{pages}{069}
  (\bibinfo{year}{2016}), \eprint{1602.07282}.

\bibitem[{\citenamefont{Farzan and Palomares-Ruiz}(2014)}]{Farzan:2014gza}
\bibinfo{author}{\bibfnamefont{Y.}~\bibnamefont{Farzan}} \bibnamefont{and}
  \bibinfo{author}{\bibfnamefont{S.}~\bibnamefont{Palomares-Ruiz}},
  \bibinfo{journal}{JCAP} \textbf{\bibinfo{volume}{1406}}, \bibinfo{pages}{014}
  (\bibinfo{year}{2014}), \eprint{1401.7019}.

\bibitem[{\citenamefont{Arg{\"u}elles et~al.}(2017)\citenamefont{Arg{\"u}elles,
  Kheirandish, and Vincent}}]{Arguelles:2017atb}
\bibinfo{author}{\bibfnamefont{C.~A.} \bibnamefont{Arg{\"u}elles}},
  \bibinfo{author}{\bibfnamefont{A.}~\bibnamefont{Kheirandish}},
  \bibnamefont{and} \bibinfo{author}{\bibfnamefont{A.~C.}
  \bibnamefont{Vincent}} (\bibinfo{year}{2017}), \eprint{1703.00451}.

\bibitem[{\citenamefont{B{\oe}hm et~al.}(2005)\citenamefont{B{\oe}hm, Mathis,
  Devriendt, and Silk}}]{Boehm:2003xr}
\bibinfo{author}{\bibfnamefont{C.}~\bibnamefont{B{\oe}hm}},
  \bibinfo{author}{\bibfnamefont{H.}~\bibnamefont{Mathis}},
  \bibinfo{author}{\bibfnamefont{J.}~\bibnamefont{Devriendt}},
  \bibnamefont{and} \bibinfo{author}{\bibfnamefont{J.}~\bibnamefont{Silk}},
  \bibinfo{journal}{Mon. Not. Roy. Astron. Soc.}
  \textbf{\bibinfo{volume}{360}}, \bibinfo{pages}{282} (\bibinfo{year}{2005}),
  \eprint{astro-ph/0309652}.

\bibitem[{\citenamefont{B{\oe}hm et~al.}(2014)\citenamefont{B{\oe}hm,
  Schewtschenko, Wilkinson, Baugh, and Pascoli}}]{Boehm:2014vja}
\bibinfo{author}{\bibfnamefont{C.}~\bibnamefont{B{\oe}hm}},
  \bibinfo{author}{\bibfnamefont{J.~A.} \bibnamefont{Schewtschenko}},
  \bibinfo{author}{\bibfnamefont{R.~J.} \bibnamefont{Wilkinson}},
  \bibinfo{author}{\bibfnamefont{C.~M.} \bibnamefont{Baugh}}, \bibnamefont{and}
  \bibinfo{author}{\bibfnamefont{S.}~\bibnamefont{Pascoli}},
  \bibinfo{journal}{Mon. Not. Roy. Astron. Soc.}
  \textbf{\bibinfo{volume}{445}}, \bibinfo{pages}{L31} (\bibinfo{year}{2014}),
  \eprint{1404.7012}.

\bibitem[{\citenamefont{Schewtschenko et~al.}(2015)\citenamefont{Schewtschenko,
  Wilkinson, Baugh, B{\oe}hm, and Pascoli}}]{Schewtschenko:2014fca}
\bibinfo{author}{\bibfnamefont{J.~A.} \bibnamefont{Schewtschenko}},
  \bibinfo{author}{\bibfnamefont{R.~J.} \bibnamefont{Wilkinson}},
  \bibinfo{author}{\bibfnamefont{C.~M.} \bibnamefont{Baugh}},
  \bibinfo{author}{\bibfnamefont{C.}~\bibnamefont{B{\oe}hm}}, \bibnamefont{and}
  \bibinfo{author}{\bibfnamefont{S.}~\bibnamefont{Pascoli}},
  \bibinfo{journal}{Mon. Not. Roy. Astron. Soc.}
  \textbf{\bibinfo{volume}{449}}, \bibinfo{pages}{3587} (\bibinfo{year}{2015}),
  \eprint{1412.4905}.

\bibitem[{\citenamefont{Schewtschenko et~al.}(2016)\citenamefont{Schewtschenko,
  Baugh, Wilkinson, B{\oe}hm, Pascoli, and Sawala}}]{Schewtschenko:2015rno}
\bibinfo{author}{\bibfnamefont{J.~A.} \bibnamefont{Schewtschenko}},
  \bibinfo{author}{\bibfnamefont{C.~M.} \bibnamefont{Baugh}},
  \bibinfo{author}{\bibfnamefont{R.~J.} \bibnamefont{Wilkinson}},
  \bibinfo{author}{\bibfnamefont{C.}~\bibnamefont{B{\oe}hm}},
  \bibinfo{author}{\bibfnamefont{S.}~\bibnamefont{Pascoli}}, \bibnamefont{and}
  \bibinfo{author}{\bibfnamefont{T.}~\bibnamefont{Sawala}},
  \bibinfo{journal}{Mon. Not. Roy. Astron. Soc.}
  \textbf{\bibinfo{volume}{461}}, \bibinfo{pages}{2282} (\bibinfo{year}{2016}),
  \eprint{1512.06774}.

\bibitem[{\citenamefont{B{\oe}hm et~al.}(2001)\citenamefont{B{\oe}hm, Fayet,
  and Schaeffer}}]{Boehm:2000gq}
\bibinfo{author}{\bibfnamefont{C.}~\bibnamefont{B{\oe}hm}},
  \bibinfo{author}{\bibfnamefont{P.}~\bibnamefont{Fayet}}, \bibnamefont{and}
  \bibinfo{author}{\bibfnamefont{R.}~\bibnamefont{Schaeffer}},
  \bibinfo{journal}{Phys. Lett.} \textbf{\bibinfo{volume}{B518}},
  \bibinfo{pages}{8} (\bibinfo{year}{2001}), \eprint{astro-ph/0012504}.

\bibitem[{\citenamefont{B{\oe}hm et~al.}(2002)\citenamefont{B{\oe}hm, Riazuelo,
  Hansen, and Schaeffer}}]{Boehm:2001hm}
\bibinfo{author}{\bibfnamefont{C.}~\bibnamefont{B{\oe}hm}},
  \bibinfo{author}{\bibfnamefont{A.}~\bibnamefont{Riazuelo}},
  \bibinfo{author}{\bibfnamefont{S.~H.} \bibnamefont{Hansen}},
  \bibnamefont{and}
  \bibinfo{author}{\bibfnamefont{R.}~\bibnamefont{Schaeffer}},
  \bibinfo{journal}{Phys. Rev.} \textbf{\bibinfo{volume}{D66}},
  \bibinfo{pages}{083505} (\bibinfo{year}{2002}), \eprint{astro-ph/0112522}.

\bibitem[{\citenamefont{B{\oe}hm and Schaeffer}(2005)}]{Boehm:2004th}
\bibinfo{author}{\bibfnamefont{C.}~\bibnamefont{B{\oe}hm}} \bibnamefont{and}
  \bibinfo{author}{\bibfnamefont{R.}~\bibnamefont{Schaeffer}},
  \bibinfo{journal}{Astron. Astrophys.} \textbf{\bibinfo{volume}{438}},
  \bibinfo{pages}{419} (\bibinfo{year}{2005}), \eprint{astro-ph/0410591}.

\bibitem[{\citenamefont{Ma}(2006)}]{Ma:2006km}
\bibinfo{author}{\bibfnamefont{E.}~\bibnamefont{Ma}}, \bibinfo{journal}{Phys.
  Rev.} \textbf{\bibinfo{volume}{D73}}, \bibinfo{pages}{077301}
  (\bibinfo{year}{2006}), \eprint{hep-ph/0601225}.

\bibitem[{\citenamefont{B{\oe}hm et~al.}(2008)\citenamefont{B{\oe}hm, Farzan,
  Hambye, Palomares-Ruiz, and Pascoli}}]{Boehm:2006mi}
\bibinfo{author}{\bibfnamefont{C.}~\bibnamefont{B{\oe}hm}},
  \bibinfo{author}{\bibfnamefont{Y.}~\bibnamefont{Farzan}},
  \bibinfo{author}{\bibfnamefont{T.}~\bibnamefont{Hambye}},
  \bibinfo{author}{\bibfnamefont{S.}~\bibnamefont{Palomares-Ruiz}},
  \bibnamefont{and} \bibinfo{author}{\bibfnamefont{S.}~\bibnamefont{Pascoli}},
  \bibinfo{journal}{Phys. Rev.} \textbf{\bibinfo{volume}{D77}},
  \bibinfo{pages}{043516} (\bibinfo{year}{2008}), \eprint{hep-ph/0612228}.

\bibitem[{\citenamefont{Farzan et~al.}(2010)\citenamefont{Farzan, Pascoli, and
  Schmidt}}]{Farzan:2010mr}
\bibinfo{author}{\bibfnamefont{Y.}~\bibnamefont{Farzan}},
  \bibinfo{author}{\bibfnamefont{S.}~\bibnamefont{Pascoli}}, \bibnamefont{and}
  \bibinfo{author}{\bibfnamefont{M.~A.} \bibnamefont{Schmidt}},
  \bibinfo{journal}{JHEP} \textbf{\bibinfo{volume}{10}}, \bibinfo{pages}{111}
  (\bibinfo{year}{2010}), \eprint{1005.5323}.

\bibitem[{\citenamefont{Farzan}(2010)}]{Farzan:2010wh}
\bibinfo{author}{\bibfnamefont{Y.}~\bibnamefont{Farzan}},
  \bibinfo{journal}{Mod. Phys. Lett.} \textbf{\bibinfo{volume}{A25}},
  \bibinfo{pages}{2111} (\bibinfo{year}{2010}), \eprint{1009.1234}.

\bibitem[{\citenamefont{Lindner et~al.}(2010)\citenamefont{Lindner, Merle, and
  Niro}}]{Lindner:2010rr}
\bibinfo{author}{\bibfnamefont{M.}~\bibnamefont{Lindner}},
  \bibinfo{author}{\bibfnamefont{A.}~\bibnamefont{Merle}}, \bibnamefont{and}
  \bibinfo{author}{\bibfnamefont{V.}~\bibnamefont{Niro}},
  \bibinfo{journal}{Phys. Rev.} \textbf{\bibinfo{volume}{D82}},
  \bibinfo{pages}{123529} (\bibinfo{year}{2010}), \eprint{1005.3116}.

\bibitem[{\citenamefont{Farzan and Heeck}(2016)}]{Farzan:2016wym}
\bibinfo{author}{\bibfnamefont{Y.}~\bibnamefont{Farzan}} \bibnamefont{and}
  \bibinfo{author}{\bibfnamefont{J.}~\bibnamefont{Heeck}},
  \bibinfo{journal}{Phys. Rev.} \textbf{\bibinfo{volume}{D94}},
  \bibinfo{pages}{053010} (\bibinfo{year}{2016}), \eprint{1607.07616}.

\bibitem[{\citenamefont{B{\oe}hm and Fayet}(2004)}]{Boehm:2003hm}
\bibinfo{author}{\bibfnamefont{C.}~\bibnamefont{B{\oe}hm}} \bibnamefont{and}
  \bibinfo{author}{\bibfnamefont{P.}~\bibnamefont{Fayet}},
  \bibinfo{journal}{Nucl. Phys.} \textbf{\bibinfo{volume}{B683}},
  \bibinfo{pages}{219} (\bibinfo{year}{2004}), \eprint{hep-ph/0305261}.

\bibitem[{\citenamefont{Giacchino et~al.}(2013)\citenamefont{Giacchino,
  Lopez-Honorez, and Tytgat}}]{Giacchino:2013bta}
\bibinfo{author}{\bibfnamefont{F.}~\bibnamefont{Giacchino}},
  \bibinfo{author}{\bibfnamefont{L.}~\bibnamefont{Lopez-Honorez}},
  \bibnamefont{and} \bibinfo{author}{\bibfnamefont{M.~H.~G.}
  \bibnamefont{Tytgat}}, \bibinfo{journal}{JCAP}
  \textbf{\bibinfo{volume}{1310}}, \bibinfo{pages}{025} (\bibinfo{year}{2013}),
  \eprint{1307.6480}.

\bibitem[{\citenamefont{Wilkinson et~al.}(2014)\citenamefont{Wilkinson,
  B{\oe}hm, and Lesgourgues}}]{Wilkinson:2014ksa}
\bibinfo{author}{\bibfnamefont{R.~J.} \bibnamefont{Wilkinson}},
  \bibinfo{author}{\bibfnamefont{C.}~\bibnamefont{B{\oe}hm}}, \bibnamefont{and}
  \bibinfo{author}{\bibfnamefont{J.}~\bibnamefont{Lesgourgues}},
  \bibinfo{journal}{JCAP} \textbf{\bibinfo{volume}{1405}}, \bibinfo{pages}{011}
  (\bibinfo{year}{2014}), \eprint{1401.7597}.

\bibitem[{\citenamefont{Di~Valentino et~al.}(2017)\citenamefont{Di~Valentino,
  Boehm, Hivon, and Bouchet}}]{DiValentino:2017oaw}
\bibinfo{author}{\bibfnamefont{E.}~\bibnamefont{Di~Valentino}},
  \bibinfo{author}{\bibfnamefont{C.}~\bibnamefont{Boehm}},
  \bibinfo{author}{\bibfnamefont{E.}~\bibnamefont{Hivon}}, \bibnamefont{and}
  \bibinfo{author}{\bibfnamefont{F.~R.} \bibnamefont{Bouchet}}
  (\bibinfo{year}{2017}), \eprint{1710.02559}.

\bibitem[{\citenamefont{Mangano et~al.}(2006)\citenamefont{Mangano, Melchiorri,
  Serra, Cooray, and Kamionkowski}}]{Mangano:2006mp}
\bibinfo{author}{\bibfnamefont{G.}~\bibnamefont{Mangano}},
  \bibinfo{author}{\bibfnamefont{A.}~\bibnamefont{Melchiorri}},
  \bibinfo{author}{\bibfnamefont{P.}~\bibnamefont{Serra}},
  \bibinfo{author}{\bibfnamefont{A.}~\bibnamefont{Cooray}}, \bibnamefont{and}
  \bibinfo{author}{\bibfnamefont{M.}~\bibnamefont{Kamionkowski}},
  \bibinfo{journal}{Phys. Rev.} \textbf{\bibinfo{volume}{D74}},
  \bibinfo{pages}{043517} (\bibinfo{year}{2006}), \eprint{astro-ph/0606190}.

\bibitem[{\citenamefont{Escudero et~al.}(2015)\citenamefont{Escudero, Mena,
  Vincent, Wilkinson, and B{\oe}hm}}]{Escudero:2015yka}
\bibinfo{author}{\bibfnamefont{M.}~\bibnamefont{Escudero}},
  \bibinfo{author}{\bibfnamefont{O.}~\bibnamefont{Mena}},
  \bibinfo{author}{\bibfnamefont{A.~C.} \bibnamefont{Vincent}},
  \bibinfo{author}{\bibfnamefont{R.~J.} \bibnamefont{Wilkinson}},
  \bibnamefont{and} \bibinfo{author}{\bibfnamefont{C.}~\bibnamefont{B{\oe}hm}},
  \bibinfo{journal}{JCAP} \textbf{\bibinfo{volume}{1509}}, \bibinfo{pages}{034}
  (\bibinfo{year}{2015}), \eprint{1505.06735}.

\bibitem[{\citenamefont{Bertoni et~al.}(2015)\citenamefont{Bertoni, Ipek,
  McKeen, and Nelson}}]{Bertoni:2014mva}
\bibinfo{author}{\bibfnamefont{B.}~\bibnamefont{Bertoni}},
  \bibinfo{author}{\bibfnamefont{S.}~\bibnamefont{Ipek}},
  \bibinfo{author}{\bibfnamefont{D.}~\bibnamefont{McKeen}}, \bibnamefont{and}
  \bibinfo{author}{\bibfnamefont{A.~E.} \bibnamefont{Nelson}},
  \bibinfo{journal}{JHEP} \textbf{\bibinfo{volume}{04}}, \bibinfo{pages}{170}
  (\bibinfo{year}{2015}), \eprint{1412.3113}.

\bibitem[{\citenamefont{Fattahi et~al.}(2016)}]{Fattahi:2016nld}
\bibinfo{author}{\bibfnamefont{A.}~\bibnamefont{Fattahi}} \bibnamefont{et~al.}
  (\bibinfo{year}{2016}), \eprint{1607.06479}.

\bibitem[{\citenamefont{Enqvist et~al.}(1992)\citenamefont{Enqvist,
  Kainulainen, and Semikoz}}]{Enqvist:1991gx}
\bibinfo{author}{\bibfnamefont{K.}~\bibnamefont{Enqvist}},
  \bibinfo{author}{\bibfnamefont{K.}~\bibnamefont{Kainulainen}},
  \bibnamefont{and} \bibinfo{author}{\bibfnamefont{V.}~\bibnamefont{Semikoz}},
  \bibinfo{journal}{Nucl. Phys.} \textbf{\bibinfo{volume}{B374}},
  \bibinfo{pages}{392} (\bibinfo{year}{1992}).

\bibitem[{\citenamefont{Kolb et~al.}(1986)\citenamefont{Kolb, Turner, and
  Walker}}]{Kolb:1986nf}
\bibinfo{author}{\bibfnamefont{E.~W.} \bibnamefont{Kolb}},
  \bibinfo{author}{\bibfnamefont{M.~S.} \bibnamefont{Turner}},
  \bibnamefont{and} \bibinfo{author}{\bibfnamefont{T.~P.}
  \bibnamefont{Walker}}, \bibinfo{journal}{Phys. Rev.}
  \textbf{\bibinfo{volume}{D34}}, \bibinfo{pages}{2197} (\bibinfo{year}{1986}).

\bibitem[{\citenamefont{Serpico and Raffelt}(2004)}]{Serpico:2004nm}
\bibinfo{author}{\bibfnamefont{P.~D.} \bibnamefont{Serpico}} \bibnamefont{and}
  \bibinfo{author}{\bibfnamefont{G.~G.} \bibnamefont{Raffelt}},
  \bibinfo{journal}{Phys. Rev.} \textbf{\bibinfo{volume}{D70}},
  \bibinfo{pages}{043526} (\bibinfo{year}{2004}), \eprint{astro-ph/0403417}.

\bibitem[{\citenamefont{Ho and Scherrer}(2013)}]{Ho:2012ug}
\bibinfo{author}{\bibfnamefont{C.~M.} \bibnamefont{Ho}} \bibnamefont{and}
  \bibinfo{author}{\bibfnamefont{R.~J.} \bibnamefont{Scherrer}},
  \bibinfo{journal}{Phys. Rev.} \textbf{\bibinfo{volume}{D87}},
  \bibinfo{pages}{023505} (\bibinfo{year}{2013}), \eprint{1208.4347}.

\bibitem[{\citenamefont{Berezhiani et~al.}(2013)\citenamefont{Berezhiani,
  Dolgov, and Tkachev}}]{Berezhiani:2012ru}
\bibinfo{author}{\bibfnamefont{Z.}~\bibnamefont{Berezhiani}},
  \bibinfo{author}{\bibfnamefont{A.}~\bibnamefont{Dolgov}}, \bibnamefont{and}
  \bibinfo{author}{\bibfnamefont{I.}~\bibnamefont{Tkachev}},
  \bibinfo{journal}{JCAP} \textbf{\bibinfo{volume}{1302}}, \bibinfo{pages}{010}
  (\bibinfo{year}{2013}), \eprint{1211.4937}.

\bibitem[{\citenamefont{Nollett and Steigman}(2014)}]{Nollett:2013pwa}
\bibinfo{author}{\bibfnamefont{K.~M.} \bibnamefont{Nollett}} \bibnamefont{and}
  \bibinfo{author}{\bibfnamefont{G.}~\bibnamefont{Steigman}},
  \bibinfo{journal}{Phys. Rev.} \textbf{\bibinfo{volume}{D89}},
  \bibinfo{pages}{083508} (\bibinfo{year}{2014}), \eprint{1312.5725}.

\bibitem[{\citenamefont{Nollett and Steigman}(2015)}]{Nollett:2014lwa}
\bibinfo{author}{\bibfnamefont{K.~M.} \bibnamefont{Nollett}} \bibnamefont{and}
  \bibinfo{author}{\bibfnamefont{G.}~\bibnamefont{Steigman}},
  \bibinfo{journal}{Phys. Rev.} \textbf{\bibinfo{volume}{D91}},
  \bibinfo{pages}{083505} (\bibinfo{year}{2015}), \eprint{1411.6005}.

\bibitem[{\citenamefont{Smith et~al.}(2012)\citenamefont{Smith, Das, and
  Zahn}}]{Smith:2011es}
\bibinfo{author}{\bibfnamefont{T.~L.} \bibnamefont{Smith}},
  \bibinfo{author}{\bibfnamefont{S.}~\bibnamefont{Das}}, \bibnamefont{and}
  \bibinfo{author}{\bibfnamefont{O.}~\bibnamefont{Zahn}},
  \bibinfo{journal}{Phys. Rev.} \textbf{\bibinfo{volume}{D85}},
  \bibinfo{pages}{023001} (\bibinfo{year}{2012}), \eprint{1105.3246}.

\bibitem[{\citenamefont{Hamann et~al.}(2011)\citenamefont{Hamann, Hannestad,
  Raffelt, and Wong}}]{Hamann:2011ge}
\bibinfo{author}{\bibfnamefont{J.}~\bibnamefont{Hamann}},
  \bibinfo{author}{\bibfnamefont{S.}~\bibnamefont{Hannestad}},
  \bibinfo{author}{\bibfnamefont{G.~G.} \bibnamefont{Raffelt}},
  \bibnamefont{and} \bibinfo{author}{\bibfnamefont{Y.~Y.~Y.}
  \bibnamefont{Wong}}, \bibinfo{journal}{JCAP} \textbf{\bibinfo{volume}{1109}},
  \bibinfo{pages}{034} (\bibinfo{year}{2011}), \eprint{1108.4136}.

\bibitem[{\citenamefont{Archidiacono et~al.}(2011)\citenamefont{Archidiacono,
  Calabrese, and Melchiorri}}]{Archidiacono:2011gq}
\bibinfo{author}{\bibfnamefont{M.}~\bibnamefont{Archidiacono}},
  \bibinfo{author}{\bibfnamefont{E.}~\bibnamefont{Calabrese}},
  \bibnamefont{and}
  \bibinfo{author}{\bibfnamefont{A.}~\bibnamefont{Melchiorri}},
  \bibinfo{journal}{Phys. Rev.} \textbf{\bibinfo{volume}{D84}},
  \bibinfo{pages}{123008} (\bibinfo{year}{2011}), \eprint{1109.2767}.

\bibitem[{\citenamefont{Hamann}(2012)}]{Hamann:2011hu}
\bibinfo{author}{\bibfnamefont{J.}~\bibnamefont{Hamann}},
  \bibinfo{journal}{JCAP} \textbf{\bibinfo{volume}{1203}}, \bibinfo{pages}{021}
  (\bibinfo{year}{2012}), \eprint{1110.4271}.

\bibitem[{\citenamefont{Nollett and Holder}(2011)}]{Nollett:2011aa}
\bibinfo{author}{\bibfnamefont{K.~M.} \bibnamefont{Nollett}} \bibnamefont{and}
  \bibinfo{author}{\bibfnamefont{G.~P.} \bibnamefont{Holder}}
  (\bibinfo{year}{2011}), \eprint{1112.2683}.

\bibitem[{\citenamefont{B{\oe}hm et~al.}(2012)\citenamefont{B{\oe}hm, Dolan,
  and McCabe}}]{Boehm:2012gr}
\bibinfo{author}{\bibfnamefont{C.}~\bibnamefont{B{\oe}hm}},
  \bibinfo{author}{\bibfnamefont{M.~J.} \bibnamefont{Dolan}}, \bibnamefont{and}
  \bibinfo{author}{\bibfnamefont{C.}~\bibnamefont{McCabe}},
  \bibinfo{journal}{JCAP} \textbf{\bibinfo{volume}{1212}}, \bibinfo{pages}{027}
  (\bibinfo{year}{2012}), \eprint{1207.0497}.

\bibitem[{\citenamefont{Steigman}(2013)}]{Steigman:2013yua}
\bibinfo{author}{\bibfnamefont{G.}~\bibnamefont{Steigman}},
  \bibinfo{journal}{Phys. Rev.} \textbf{\bibinfo{volume}{D87}},
  \bibinfo{pages}{103517} (\bibinfo{year}{2013}), \eprint{1303.0049}.

\bibitem[{\citenamefont{Archidiacono et~al.}(2013)\citenamefont{Archidiacono,
  Giusarma, Melchiorri, and Mena}}]{Archidiacono:2013lva}
\bibinfo{author}{\bibfnamefont{M.}~\bibnamefont{Archidiacono}},
  \bibinfo{author}{\bibfnamefont{E.}~\bibnamefont{Giusarma}},
  \bibinfo{author}{\bibfnamefont{A.}~\bibnamefont{Melchiorri}},
  \bibnamefont{and} \bibinfo{author}{\bibfnamefont{O.}~\bibnamefont{Mena}},
  \bibinfo{journal}{Phys. Rev.} \textbf{\bibinfo{volume}{D87}},
  \bibinfo{pages}{103519} (\bibinfo{year}{2013}), \eprint{1303.0143}.

\bibitem[{\citenamefont{B{\oe}hm et~al.}(2013)\citenamefont{B{\oe}hm, Dolan,
  and McCabe}}]{Boehm:2013jpa}
\bibinfo{author}{\bibfnamefont{C.}~\bibnamefont{B{\oe}hm}},
  \bibinfo{author}{\bibfnamefont{M.~J.} \bibnamefont{Dolan}}, \bibnamefont{and}
  \bibinfo{author}{\bibfnamefont{C.}~\bibnamefont{McCabe}},
  \bibinfo{journal}{JCAP} \textbf{\bibinfo{volume}{1308}}, \bibinfo{pages}{041}
  (\bibinfo{year}{2013}), \eprint{1303.6270}.

\bibitem[{\citenamefont{Di~Valentino et~al.}(2013)\citenamefont{Di~Valentino,
  Melchiorri, and Mena}}]{DiValentino:2013qma}
\bibinfo{author}{\bibfnamefont{E.}~\bibnamefont{Di~Valentino}},
  \bibinfo{author}{\bibfnamefont{A.}~\bibnamefont{Melchiorri}},
  \bibnamefont{and} \bibinfo{author}{\bibfnamefont{O.}~\bibnamefont{Mena}},
  \bibinfo{journal}{JCAP} \textbf{\bibinfo{volume}{1311}}, \bibinfo{pages}{018}
  (\bibinfo{year}{2013}), \eprint{1304.5981}.

\bibitem[{\citenamefont{Di~Valentino et~al.}(2016)\citenamefont{Di~Valentino,
  Gariazzo, Gerbino, Giusarma, and Mena}}]{DiValentino:2016ikp}
\bibinfo{author}{\bibfnamefont{E.}~\bibnamefont{Di~Valentino}},
  \bibinfo{author}{\bibfnamefont{S.}~\bibnamefont{Gariazzo}},
  \bibinfo{author}{\bibfnamefont{M.}~\bibnamefont{Gerbino}},
  \bibinfo{author}{\bibfnamefont{E.}~\bibnamefont{Giusarma}}, \bibnamefont{and}
  \bibinfo{author}{\bibfnamefont{O.}~\bibnamefont{Mena}},
  \bibinfo{journal}{Phys. Rev.} \textbf{\bibinfo{volume}{D93}},
  \bibinfo{pages}{083523} (\bibinfo{year}{2016}), \eprint{1601.07557}.

\bibitem[{\citenamefont{Wilkinson et~al.}(2016)\citenamefont{Wilkinson,
  Vincent, B{\oe}hm, and McCabe}}]{Wilkinson:2016gsy}
\bibinfo{author}{\bibfnamefont{R.~J.} \bibnamefont{Wilkinson}},
  \bibinfo{author}{\bibfnamefont{A.~C.} \bibnamefont{Vincent}},
  \bibinfo{author}{\bibfnamefont{C.}~\bibnamefont{B{\oe}hm}}, \bibnamefont{and}
  \bibinfo{author}{\bibfnamefont{C.}~\bibnamefont{McCabe}},
  \bibinfo{journal}{Phys. Rev.} \textbf{\bibinfo{volume}{D94}},
  \bibinfo{pages}{103525} (\bibinfo{year}{2016}), \eprint{1602.01114}.

\bibitem[{\citenamefont{Frankiewicz}(2015)}]{Frankiewicz:2015zma}
\bibinfo{author}{\bibfnamefont{K.}~\bibnamefont{Frankiewicz}}
  (\bibinfo{collaboration}{Super-Kamiokande}), in
  \emph{\bibinfo{booktitle}{{Proceedings, Meeting of the APS Division of
  Particles and Fields (DPF 2015): Ann Arbor, Michigan, USA, 4-8 Aug 2015}}}
  (\bibinfo{year}{2015}), \eprint{1510.07999}.

\bibitem[{\citenamefont{Navarro et~al.}(1996)\citenamefont{Navarro, Frenk, and
  White}}]{Navarro:1995iw}
\bibinfo{author}{\bibfnamefont{J.~F.} \bibnamefont{Navarro}},
  \bibinfo{author}{\bibfnamefont{C.~S.} \bibnamefont{Frenk}}, \bibnamefont{and}
  \bibinfo{author}{\bibfnamefont{S.~D.~M.} \bibnamefont{White}},
  \bibinfo{journal}{Astrophys. J.} \textbf{\bibinfo{volume}{462}},
  \bibinfo{pages}{563} (\bibinfo{year}{1996}), \eprint{astro-ph/9508025}.

\bibitem[{\citenamefont{Kravtsov et~al.}(1998)\citenamefont{Kravtsov, Klypin,
  Bullock, and Primack}}]{Kravtsov:1997dp}
\bibinfo{author}{\bibfnamefont{A.~V.} \bibnamefont{Kravtsov}},
  \bibinfo{author}{\bibfnamefont{A.~A.} \bibnamefont{Klypin}},
  \bibinfo{author}{\bibfnamefont{J.~S.} \bibnamefont{Bullock}},
  \bibnamefont{and} \bibinfo{author}{\bibfnamefont{J.~R.}
  \bibnamefont{Primack}}, \bibinfo{journal}{Astrophys. J.}
  \textbf{\bibinfo{volume}{502}}, \bibinfo{pages}{48} (\bibinfo{year}{1998}),
  \eprint{astro-ph/9708176}.

\bibitem[{\citenamefont{Moore et~al.}(1999)\citenamefont{Moore, Quinn,
  Governato, Stadel, and Lake}}]{Moore:1999gc}
\bibinfo{author}{\bibfnamefont{B.}~\bibnamefont{Moore}},
  \bibinfo{author}{\bibfnamefont{T.~R.} \bibnamefont{Quinn}},
  \bibinfo{author}{\bibfnamefont{F.}~\bibnamefont{Governato}},
  \bibinfo{author}{\bibfnamefont{J.}~\bibnamefont{Stadel}}, \bibnamefont{and}
  \bibinfo{author}{\bibfnamefont{G.}~\bibnamefont{Lake}},
  \bibinfo{journal}{Mon. Not. Roy. Astron. Soc.}
  \textbf{\bibinfo{volume}{310}}, \bibinfo{pages}{1147} (\bibinfo{year}{1999}),
  \eprint{astro-ph/9903164}.

\bibitem[{\citenamefont{Bays et~al.}(2012)}]{Bays:2011si}
\bibinfo{author}{\bibfnamefont{K.}~\bibnamefont{Bays}} \bibnamefont{et~al.}
  (\bibinfo{collaboration}{Super-Kamiokande}), \bibinfo{journal}{Phys. Rev.}
  \textbf{\bibinfo{volume}{D85}}, \bibinfo{pages}{052007}
  (\bibinfo{year}{2012}), \eprint{1111.5031}.

\bibitem[{\citenamefont{Hosaka et~al.}(2006)}]{Hosaka:2005um}
\bibinfo{author}{\bibfnamefont{J.}~\bibnamefont{Hosaka}} \bibnamefont{et~al.}
  (\bibinfo{collaboration}{Super-Kamiokande}), \bibinfo{journal}{Phys. Rev.}
  \textbf{\bibinfo{volume}{D73}}, \bibinfo{pages}{112001}
  (\bibinfo{year}{2006}), \eprint{hep-ex/0508053}.

\bibitem[{\citenamefont{Cravens et~al.}(2008)}]{Cravens:2008aa}
\bibinfo{author}{\bibfnamefont{J.~P.} \bibnamefont{Cravens}}
  \bibnamefont{et~al.} (\bibinfo{collaboration}{Super-Kamiokande}),
  \bibinfo{journal}{Phys. Rev.} \textbf{\bibinfo{volume}{D78}},
  \bibinfo{pages}{032002} (\bibinfo{year}{2008}), \eprint{0803.4312}.

\bibitem[{\citenamefont{Abe et~al.}(2011)}]{Abe:2010hy}
\bibinfo{author}{\bibfnamefont{K.}~\bibnamefont{Abe}} \bibnamefont{et~al.}
  (\bibinfo{collaboration}{Super-Kamiokande}), \bibinfo{journal}{Phys. Rev.}
  \textbf{\bibinfo{volume}{D83}}, \bibinfo{pages}{052010}
  (\bibinfo{year}{2011}), \eprint{1010.0118}.

\bibitem[{\citenamefont{Bernal et~al.}(2013)\citenamefont{Bernal,
  Mart{\'{\i}}n-Albo, and Palomares-Ruiz}}]{Bernal:2012qh}
\bibinfo{author}{\bibfnamefont{N.}~\bibnamefont{Bernal}},
  \bibinfo{author}{\bibfnamefont{J.}~\bibnamefont{Mart{\'{\i}}n-Albo}},
  \bibnamefont{and}
  \bibinfo{author}{\bibfnamefont{S.}~\bibnamefont{Palomares-Ruiz}},
  \bibinfo{journal}{JCAP} \textbf{\bibinfo{volume}{1308}}, \bibinfo{pages}{011}
  (\bibinfo{year}{2013}), \eprint{1208.0834}.

\bibitem[{\citenamefont{Bellini et~al.}(2011)}]{Bellini:2010gn}
\bibinfo{author}{\bibfnamefont{G.}~\bibnamefont{Bellini}} \bibnamefont{et~al.}
  (\bibinfo{collaboration}{Borexino}), \bibinfo{journal}{Phys. Lett.}
  \textbf{\bibinfo{volume}{B696}}, \bibinfo{pages}{191} (\bibinfo{year}{2011}),
  \eprint{1010.0029}.

\bibitem[{\citenamefont{Diamanti et~al.}(2014)\citenamefont{Diamanti,
  Lopez-Honorez, Mena, Palomares-Ruiz, and Vincent}}]{Diamanti:2013bia}
\bibinfo{author}{\bibfnamefont{R.}~\bibnamefont{Diamanti}},
  \bibinfo{author}{\bibfnamefont{L.}~\bibnamefont{Lopez-Honorez}},
  \bibinfo{author}{\bibfnamefont{O.}~\bibnamefont{Mena}},
  \bibinfo{author}{\bibfnamefont{S.}~\bibnamefont{Palomares-Ruiz}},
  \bibnamefont{and} \bibinfo{author}{\bibfnamefont{A.~C.}
  \bibnamefont{Vincent}}, \bibinfo{journal}{JCAP}
  \textbf{\bibinfo{volume}{1402}}, \bibinfo{pages}{017} (\bibinfo{year}{2014}),
  \eprint{1308.2578}.

\bibitem[{\citenamefont{Barr et~al.}(1990)\citenamefont{Barr, Chivukula, and
  Farhi}}]{Barr:1990ca}
\bibinfo{author}{\bibfnamefont{S.~M.} \bibnamefont{Barr}},
  \bibinfo{author}{\bibfnamefont{R.~S.} \bibnamefont{Chivukula}},
  \bibnamefont{and} \bibinfo{author}{\bibfnamefont{E.}~\bibnamefont{Farhi}},
  \bibinfo{journal}{Phys. Lett.} \textbf{\bibinfo{volume}{B241}},
  \bibinfo{pages}{387} (\bibinfo{year}{1990}).

\bibitem[{\citenamefont{Barr}(1991)}]{Barr:1991qn}
\bibinfo{author}{\bibfnamefont{S.~M.} \bibnamefont{Barr}},
  \bibinfo{journal}{Phys. Rev.} \textbf{\bibinfo{volume}{D44}},
  \bibinfo{pages}{3062} (\bibinfo{year}{1991}).

\bibitem[{\citenamefont{Kaplan}(1992)}]{Kaplan:1991ah}
\bibinfo{author}{\bibfnamefont{D.~B.} \bibnamefont{Kaplan}},
  \bibinfo{journal}{Phys. Rev. Lett.} \textbf{\bibinfo{volume}{68}},
  \bibinfo{pages}{741} (\bibinfo{year}{1992}).

\bibitem[{\citenamefont{Jungman et~al.}(1996)\citenamefont{Jungman,
  Kamionkowski, and Griest}}]{Jungman:1995df}
\bibinfo{author}{\bibfnamefont{G.}~\bibnamefont{Jungman}},
  \bibinfo{author}{\bibfnamefont{M.}~\bibnamefont{Kamionkowski}},
  \bibnamefont{and} \bibinfo{author}{\bibfnamefont{K.}~\bibnamefont{Griest}},
  \bibinfo{journal}{Phys. Rept.} \textbf{\bibinfo{volume}{267}},
  \bibinfo{pages}{195} (\bibinfo{year}{1996}), \eprint{hep-ph/9506380}.

\bibitem[{\citenamefont{B{\oe}hm et~al.}(2000)\citenamefont{B{\oe}hm, Djouadi,
  and Drees}}]{Boehm:1999bj}
\bibinfo{author}{\bibfnamefont{C.}~\bibnamefont{B{\oe}hm}},
  \bibinfo{author}{\bibfnamefont{A.}~\bibnamefont{Djouadi}}, \bibnamefont{and}
  \bibinfo{author}{\bibfnamefont{M.}~\bibnamefont{Drees}},
  \bibinfo{journal}{Phys. Rev.} \textbf{\bibinfo{volume}{D62}},
  \bibinfo{pages}{035012} (\bibinfo{year}{2000}), \eprint{hep-ph/9911496}.

\bibitem[{\citenamefont{Bagnaschi et~al.}(2015)}]{Bagnaschi:2015eha}
\bibinfo{author}{\bibfnamefont{E.~A.} \bibnamefont{Bagnaschi}}
  \bibnamefont{et~al.}, \bibinfo{journal}{Eur. Phys. J.}
  \textbf{\bibinfo{volume}{C75}}, \bibinfo{pages}{500} (\bibinfo{year}{2015}),
  \eprint{1508.01173}.

\bibitem[{\citenamefont{Goudelis et~al.}(2013)\citenamefont{Goudelis, Herrmann,
  and St{\aa}l}}]{Goudelis:2013uca}
\bibinfo{author}{\bibfnamefont{A.}~\bibnamefont{Goudelis}},
  \bibinfo{author}{\bibfnamefont{B.}~\bibnamefont{Herrmann}}, \bibnamefont{and}
  \bibinfo{author}{\bibfnamefont{O.}~\bibnamefont{St{\aa}l}},
  \bibinfo{journal}{JHEP} \textbf{\bibinfo{volume}{09}}, \bibinfo{pages}{106}
  (\bibinfo{year}{2013}), \eprint{1303.3010}.

\bibitem[{\citenamefont{Kopp et~al.}(2014)\citenamefont{Kopp, Michaels, and
  Smirnov}}]{Kopp:2014tsa}
\bibinfo{author}{\bibfnamefont{J.}~\bibnamefont{Kopp}},
  \bibinfo{author}{\bibfnamefont{L.}~\bibnamefont{Michaels}}, \bibnamefont{and}
  \bibinfo{author}{\bibfnamefont{J.}~\bibnamefont{Smirnov}},
  \bibinfo{journal}{JCAP} \textbf{\bibinfo{volume}{1404}}, \bibinfo{pages}{022}
  (\bibinfo{year}{2014}), \eprint{1401.6457}.

\bibitem[{\citenamefont{Albornoz~Vasquez
  et~al.}(2011)\citenamefont{Albornoz~Vasquez, Boehm, and
  Idarraga}}]{Vasquez:2009kq}
\bibinfo{author}{\bibfnamefont{D.}~\bibnamefont{Albornoz~Vasquez}},
  \bibinfo{author}{\bibfnamefont{C.}~\bibnamefont{Boehm}}, \bibnamefont{and}
  \bibinfo{author}{\bibfnamefont{J.}~\bibnamefont{Idarraga}},
  \bibinfo{journal}{Phys. Rev.} \textbf{\bibinfo{volume}{D83}},
  \bibinfo{pages}{115017} (\bibinfo{year}{2011}), \eprint{0912.5373}.

\bibitem[{\citenamefont{B{\oe}hm and Silk}(2008)}]{Boehm:2007na}
\bibinfo{author}{\bibfnamefont{C.}~\bibnamefont{B{\oe}hm}} \bibnamefont{and}
  \bibinfo{author}{\bibfnamefont{J.}~\bibnamefont{Silk}},
  \bibinfo{journal}{Phys. Lett.} \textbf{\bibinfo{volume}{B661}},
  \bibinfo{pages}{287} (\bibinfo{year}{2008}), \eprint{0708.2768}.

\bibitem[{\citenamefont{Akeroyd et~al.}(2000)\citenamefont{Akeroyd, Arhrib, and
  Naimi}}]{Akeroyd:2000wc}
\bibinfo{author}{\bibfnamefont{A.~G.} \bibnamefont{Akeroyd}},
  \bibinfo{author}{\bibfnamefont{A.}~\bibnamefont{Arhrib}}, \bibnamefont{and}
  \bibinfo{author}{\bibfnamefont{E.-M.} \bibnamefont{Naimi}},
  \bibinfo{journal}{Phys. Lett.} \textbf{\bibinfo{volume}{B490}},
  \bibinfo{pages}{119} (\bibinfo{year}{2000}), \eprint{hep-ph/0006035}.

\bibitem[{\citenamefont{Baer et~al.}(2015)\citenamefont{Baer, Choi, Kim, and
  Roszkowski}}]{Baer:2014eja}
\bibinfo{author}{\bibfnamefont{H.}~\bibnamefont{Baer}},
  \bibinfo{author}{\bibfnamefont{K.-Y.} \bibnamefont{Choi}},
  \bibinfo{author}{\bibfnamefont{J.~E.} \bibnamefont{Kim}}, \bibnamefont{and}
  \bibinfo{author}{\bibfnamefont{L.}~\bibnamefont{Roszkowski}},
  \bibinfo{journal}{Phys. Rept.} \textbf{\bibinfo{volume}{555}},
  \bibinfo{pages}{1} (\bibinfo{year}{2015}), \eprint{1407.0017}.

\bibitem[{\citenamefont{Williams et~al.}(2012)\citenamefont{Williams, B{\oe}hm,
  West, and Albornoz~Vasquez}}]{Williams:2012pz}
\bibinfo{author}{\bibfnamefont{A.~J.} \bibnamefont{Williams}},
  \bibinfo{author}{\bibfnamefont{C.}~\bibnamefont{B{\oe}hm}},
  \bibinfo{author}{\bibfnamefont{S.~M.} \bibnamefont{West}}, \bibnamefont{and}
  \bibinfo{author}{\bibfnamefont{D.}~\bibnamefont{Albornoz~Vasquez}},
  \bibinfo{journal}{Phys. Rev.} \textbf{\bibinfo{volume}{D86}},
  \bibinfo{pages}{055018} (\bibinfo{year}{2012}), \eprint{1204.3727}.

\bibitem[{\citenamefont{Boehm et~al.}(2004)\citenamefont{Boehm, Ensslin, and
  Silk}}]{Boehm:2002yz}
\bibinfo{author}{\bibfnamefont{C.}~\bibnamefont{Boehm}},
  \bibinfo{author}{\bibfnamefont{T.~A.} \bibnamefont{Ensslin}},
  \bibnamefont{and} \bibinfo{author}{\bibfnamefont{J.}~\bibnamefont{Silk}},
  \bibinfo{journal}{J. Phys.} \textbf{\bibinfo{volume}{G30}},
  \bibinfo{pages}{279} (\bibinfo{year}{2004}), \eprint{astro-ph/0208458}.

\bibitem[{\citenamefont{Chen and Kamionkowski}(2004)}]{Chen:2003gz}
\bibinfo{author}{\bibfnamefont{X.-L.} \bibnamefont{Chen}} \bibnamefont{and}
  \bibinfo{author}{\bibfnamefont{M.}~\bibnamefont{Kamionkowski}},
  \bibinfo{journal}{Phys.Rev.} \textbf{\bibinfo{volume}{D70}},
  \bibinfo{pages}{043502} (\bibinfo{year}{2004}), \eprint{astro-ph/0310473}.

\bibitem[{\citenamefont{Padmanabhan and Finkbeiner}(2005)}]{Padmanabhan:2005es}
\bibinfo{author}{\bibfnamefont{N.}~\bibnamefont{Padmanabhan}} \bibnamefont{and}
  \bibinfo{author}{\bibfnamefont{D.~P.} \bibnamefont{Finkbeiner}},
  \bibinfo{journal}{Phys.Rev.} \textbf{\bibinfo{volume}{D72}},
  \bibinfo{pages}{023508} (\bibinfo{year}{2005}), \eprint{astro-ph/0503486}.

\bibitem[{\citenamefont{Mapelli et~al.}(2006)\citenamefont{Mapelli, Ferrara,
  and Pierpaoli}}]{Mapelli:2006ej}
\bibinfo{author}{\bibfnamefont{M.}~\bibnamefont{Mapelli}},
  \bibinfo{author}{\bibfnamefont{A.}~\bibnamefont{Ferrara}}, \bibnamefont{and}
  \bibinfo{author}{\bibfnamefont{E.}~\bibnamefont{Pierpaoli}},
  \bibinfo{journal}{Mon.Not.Roy.Astron.Soc.} \textbf{\bibinfo{volume}{369}},
  \bibinfo{pages}{1719} (\bibinfo{year}{2006}), \eprint{astro-ph/0603237}.

\bibitem[{\citenamefont{Zhang et~al.}(2006)\citenamefont{Zhang, Chen, Lei, and
  Si}}]{Zhang:2006fr}
\bibinfo{author}{\bibfnamefont{L.}~\bibnamefont{Zhang}},
  \bibinfo{author}{\bibfnamefont{X.-L.} \bibnamefont{Chen}},
  \bibinfo{author}{\bibfnamefont{Y.-A.} \bibnamefont{Lei}}, \bibnamefont{and}
  \bibinfo{author}{\bibfnamefont{Z.-G.} \bibnamefont{Si}},
  \bibinfo{journal}{Phys.Rev.} \textbf{\bibinfo{volume}{D74}},
  \bibinfo{pages}{103519} (\bibinfo{year}{2006}), \eprint{astro-ph/0603425}.

\bibitem[{\citenamefont{Ripamonti et~al.}(2007)\citenamefont{Ripamonti,
  Mapelli, and Ferrara}}]{Ripamonti:2006gq}
\bibinfo{author}{\bibfnamefont{E.}~\bibnamefont{Ripamonti}},
  \bibinfo{author}{\bibfnamefont{M.}~\bibnamefont{Mapelli}}, \bibnamefont{and}
  \bibinfo{author}{\bibfnamefont{A.}~\bibnamefont{Ferrara}},
  \bibinfo{journal}{Mon.Not.Roy.Astron.Soc.} \textbf{\bibinfo{volume}{374}},
  \bibinfo{pages}{1067} (\bibinfo{year}{2007}), \eprint{astro-ph/0606482}.

\bibitem[{\citenamefont{Chuzhoy}(2008)}]{Chuzhoy:2007fg}
\bibinfo{author}{\bibfnamefont{L.}~\bibnamefont{Chuzhoy}},
  \bibinfo{journal}{Astrophys. J.} \textbf{\bibinfo{volume}{679}},
  \bibinfo{pages}{L65} (\bibinfo{year}{2008}), \eprint{0710.1856}.

\bibitem[{\citenamefont{Finkbeiner et~al.}(2008)\citenamefont{Finkbeiner,
  Padmanabhan, and Weiner}}]{Finkbeiner:2008gw}
\bibinfo{author}{\bibfnamefont{D.~P.} \bibnamefont{Finkbeiner}},
  \bibinfo{author}{\bibfnamefont{N.}~\bibnamefont{Padmanabhan}},
  \bibnamefont{and} \bibinfo{author}{\bibfnamefont{N.}~\bibnamefont{Weiner}},
  \bibinfo{journal}{Phys.Rev.} \textbf{\bibinfo{volume}{D78}},
  \bibinfo{pages}{063530} (\bibinfo{year}{2008}), \eprint{0805.3531}.

\bibitem[{\citenamefont{Natarajan and Schwarz}(2008)}]{Natarajan:2008pk}
\bibinfo{author}{\bibfnamefont{A.}~\bibnamefont{Natarajan}} \bibnamefont{and}
  \bibinfo{author}{\bibfnamefont{D.~J.} \bibnamefont{Schwarz}},
  \bibinfo{journal}{Phys.Rev.} \textbf{\bibinfo{volume}{D78}},
  \bibinfo{pages}{103524} (\bibinfo{year}{2008}), \eprint{0805.3945}.

\bibitem[{\citenamefont{Natarajan and Schwarz}(2009)}]{Natarajan:2009bm}
\bibinfo{author}{\bibfnamefont{A.}~\bibnamefont{Natarajan}} \bibnamefont{and}
  \bibinfo{author}{\bibfnamefont{D.~J.} \bibnamefont{Schwarz}},
  \bibinfo{journal}{Phys.Rev.} \textbf{\bibinfo{volume}{D80}},
  \bibinfo{pages}{043529} (\bibinfo{year}{2009}), \eprint{0903.4485}.

\bibitem[{\citenamefont{Galli et~al.}(2009)\citenamefont{Galli, Iocco, Bertone,
  and Melchiorri}}]{Galli:2009zc}
\bibinfo{author}{\bibfnamefont{S.}~\bibnamefont{Galli}},
  \bibinfo{author}{\bibfnamefont{F.}~\bibnamefont{Iocco}},
  \bibinfo{author}{\bibfnamefont{G.}~\bibnamefont{Bertone}}, \bibnamefont{and}
  \bibinfo{author}{\bibfnamefont{A.}~\bibnamefont{Melchiorri}},
  \bibinfo{journal}{Phys.Rev.} \textbf{\bibinfo{volume}{D80}},
  \bibinfo{pages}{023505} (\bibinfo{year}{2009}), \eprint{0905.0003}.

\bibitem[{\citenamefont{Slatyer et~al.}(2009)\citenamefont{Slatyer,
  Padmanabhan, and Finkbeiner}}]{Slatyer:2009yq}
\bibinfo{author}{\bibfnamefont{T.~R.} \bibnamefont{Slatyer}},
  \bibinfo{author}{\bibfnamefont{N.}~\bibnamefont{Padmanabhan}},
  \bibnamefont{and} \bibinfo{author}{\bibfnamefont{D.~P.}
  \bibnamefont{Finkbeiner}}, \bibinfo{journal}{Phys.Rev.}
  \textbf{\bibinfo{volume}{D80}}, \bibinfo{pages}{043526}
  (\bibinfo{year}{2009}), \eprint{0906.1197}.

\bibitem[{\citenamefont{Cirelli et~al.}(2009)\citenamefont{Cirelli, Iocco, and
  Panci}}]{Cirelli:2009bb}
\bibinfo{author}{\bibfnamefont{M.}~\bibnamefont{Cirelli}},
  \bibinfo{author}{\bibfnamefont{F.}~\bibnamefont{Iocco}}, \bibnamefont{and}
  \bibinfo{author}{\bibfnamefont{P.}~\bibnamefont{Panci}},
  \bibinfo{journal}{JCAP} \textbf{\bibinfo{volume}{0910}}, \bibinfo{pages}{009}
  (\bibinfo{year}{2009}), \eprint{0907.0719}.

\bibitem[{\citenamefont{Kanzaki et~al.}(2010)\citenamefont{Kanzaki, Kawasaki,
  and Nakayama}}]{Kanzaki:2009hf}
\bibinfo{author}{\bibfnamefont{T.}~\bibnamefont{Kanzaki}},
  \bibinfo{author}{\bibfnamefont{M.}~\bibnamefont{Kawasaki}}, \bibnamefont{and}
  \bibinfo{author}{\bibfnamefont{K.}~\bibnamefont{Nakayama}},
  \bibinfo{journal}{Prog.Theor.Phys.} \textbf{\bibinfo{volume}{123}},
  \bibinfo{pages}{853} (\bibinfo{year}{2010}), \eprint{0907.3985}.

\bibitem[{\citenamefont{Chluba}(2010)}]{Chluba:2009uv}
\bibinfo{author}{\bibfnamefont{J.}~\bibnamefont{Chluba}},
  \bibinfo{journal}{Mon.Not.Roy.Astron.Soc.} \textbf{\bibinfo{volume}{402}},
  \bibinfo{pages}{1195} (\bibinfo{year}{2010}), \eprint{0910.3663}.

\bibitem[{\citenamefont{Valdes et~al.}(2010)\citenamefont{Valdes, Evoli, and
  Ferrara}}]{Valdes:2009cq}
\bibinfo{author}{\bibfnamefont{M.}~\bibnamefont{Valdes}},
  \bibinfo{author}{\bibfnamefont{C.}~\bibnamefont{Evoli}}, \bibnamefont{and}
  \bibinfo{author}{\bibfnamefont{A.}~\bibnamefont{Ferrara}},
  \bibinfo{journal}{Mon.Not.Roy.Astron.Soc.} \textbf{\bibinfo{volume}{404}},
  \bibinfo{pages}{1569} (\bibinfo{year}{2010}), \eprint{0911.1125}.

\bibitem[{\citenamefont{Natarajan and Schwarz}(2010)}]{Natarajan:2010dc}
\bibinfo{author}{\bibfnamefont{A.}~\bibnamefont{Natarajan}} \bibnamefont{and}
  \bibinfo{author}{\bibfnamefont{D.~J.} \bibnamefont{Schwarz}},
  \bibinfo{journal}{Phys.Rev.} \textbf{\bibinfo{volume}{D81}},
  \bibinfo{pages}{123510} (\bibinfo{year}{2010}), \eprint{1002.4405}.

\bibitem[{\citenamefont{Galli et~al.}(2011)\citenamefont{Galli, Iocco, Bertone,
  and Melchiorri}}]{Galli:2011rz}
\bibinfo{author}{\bibfnamefont{S.}~\bibnamefont{Galli}},
  \bibinfo{author}{\bibfnamefont{F.}~\bibnamefont{Iocco}},
  \bibinfo{author}{\bibfnamefont{G.}~\bibnamefont{Bertone}}, \bibnamefont{and}
  \bibinfo{author}{\bibfnamefont{A.}~\bibnamefont{Melchiorri}},
  \bibinfo{journal}{Phys.Rev.} \textbf{\bibinfo{volume}{D84}},
  \bibinfo{pages}{027302} (\bibinfo{year}{2011}), \eprint{1106.1528}.

\bibitem[{\citenamefont{Hutsi et~al.}(2011)\citenamefont{Hutsi, Chluba, Hektor,
  and Raidal}}]{Hutsi:2011vx}
\bibinfo{author}{\bibfnamefont{G.}~\bibnamefont{Hutsi}},
  \bibinfo{author}{\bibfnamefont{J.}~\bibnamefont{Chluba}},
  \bibinfo{author}{\bibfnamefont{A.}~\bibnamefont{Hektor}}, \bibnamefont{and}
  \bibinfo{author}{\bibfnamefont{M.}~\bibnamefont{Raidal}},
  \bibinfo{journal}{Astron.Astrophys.} \textbf{\bibinfo{volume}{535}},
  \bibinfo{pages}{A26} (\bibinfo{year}{2011}), \eprint{1103.2766}.

\bibitem[{\citenamefont{Evoli et~al.}(2012)\citenamefont{Evoli, Valdes,
  Ferrara, and Yoshida}}]{Evoli:2012zz}
\bibinfo{author}{\bibfnamefont{C.}~\bibnamefont{Evoli}},
  \bibinfo{author}{\bibfnamefont{M.}~\bibnamefont{Valdes}},
  \bibinfo{author}{\bibfnamefont{A.}~\bibnamefont{Ferrara}}, \bibnamefont{and}
  \bibinfo{author}{\bibfnamefont{N.}~\bibnamefont{Yoshida}},
  \bibinfo{journal}{Mon.Not.Roy.Astron.Soc.} \textbf{\bibinfo{volume}{422}},
  \bibinfo{pages}{420} (\bibinfo{year}{2012}).

\bibitem[{\citenamefont{Giesen et~al.}(2012)\citenamefont{Giesen, Lesgourgues,
  Audren, and Ali-Haimoud}}]{Giesen:2012rp}
\bibinfo{author}{\bibfnamefont{G.}~\bibnamefont{Giesen}},
  \bibinfo{author}{\bibfnamefont{J.}~\bibnamefont{Lesgourgues}},
  \bibinfo{author}{\bibfnamefont{B.}~\bibnamefont{Audren}}, \bibnamefont{and}
  \bibinfo{author}{\bibfnamefont{Y.}~\bibnamefont{Ali-Haimoud}},
  \bibinfo{journal}{JCAP} \textbf{\bibinfo{volume}{1212}}, \bibinfo{pages}{008}
  (\bibinfo{year}{2012}), \eprint{1209.0247}.

\bibitem[{\citenamefont{Evoli et~al.}(2013)\citenamefont{Evoli, Pandolfi, and
  Ferrara}}]{Evoli:2012qh}
\bibinfo{author}{\bibfnamefont{C.}~\bibnamefont{Evoli}},
  \bibinfo{author}{\bibfnamefont{S.}~\bibnamefont{Pandolfi}}, \bibnamefont{and}
  \bibinfo{author}{\bibfnamefont{A.}~\bibnamefont{Ferrara}},
  \bibinfo{journal}{Mon. Not. Roy. Astron. Soc.}
  \textbf{\bibinfo{volume}{433}}, \bibinfo{pages}{1736} (\bibinfo{year}{2013}),
  \eprint{1210.6845}.

\bibitem[{\citenamefont{Slatyer}(2013)}]{Slatyer:2012yq}
\bibinfo{author}{\bibfnamefont{T.~R.} \bibnamefont{Slatyer}},
  \bibinfo{journal}{Phys. Rev.} \textbf{\bibinfo{volume}{D87}},
  \bibinfo{pages}{123513} (\bibinfo{year}{2013}), \eprint{1211.0283}.

\bibitem[{\citenamefont{Frey and Reid}(2013)}]{Frey:2013wh}
\bibinfo{author}{\bibfnamefont{A.~R.} \bibnamefont{Frey}} \bibnamefont{and}
  \bibinfo{author}{\bibfnamefont{N.~B.} \bibnamefont{Reid}},
  \bibinfo{journal}{Phys. Rev.} \textbf{\bibinfo{volume}{D87}},
  \bibinfo{pages}{103508} (\bibinfo{year}{2013}), \eprint{1301.0819}.

\bibitem[{\citenamefont{Cline and Scott}(2013)}]{Cline:2013fm}
\bibinfo{author}{\bibfnamefont{J.~M.} \bibnamefont{Cline}} \bibnamefont{and}
  \bibinfo{author}{\bibfnamefont{P.}~\bibnamefont{Scott}},
  \bibinfo{journal}{JCAP} \textbf{\bibinfo{volume}{1303}}, \bibinfo{pages}{044}
  (\bibinfo{year}{2013}), \bibinfo{note}{[Erratum: JCAP1305,E01(2013)]},
  \eprint{1301.5908}.

\bibitem[{\citenamefont{Weniger et~al.}(2013)\citenamefont{Weniger, Serpico,
  Iocco, and Bertone}}]{Weniger:2013hja}
\bibinfo{author}{\bibfnamefont{C.}~\bibnamefont{Weniger}},
  \bibinfo{author}{\bibfnamefont{P.~D.} \bibnamefont{Serpico}},
  \bibinfo{author}{\bibfnamefont{F.}~\bibnamefont{Iocco}}, \bibnamefont{and}
  \bibinfo{author}{\bibfnamefont{G.}~\bibnamefont{Bertone}},
  \bibinfo{journal}{Phys. Rev.} \textbf{\bibinfo{volume}{D87}},
  \bibinfo{pages}{123008} (\bibinfo{year}{2013}), \eprint{1303.0942}.

\bibitem[{\citenamefont{Lopez-Honorez et~al.}(2013)\citenamefont{Lopez-Honorez,
  Mena, Palomares-Ruiz, and Vincent}}]{Lopez-Honorez:2013lcm}
\bibinfo{author}{\bibfnamefont{L.}~\bibnamefont{Lopez-Honorez}},
  \bibinfo{author}{\bibfnamefont{O.}~\bibnamefont{Mena}},
  \bibinfo{author}{\bibfnamefont{S.}~\bibnamefont{Palomares-Ruiz}},
  \bibnamefont{and} \bibinfo{author}{\bibfnamefont{A.~C.}
  \bibnamefont{Vincent}}, \bibinfo{journal}{JCAP}
  \textbf{\bibinfo{volume}{1307}}, \bibinfo{pages}{046} (\bibinfo{year}{2013}),
  \eprint{1303.5094}.

\bibitem[{\citenamefont{Madhavacheril et~al.}(2014)\citenamefont{Madhavacheril,
  Sehgal, and Slatyer}}]{Madhavacheril:2013cna}
\bibinfo{author}{\bibfnamefont{M.~S.} \bibnamefont{Madhavacheril}},
  \bibinfo{author}{\bibfnamefont{N.}~\bibnamefont{Sehgal}}, \bibnamefont{and}
  \bibinfo{author}{\bibfnamefont{T.~R.} \bibnamefont{Slatyer}},
  \bibinfo{journal}{Phys. Rev.} \textbf{\bibinfo{volume}{D89}},
  \bibinfo{pages}{103508} (\bibinfo{year}{2014}), \eprint{1310.3815}.

\bibitem[{\citenamefont{Ade et~al.}(2016)}]{Ade:2015xua}
\bibinfo{author}{\bibfnamefont{P.~A.~R.} \bibnamefont{Ade}}
  \bibnamefont{et~al.} (\bibinfo{collaboration}{Planck}),
  \bibinfo{journal}{Astron. Astrophys.} \textbf{\bibinfo{volume}{594}},
  \bibinfo{pages}{A13} (\bibinfo{year}{2016}), \eprint{1502.01589}.

\bibitem[{\citenamefont{Slatyer}(2016{\natexlab{a}})}]{Slatyer:2015jla}
\bibinfo{author}{\bibfnamefont{T.~R.} \bibnamefont{Slatyer}},
  \bibinfo{journal}{Phys. Rev.} \textbf{\bibinfo{volume}{D93}},
  \bibinfo{pages}{023527} (\bibinfo{year}{2016}{\natexlab{a}}),
  \eprint{1506.03811}.

\bibitem[{\citenamefont{Slatyer}(2016{\natexlab{b}})}]{Slatyer:2015kla}
\bibinfo{author}{\bibfnamefont{T.~R.} \bibnamefont{Slatyer}},
  \bibinfo{journal}{Phys. Rev.} \textbf{\bibinfo{volume}{D93}},
  \bibinfo{pages}{023521} (\bibinfo{year}{2016}{\natexlab{b}}),
  \eprint{1506.03812}.

\bibitem[{\citenamefont{Poulin et~al.}(2015)\citenamefont{Poulin, Serpico, and
  Lesgourgues}}]{Poulin:2015pna}
\bibinfo{author}{\bibfnamefont{V.}~\bibnamefont{Poulin}},
  \bibinfo{author}{\bibfnamefont{P.~D.} \bibnamefont{Serpico}},
  \bibnamefont{and}
  \bibinfo{author}{\bibfnamefont{J.}~\bibnamefont{Lesgourgues}},
  \bibinfo{journal}{JCAP} \textbf{\bibinfo{volume}{1512}}, \bibinfo{pages}{041}
  (\bibinfo{year}{2015}), \eprint{1508.01370}.

\bibitem[{\citenamefont{Kawasaki et~al.}(2016)\citenamefont{Kawasaki, Nakayama,
  and Sekiguchi}}]{Kawasaki:2015peu}
\bibinfo{author}{\bibfnamefont{M.}~\bibnamefont{Kawasaki}},
  \bibinfo{author}{\bibfnamefont{K.}~\bibnamefont{Nakayama}}, \bibnamefont{and}
  \bibinfo{author}{\bibfnamefont{T.}~\bibnamefont{Sekiguchi}},
  \bibinfo{journal}{Phys. Lett.} \textbf{\bibinfo{volume}{B756}},
  \bibinfo{pages}{212} (\bibinfo{year}{2016}), \eprint{1512.08015}.

\bibitem[{\citenamefont{Liu et~al.}(2016)\citenamefont{Liu, Slatyer, and
  Zavala}}]{Liu:2016cnk}
\bibinfo{author}{\bibfnamefont{H.}~\bibnamefont{Liu}},
  \bibinfo{author}{\bibfnamefont{T.~R.} \bibnamefont{Slatyer}},
  \bibnamefont{and} \bibinfo{author}{\bibfnamefont{J.}~\bibnamefont{Zavala}},
  \bibinfo{journal}{Phys. Rev.} \textbf{\bibinfo{volume}{D94}},
  \bibinfo{pages}{063507} (\bibinfo{year}{2016}), \eprint{1604.02457}.

\bibitem[{\citenamefont{Oldengott et~al.}(2016)\citenamefont{Oldengott,
  Boriero, and Schwarz}}]{Oldengott:2016yjc}
\bibinfo{author}{\bibfnamefont{I.~M.} \bibnamefont{Oldengott}},
  \bibinfo{author}{\bibfnamefont{D.}~\bibnamefont{Boriero}}, \bibnamefont{and}
  \bibinfo{author}{\bibfnamefont{D.~J.} \bibnamefont{Schwarz}},
  \bibinfo{journal}{JCAP} \textbf{\bibinfo{volume}{1608}}, \bibinfo{pages}{054}
  (\bibinfo{year}{2016}), \eprint{1605.03928}.

\bibitem[{\citenamefont{Slatyer and Wu}(2017)}]{Slatyer:2016qyl}
\bibinfo{author}{\bibfnamefont{T.~R.} \bibnamefont{Slatyer}} \bibnamefont{and}
  \bibinfo{author}{\bibfnamefont{C.-L.} \bibnamefont{Wu}},
  \bibinfo{journal}{Phys. Rev.} \textbf{\bibinfo{volume}{D95}},
  \bibinfo{pages}{023010} (\bibinfo{year}{2017}), \eprint{1610.06933}.

\bibitem[{\citenamefont{Farzan}(2015)}]{Farzan:2014aca}
\bibinfo{author}{\bibfnamefont{Y.}~\bibnamefont{Farzan}},
  \bibinfo{journal}{JHEP} \textbf{\bibinfo{volume}{05}}, \bibinfo{pages}{029}
  (\bibinfo{year}{2015}), \eprint{1412.6283}.

\bibitem[{\citenamefont{Arhrib et~al.}(2016)\citenamefont{Arhrib, B{\oe}hm, Ma,
  and Yuan}}]{Arhrib:2015dez}
\bibinfo{author}{\bibfnamefont{A.}~\bibnamefont{Arhrib}},
  \bibinfo{author}{\bibfnamefont{C.}~\bibnamefont{B{\oe}hm}},
  \bibinfo{author}{\bibfnamefont{E.}~\bibnamefont{Ma}}, \bibnamefont{and}
  \bibinfo{author}{\bibfnamefont{T.-C.} \bibnamefont{Yuan}},
  \bibinfo{journal}{JCAP} \textbf{\bibinfo{volume}{1604}}, \bibinfo{pages}{049}
  (\bibinfo{year}{2016}), \eprint{1512.08796}.

\bibitem[{\citenamefont{Hut}(1977)}]{Hut:1977zn}
\bibinfo{author}{\bibfnamefont{P.}~\bibnamefont{Hut}}, \bibinfo{journal}{Phys.
  Lett.} \textbf{\bibinfo{volume}{69B}}, \bibinfo{pages}{85}
  (\bibinfo{year}{1977}).

\bibitem[{\citenamefont{Lee and Weinberg}(1977)}]{Lee:1977ua}
\bibinfo{author}{\bibfnamefont{B.~W.} \bibnamefont{Lee}} \bibnamefont{and}
  \bibinfo{author}{\bibfnamefont{S.}~\bibnamefont{Weinberg}},
  \bibinfo{journal}{Phys. Rev. Lett.} \textbf{\bibinfo{volume}{39}},
  \bibinfo{pages}{165} (\bibinfo{year}{1977}).

\bibitem[{\citenamefont{Tulin et~al.}(2013)\citenamefont{Tulin, Yu, and
  Zurek}}]{Tulin:2013teo}
\bibinfo{author}{\bibfnamefont{S.}~\bibnamefont{Tulin}},
  \bibinfo{author}{\bibfnamefont{H.-B.} \bibnamefont{Yu}}, \bibnamefont{and}
  \bibinfo{author}{\bibfnamefont{K.~M.} \bibnamefont{Zurek}},
  \bibinfo{journal}{Phys. Rev.} \textbf{\bibinfo{volume}{D87}},
  \bibinfo{pages}{115007} (\bibinfo{year}{2013}), \eprint{1302.3898}.

\bibitem[{\citenamefont{van~den Aarssen et~al.}(2012)\citenamefont{van~den
  Aarssen, Bringmann, and Pfrommer}}]{Aarssen:2012fx}
\bibinfo{author}{\bibfnamefont{L.~G.} \bibnamefont{van~den Aarssen}},
  \bibinfo{author}{\bibfnamefont{T.}~\bibnamefont{Bringmann}},
  \bibnamefont{and} \bibinfo{author}{\bibfnamefont{C.}~\bibnamefont{Pfrommer}},
  \bibinfo{journal}{Phys. Rev. Lett.} \textbf{\bibinfo{volume}{109}},
  \bibinfo{pages}{231301} (\bibinfo{year}{2012}), \eprint{1205.5809}.

\bibitem[{\citenamefont{Miranda et~al.}(2015)\citenamefont{Miranda, Moura, and
  Parada}}]{Miranda:2013wla}
\bibinfo{author}{\bibfnamefont{O.~G.} \bibnamefont{Miranda}},
  \bibinfo{author}{\bibfnamefont{C.~A.} \bibnamefont{Moura}}, \bibnamefont{and}
  \bibinfo{author}{\bibfnamefont{A.}~\bibnamefont{Parada}},
  \bibinfo{journal}{Phys. Lett.} \textbf{\bibinfo{volume}{B744}},
  \bibinfo{pages}{55} (\bibinfo{year}{2015}), \eprint{1308.1408}.

\bibitem[{\citenamefont{Bringmann et~al.}(2017)\citenamefont{Bringmann,
  Kahlhoefer, Schmidt-Hoberg, and Walia}}]{Bringmann:2016din}
\bibinfo{author}{\bibfnamefont{T.}~\bibnamefont{Bringmann}},
  \bibinfo{author}{\bibfnamefont{F.}~\bibnamefont{Kahlhoefer}},
  \bibinfo{author}{\bibfnamefont{K.}~\bibnamefont{Schmidt-Hoberg}},
  \bibnamefont{and} \bibinfo{author}{\bibfnamefont{P.}~\bibnamefont{Walia}},
  \bibinfo{journal}{Phys. Rev. Lett.} \textbf{\bibinfo{volume}{118}},
  \bibinfo{pages}{141802} (\bibinfo{year}{2017}), \eprint{1612.00845}.

\bibitem[{\citenamefont{Ibarra et~al.}(2012)\citenamefont{Ibarra, Lopez~Gehler,
  and Pato}}]{Ibarra:2012dw}
\bibinfo{author}{\bibfnamefont{A.}~\bibnamefont{Ibarra}},
  \bibinfo{author}{\bibfnamefont{S.}~\bibnamefont{Lopez~Gehler}},
  \bibnamefont{and} \bibinfo{author}{\bibfnamefont{M.}~\bibnamefont{Pato}},
  \bibinfo{journal}{JCAP} \textbf{\bibinfo{volume}{1207}}, \bibinfo{pages}{043}
  (\bibinfo{year}{2012}), \eprint{1205.0007}.

\bibitem[{\citenamefont{Feng et~al.}(2010)\citenamefont{Feng, Kaplinghat, and
  Yu}}]{Feng:2010zp}
\bibinfo{author}{\bibfnamefont{J.~L.} \bibnamefont{Feng}},
  \bibinfo{author}{\bibfnamefont{M.}~\bibnamefont{Kaplinghat}},
  \bibnamefont{and} \bibinfo{author}{\bibfnamefont{H.-B.} \bibnamefont{Yu}},
  \bibinfo{journal}{Phys. Rev.} \textbf{\bibinfo{volume}{D82}},
  \bibinfo{pages}{083525} (\bibinfo{year}{2010}), \eprint{1005.4678}.

\bibitem[{\citenamefont{Hannestad and Tram}(2011)}]{Hannestad:2010zt}
\bibinfo{author}{\bibfnamefont{S.}~\bibnamefont{Hannestad}} \bibnamefont{and}
  \bibinfo{author}{\bibfnamefont{T.}~\bibnamefont{Tram}},
  \bibinfo{journal}{JCAP} \textbf{\bibinfo{volume}{1101}}, \bibinfo{pages}{016}
  (\bibinfo{year}{2011}), \eprint{1008.1511}.

\bibitem[{\citenamefont{Hisano et~al.}(2004)\citenamefont{Hisano, Matsumoto,
  and Nojiri}}]{Hisano:2003ec}
\bibinfo{author}{\bibfnamefont{J.}~\bibnamefont{Hisano}},
  \bibinfo{author}{\bibfnamefont{S.}~\bibnamefont{Matsumoto}},
  \bibnamefont{and} \bibinfo{author}{\bibfnamefont{M.~M.}
  \bibnamefont{Nojiri}}, \bibinfo{journal}{Phys. Rev. Lett.}
  \textbf{\bibinfo{volume}{92}}, \bibinfo{pages}{031303}
  (\bibinfo{year}{2004}), \eprint{hep-ph/0307216}.

\bibitem[{\citenamefont{Das and Dasgupta}(2017)}]{Das:2016ced}
\bibinfo{author}{\bibfnamefont{A.}~\bibnamefont{Das}} \bibnamefont{and}
  \bibinfo{author}{\bibfnamefont{B.}~\bibnamefont{Dasgupta}},
  \bibinfo{journal}{Phys. Rev. Lett.} \textbf{\bibinfo{volume}{118}},
  \bibinfo{pages}{251101} (\bibinfo{year}{2017}), \eprint{1611.04606}.

\bibitem[{\citenamefont{El~Aisati et~al.}(2017)\citenamefont{El~Aisati,
  Garcia-Cely, Hambye, and Vanderheyden}}]{ElAisati:2017ppn}
\bibinfo{author}{\bibfnamefont{C.}~\bibnamefont{El~Aisati}},
  \bibinfo{author}{\bibfnamefont{C.}~\bibnamefont{Garcia-Cely}},
  \bibinfo{author}{\bibfnamefont{T.}~\bibnamefont{Hambye}}, \bibnamefont{and}
  \bibinfo{author}{\bibfnamefont{L.}~\bibnamefont{Vanderheyden}},
  \bibinfo{journal}{JCAP} \textbf{\bibinfo{volume}{1710}}, \bibinfo{pages}{021}
  (\bibinfo{year}{2017}), \eprint{1706.06600}.

\bibitem[{\citenamefont{Wells}(1994)}]{Wells:1994qy}
\bibinfo{author}{\bibfnamefont{J.~D.} \bibnamefont{Wells}}
  (\bibinfo{year}{1994}), \eprint{hep-ph/9404219}.

\bibitem[{\citenamefont{Berlin and Blinov}(2017)}]{Berlin:2017ftj}
\bibinfo{author}{\bibfnamefont{A.}~\bibnamefont{Berlin}} \bibnamefont{and}
  \bibinfo{author}{\bibfnamefont{N.}~\bibnamefont{Blinov}}
  (\bibinfo{year}{2017}), \eprint{1706.07046}.

\bibitem[{\citenamefont{Denner et~al.}(1992)\citenamefont{Denner, Eck, Hahn,
  and Kublbeck}}]{Denner:1992vza}
\bibinfo{author}{\bibfnamefont{A.}~\bibnamefont{Denner}},
  \bibinfo{author}{\bibfnamefont{H.}~\bibnamefont{Eck}},
  \bibinfo{author}{\bibfnamefont{O.}~\bibnamefont{Hahn}}, \bibnamefont{and}
  \bibinfo{author}{\bibfnamefont{J.}~\bibnamefont{Kublbeck}},
  \bibinfo{journal}{Nucl. Phys.} \textbf{\bibinfo{volume}{B387}},
  \bibinfo{pages}{467} (\bibinfo{year}{1992}).

\bibitem[{\citenamefont{Haber and Kane}(1985)}]{Haber:1984rc}
\bibinfo{author}{\bibfnamefont{H.~E.} \bibnamefont{Haber}} \bibnamefont{and}
  \bibinfo{author}{\bibfnamefont{G.~L.} \bibnamefont{Kane}},
  \bibinfo{journal}{Phys. Rept.} \textbf{\bibinfo{volume}{117}},
  \bibinfo{pages}{75} (\bibinfo{year}{1985}).

\bibitem[{\citenamefont{Berestetskii et~al.}(1971)\citenamefont{Berestetskii,
  Lifshitz, and Pitaevskii}}]{Landau}
\bibinfo{author}{\bibnamefont{Berestetskii}},
  \bibinfo{author}{\bibnamefont{Lifshitz}}, \bibnamefont{and}
  \bibinfo{author}{\bibnamefont{Pitaevskii}},
  \emph{\bibinfo{title}{{Relativistic Quantum Theory}}},
  vol.~\bibinfo{volume}{4} of \emph{\bibinfo{series}{{Course of Theoretical
  Physics}}} (\bibinfo{publisher}{Pergamon}, \bibinfo{year}{1971}).

\bibitem[{\citenamefont{Servant and Tait}(2002)}]{Servant:2002hb}
\bibinfo{author}{\bibfnamefont{G.}~\bibnamefont{Servant}} \bibnamefont{and}
  \bibinfo{author}{\bibfnamefont{T.~M.~P.} \bibnamefont{Tait}},
  \bibinfo{journal}{New J. Phys.} \textbf{\bibinfo{volume}{4}},
  \bibinfo{pages}{99} (\bibinfo{year}{2002}), \eprint{hep-ph/0209262}.

\bibitem[{\citenamefont{Profumo et~al.}(2006)\citenamefont{Profumo, Sigurdson,
  and Kamionkowski}}]{Profumo:2006bv}
\bibinfo{author}{\bibfnamefont{S.}~\bibnamefont{Profumo}},
  \bibinfo{author}{\bibfnamefont{K.}~\bibnamefont{Sigurdson}},
  \bibnamefont{and}
  \bibinfo{author}{\bibfnamefont{M.}~\bibnamefont{Kamionkowski}},
  \bibinfo{journal}{Phys. Rev. Lett.} \textbf{\bibinfo{volume}{97}},
  \bibinfo{pages}{031301} (\bibinfo{year}{2006}), \eprint{astro-ph/0603373}.

\end{thebibliography}
\vspace{1cm}

\appendix

\section{Annihilation cross section}

The thermally averaged annihilation cross section, multiplied by the relative DM velocity $v_\text{r}$, has been calculated for each scenario to order $\mathcal{O}(v^2_{\rm{CM}})$ in the center-of-mass (CM) velocity. All our calculations agree with the results presented in Appendix 9.1 of Ref.~\cite{Boehm:2003hm} in the limit of $m_f \rightarrow 0$ and $C_lC_r\rightarrow 0$, which in their notation, corresponds to massless left-handed neutrinos and right-handed antineutrinos in the final state, i.e., massless chiral fermions in the final state. 

The annihilation cross section can be written as
\begin{equation}
\sigma v_{\rm r} = \frac{1}{16 \ \pi \ s} \ \abs{\mathcal{M}}^2 = \frac{1}{64 \ \pi \ m^2_{\rm DM}} \ \abs{\mathcal{M}}^2 \ -  \ \frac{v^2_{\rm{CM}}}{64 \ \pi \ m^2_{\rm DM}} \, \abs{\mathcal{M}}^2 + \mathcal{O}\left(v^4_{\rm{CM}}\right) ~,  
\label{Eq1}
\end{equation}
where $s$ is the Mandelstam variable and $\abs{\mathcal{M}}^2$ is the squared amplitude averaged over initial state spins and summed over final state spins, integrated over the solid angle.

The expansions of the Mandelstam variables in the CM frame for massless neutrinos are given by
\begin{widetext}
\begin{align}
s = & \, 4 \, m^2_{\rm DM} \, (1 + v^2_{\rm CM}) + \mathcal{O}\left(v^4_{\rm CM}\right) ~, \nonumber \\
t = & - m^2_{\rm DM} \, (1 + 2 \, v^2_{\rm CM}) + 2 \, m^2_{\rm DM} \, \cos\theta v_{\rm CM} \, (1 + \, v^2_{\rm CM}) + \mathcal{O}\left(v^4_{\rm CM}\right) ~, \nonumber \\
u = & - m^2_{\rm DM} \, (1 + 2 \, v^2_{\rm CM}) - 2 \, m^2_{\rm DM} \, \cos\theta v_{\rm CM} \, \left(1 + v^2_{\rm CM}\right) + \mathcal{O}\left(v^4_{\rm CM}\right) ~, \nonumber \\
(s - m_{\rm M}^2)^{-2} = & \, (4 \, m^2_{\rm DM} - m^2_{\rm M})^{-2} - 8 \, m^2_{\rm DM} \, v^2_{\rm CM} \, (4 \, m^2_{\rm DM} - m^2_{\rm M})^{-3}  + \mathcal{O}\left(v^4_{\rm CM}\right) ~, \nonumber \\
(t - m_{\rm M}^2)^{-2} = & \, (- m^2_{\rm DM} - m^2_{\rm M})^{-2} -\frac{4 \, m^2_{\rm DM} \, v_{\rm CM} \, \cos\theta}{(- m^2_{\rm DM} -m^2_{\rm M})^{3}} + \, \frac{4 \, m^2_{\rm DM} \, v^2_{\rm CM}}{(- m^2_{\rm DM} - m^2_{\rm M})^{3}} + \frac{12 \, m^4_{\rm DM} \, \cos^2\theta \, v^2_{\rm CM}}{(- m^2_{\rm DM} - m^2_{\rm M})^4} + \mathcal{O}\left(v^3_{\rm CM}\right) ~, \nonumber \\
(u - m_{\rm M}^2)^{-2} = & \, (- m^2_{\rm DM} - m^2_{\rm M})^{-2} + \frac{4 \, m^2_{\rm DM} \, v_{\rm CM} \, \cos\theta}{(- m^2_{\rm DM} -m^2_{\rm M})^{3}} + \, \frac{4 \, m^2_{\rm DM} \, v^2_{\rm CM}}{(- m^2_{\rm DM} -m^2_{\rm M})^{3}} + \frac{12 \, m^4_{\rm DM} \, \cos^2\theta \, v^2_{\rm CM}}{(- m^2_{\rm DM} - m^2_{\rm M})^4} + \mathcal{O}\left(v^3_{\rm CM}\right) ~, \nonumber \\
(t - m_{\rm M}^2)^{-1} \, (u - m_{\rm M}^2)^{-1} = & \, (- m^2_{\rm DM} -m^2_{\rm M})^{-2} + \frac{4 \, m^2_{\rm DM} \, v^2_{\rm CM}}{(- m^2_{\rm DM} - m^2_{\rm M})^3} + \frac{4 \, m^4_{\rm DM} \, v^2_{\rm CM} \, \cos^2\theta}{(- m^2_{\rm DM} - m^2_{\rm M})^4} + \mathcal{O}\left(v^3_{\rm CM}\right) ~, 
\end{align}
\end{widetext}
where $m_{\rm{M}}$ denotes the mass of the mediator. These expressions are in agreement with Ref.~\cite{Boehm:2003hm, Wells:1994qy} in the limit of massless chiral outgoing particles. 

By expanding the cross section as
\begin{equation}
\sigma v_{\rm r} = a + b \, v^2_{\rm CM} + d \, v^4_{\rm CM} ~, 
\end{equation}
we can take the thermal average so that 
\begin{equation}
\ev{\sigma v_{\rm r}} = a + \frac{9}{4} \, b \, v^2_{\rm CM} + \frac{135}{32} \, d v^4_{\rm CM} ~, 
\end{equation}
where equipartition of energy in the non-relativistic limit is assumed~\cite{Berlin:2017ftj}.

In the calculation of the averaged square amplitude, in scenarios with a vector mediator, the momentum-dependent term in the propagator vanishes by virtue of the Dirac equation for massless particles and hence, the massive spin-1 propagator reduces to $\frac{i(g_{\mu\nu} - p_\mu p_\nu/m^2_{\rm{Z}'})}{p^2 - m^2_{\rm{Z}'}} \rightarrow \frac{i \, g_{\mu\nu}}{p^2 - m^2_{\rm{Z}'}}$. In addition, all the calculations involving Majorana fermions are performed using the Majorana Feynman rules~\cite{Denner:1992vza, Haber:1984rc}.

\section{Elastic scattering cross section}
\label{App:el} 

Without loss of generality, in the calculations for the elastic scattering, we assume the velocity of the incident DM particle to be along the $z$-axis. The incoming neutrino is then denoted by $p_1$ whereas the outgoing neutrino and DM particles are labeled as $k_1$ and $k_2$, respectively. The geometry of the system is then
\begin{align}
p_1 = &E_{\nu_1}(1, \sin\theta \, \cos\phi, \sin\phi \, \sin\theta,\cos\theta) ~, \nonumber\\
p_2 = &E_{\rm DM_1}(1, 0, 0, \beta) ~, \nonumber\\
k_1 = &E_{\nu_2}(1, \sin\theta' \, \cos\phi', \sin\phi' \, \sin\theta', \cos\theta') ~,  \nonumber\\
k_2 = &p_1 + p_2 - k_1 ~, 
\label{4mom} 
\end{align}
where $\beta = \frac{\abs{\vec{p}_{\rm DM_1}}}{E_{\rm DM_1}}$. Moreover, $\theta$ is the angle between the incident neutrino and the incoming DM particle, $\theta'$ is the angle between the incident DM particle and the outgoing neutrino. In addition, $\phi$ and $\phi'$ are the angles between the direction of the incoming DM particle and the incoming neutrino, and the angle between the direction of the incoming DM particle and the outgoing neutrino in the $x-y$ plane, respectively. 

From the 4-momentum expressions obtained in Eq.~(\ref{4mom}), we can define the Mandelstam variables as
\begin{align}
s = &(p_1 + p_2)^2 = m_{\rm DM}^2 + 2 \, (p_1 \cdot p_2) = 
  m^2_{\rm DM} + 2 \, E_{\rm DM_1} \, E_{\nu_1}(1-\beta \, \mu) \label{s}  ~, \nonumber\\
u = &(p_1 - k_2)^2 = (p_2 - k_1)^2 = m_{\rm DM}^2 - 2 \, (p_2 \cdot k_1) = m^2_{\rm DM} - 2 \, E_{\rm DM_1} \, E_{\nu_2}(1-\beta\mu') ~, \nonumber\\
t = &(p_1 - k_1)^2 = - 2 \, (p_1 \cdot k_1) = - 2 \, E_{\nu_1} \, E_{\nu_2} \, (1 - \Delta(\mu,\mu')) ~, \nonumber \\ 
t = &(p_2 - k_2)^2 = 2 \, m_{\rm DM}^2 - 2 \, (p_2 \cdot k_2) = 2 \, E_{\rm DM_1} \, \left[E_{\nu_2} \, (1 - \beta\mu') - E_{\nu_1} \, (1 - \beta\mu)\right] ~,
\end{align}
where $E_{\nu_1/\nu_2}$ and $E_{\rm DM_1/\rm DM_2}$ refer to the incoming/outgoing neutrino and DM energies, respectively, $\mu \equiv \cos\theta$ and $\mu' \equiv \cos\theta'$. Finally, $\Delta(\mu,\mu')$ is the cosine of the angle between the incoming and outgoing neutrinos and it is related to the other two cosines as
\begin{equation}
\Delta(\mu, \mu') = \mu \, \mu' + \sqrt{1 - \mu^2} \, \sqrt{1 - \mu'^2} \, \cos(\phi - \phi') ~, 
\end{equation}
with the definitions of $\phi$ and $\phi'$ given above.

In order to calculate the elastic scattering, we start from the Lorentz invariant cross section in the limit where one of the incoming particles is massless (the neutrino in this case):
\begin{equation}
\frac{d\sigma_{\rm el}}{dt} = \frac{1}{16 \, \pi \, (s - m_{\rm{DM}}^2)^2} \, \abs{\mathcal{M}(t,s)}^2 ~, 
\end{equation}
with $t \leq 0$. One can express $t$ in terms of $s$ and $u$, and make a change of variables $u = 2 \, m_{\rm{DM}}^2 - s - t$ for a given initial energy of the incoming particles (so that $s$ is fixed). We then get
\begin{equation}
\frac{d\sigma_{\rm el}}{du} = -\frac{1}{16 \, \pi \, (s - m_{\rm{DM}}^2)^2} \, \abs{\mathcal{M}(u,s)}^2 ~,
\label{eq:dsdu}
\end{equation}
where $2 \, m^2_{\rm{DM}} - s \leq u \leq \frac{m^4_{\rm{DM}}}{s}$,  since for $t\leq 0$, $s \, u\leq m^4_{\rm{DM}}$ and $s\geq m^2_{\rm{DM}}$. These relations are frame independent and can be entirely derived from conservation laws when one of the incoming particles is massless~\cite[Chapter VII ]{Landau}. Note that the minus sign comes from the change of variable.

In order to further simplify the analytic calculations, we make another change of variables, $y = \frac{(s - m^2_{\rm DM})}{m^2_{\rm DM}}$ and $w = \frac{(m^2_{\rm DM} - u)}{m^2_{\rm DM}}$. We then integrate Eq.~(\ref{eq:dsdu}) with integration limits $\frac{y}{(y + 1)}\leq w \leq y$, so that we get the full cross section in terms of $y$. Note that using Eq.~(\ref{s}), $y = \frac{2 \, E_{\rm DM} \, E_{\nu} \, (1 - \beta \,  \mu)}{m^2_{\rm DM}}$ and so, at low energies $\beta \rightarrow 0$ and $y\sim \frac{2 \, E_{\nu}}{m_{\rm DM}}$, which we use to calculate our results. By keeping a general frame, one can quickly read off the results for different frames of reference. For example, in the lab frame, $y$ also reduces to $y\sim \frac{2 \, E_{\nu}}{m_{\rm DM}}$, since $p_{\rm DM} = 0$ and $E_{\rm DM} = m_{\rm DM}$. Similarly, in the CM frame, $\mu = -1$ and so, $y = \frac{2 \, E_{\rm DM} \, E_{\nu} \, (1 + \beta \, \mu)}{m^2_{\rm DM}}$. We now proceed to give the full elastic scattering expressions for the different scenarios.

\subsection{Scalar DM coupled to a fermion mediator}

The full elastic scattering cross section for a complex DM coupled to a Dirac mediator is given by
\begin{equation}
\sigma_{\rm el} = \frac{g^4}{16 \, \pi} \, \left(\frac{y + 1}{m^2_{\rm{DM}} \, y^2} \, \log\left[1 + \frac{m^2_{\rm{DM}} \, y^2}{m^2_{\rm{N}} \, (1 + y) - m^2_{\rm{DM}}}\right] \, - \right.  \left. \frac{1}{m^2_{\rm{DM}} \, (y - 1) + m^2_{\rm{N}}}\right) ~, 
\label{ElCDMDiM}
\end{equation}
so that, in the low $E_\nu$ limit, it has the form
\begin{equation}
\sigma_{\rm el} = \frac{g^4 \, m_{\rm{DM}}^2 \, y^2}{32 \, \pi \, (m_{\rm{N}}^2 - m_{\rm{DM}}^2)^2} + \mathcal{O}\left(y^3\right) \propto E^2_\nu ~,
\end{equation}
whereas at high neutrino energies, the elastic scattering cross section becomes
\begin{equation}
\sigma_{\rm el} = \frac{g^4}{16 \, \pi \,  m^2_{\rm{DM}} \, y} \, \left(\log\left[\frac{m^2_{\rm{DM}} \, y}{m^2_{\rm{N}}}\right] - 1 \right) + \mathcal{O}\left(\frac{1}{y^2}\right) \rightarrow 0 ~.
\end{equation}

For a real DM candidate, the elastic cross section is
\begin{align}
\sigma_\text{el}& = \frac{g^4}{32\pi}\left( - \frac{\left(m_{\rm{DM}}^4 \, \left(y \, (10 - 3 \, y \, (y-1)\right) + 6) - m_{\rm{DM}}^2 \, m_{\rm{N}}^2 \, (y+2)\, (5y+6) + 6 \, m_{\rm{N}}^4 \, (y+1)\right)}{(y+1) \, (m_{\rm{DM}}^2 \, (y-1) + m_{\rm{N}}^2)\, (m_{\rm{N}}^2 - m_{\rm{DM}}^2 \, (y+1))^2} \right. \nonumber \\ 
& - \, \left. \frac{2 \, \left(m_{\rm{DM}}^2 \, (y \, (y + 2) + 3) - 3 \, m_{\rm{N}}^2 \, (y + 1)\right)}{m_{\rm{DM}}^2 \, x^2 \, (m_{\rm{N}}^2 - m_{\rm{DM}}^2 \, (y+1))} \log\left[1 + \frac{m^2_{\rm{DM}} \, y^2}{m^2_{\rm{N}} \, (1+y) -m^2_{\rm{DM}}}\right] \right) ~, 
\end{align}
which, for low neutrino energies, can be approximated as
\begin{equation}
\sigma_\text{el} = \frac{g^4 \, m_{\rm{DM}}^6 \, y^4}{8 \, \pi \, (m^2_{\rm{DM}} - m_{\rm{N}}^2)^4} + \mathcal{O}(y^5) \propto E^4_\nu ~, 
\end{equation}
and in the high-energy limit as
\begin{equation}
\sigma_\text{el} = \frac{g^4}{16 \, \pi \, m^2_{\rm{DM}} \, y}\left(\log\left[\frac{m^2_{\rm{DM}} \, y}{m^2_{\rm{N}}}\right] - \frac{3}{2}\right) + \mathcal{O}\left(\frac{1}{y^2}\right) \rightarrow 0 ~.
\end{equation}

For the case of degenerate fermion mediator and scalar DM, we get
\begin{equation}
\sigma_\text{el} = \frac{g^4}{16 \, \pi \, m^2_{\rm{DM}}} \, \frac{(y+1)\log\left(y+1\right) - y}{y^2} ~, 
\end{equation}
for complex DM and 
\begin{equation}
\sigma_\text{el} = \frac{g^4}{32 \, \pi \, m^2_{\rm{DM}}} \, \frac{2 \, (y^2 - 1) \log\left(y+1\right) + y \, (3 \, y + 2)}{y^2 \, (y+1)} ~, 
\end{equation}
for scalar DM. Consequently, in the high-energy limit both cross sections tend to zero, whereas in the low-energy limit the cross section is independent of the temperature
\begin{equation}
\sigma_\text{el} = A \, \frac{g^4}{\pi \, m^2_{\rm{DM}}} ~, 
\end{equation}
with $A = \frac{1}{32}$ and $A = \frac{1}{8}$ for complex and real DM, respectively.

\subsection{Fermion DM coupled to a scalar mediator}
If the DM candidate is a Dirac fermion, the elastic scattering occurs only via the $u$- channel diagram,
\begin{align}
\sigma_\text{el}& = \frac{g^4}{32 \, \pi \, m_{\rm{DM}}^4 \, y^2}\left(\frac{m^2_{\rm{DM}} \, y^2}{y + 1} - \frac{(m_{\rm{DM}}^2 \, -m_{\phi}^2)^2}{m_{\rm{DM}}^2 \, (y-1) + m_{\phi}^2} - \right. \left. \frac{(y+1) \, (m_{\rm{DM}}^2 - m_{\phi}^2)^2}{m^2_{\rm{DM}} -m^2_\phi \, (y + 1)} \, \right. \nonumber\\
& + \left. 2 \, (m^2_{\rm{DM}} - m_{\phi}^2) \, \log\left[1 + \frac{m^2_{\rm{DM}} \, y^2}{m^2_{\phi} \, (1 + y) - m^2_{\rm{DM}}}\right]\right)  ~.
\label{ElFerDMScalarM}
\end{align}

In the low-$E_\nu$ limit, the cross section is approximated as
\begin{equation}
\sigma_\text{el}  =  \frac{g^4 \, m^2_{\rm{DM}} \, y^2}{32 \, \pi \, (m^2_{\rm{DM}} - m^2_\phi)^2} \, \left(1 + \frac{2 \, m_\phi^2 \, y}{m^2_{\rm{DM}} - m^2_\phi}\right) \, + \, \mathcal{O}\left(y^4\right) \propto E_{\nu}^2 ~,
\end{equation}
while in the high-$E_\nu$ limit,
\begin{equation}
\sigma_\text{el} = \frac{g^4}{32 \, \pi \, m_{\rm{DM}}^2 \, y}  + \mathcal{O}\left(\frac{1}{y^2}\right)  \rightarrow 0 ~.
\end{equation}

On the other hand, if the DM candidate is a Majorana particle, the elastic cross section is given by
\begin{eqnarray}
\sigma_{\rm el} & = & \frac{g^4}{32 \, \pi \, m^4_{\rm{DM}} \, y^2} \, \left( \frac{m^6_{\rm{DM}} \, y^4}{(y + 1) \, (m^2_\phi -(y + 1) \, m^2_{\rm{DM}})^2} \, - \right. \left. \frac{(m_{\rm{DM}}^2 - m_{\phi}^2)^2}{m_{\rm{DM}}^2 \, (y - 1) + m_{\phi}^2} + \frac{m^2_{\rm{DM}} \, y^2}{y + 1} \, \right. \nonumber \\
& & - \, \left. \frac{(y + 1) \, (m_{\rm{DM}}^2 - m_{\phi}^2)^2}{m^2_{\rm{DM}} -m^2_\phi \, (y + 1)} \, + \right.  \left. 2 \, (m^2_{\rm{DM}} - m_{\phi}^2) \, \log\left[1 + \frac{m^2_{\rm{DM}} \, y^2}{m^2_{\phi} \, (1 + y) - m^2_{\rm{DM}}}\right]\right) ~.
\label{ElMajoranaDMScalarM}
\end{eqnarray}

For low neutrino energies, the cross section reads
\begin{equation}
\sigma_{\rm el} = \frac{g^4 \, m^2_{\rm{DM}} \, y^2}{16 \, \pi \, (m^2_{\rm{DM}} - m^2_\phi)^2} \, \left(1 - \frac{3 \, y}{2}\right)  + \mathcal{O}\left(y^4\right) \propto E_{\nu}^2 ~, 
\end{equation}
while in the high-$E_\nu$ limit,
\begin{equation}
\sigma_{\rm el} = \frac{g^4}{16 \, \pi \, m_{\rm{DM}}^2 \, y}  + \mathcal{O}\left(\frac{1}{y^2}\right) \rightarrow 0 ~.
\end{equation}

In this scenario, different results are obtained if the scalar mediator and the fermion DM candidate are degenerated in mass:
\begin{equation}
\sigma_{\rm el} = A \, \frac{g^4}{\pi \, m^2_{\rm{DM}}} \, \left(\frac{1}{1 + y}\right) ~,
\end{equation}
with $A = \frac{1}{32}$ and $A = \frac{1}{16}$ for Dirac and Majorana DM, respectively. This implies that in the mass degenerated regime, the low-energy neutrino energy limits are $\sigma_{\rm el} = \frac{g^4}{32 \, \pi \, m^2_{\rm{DM}}}$ for Dirac DM and $\sigma_{\rm el} = \frac{g^4}{16 \, \pi \, m^2_{\rm{DM}}}$ for Majorana DM, while they are both zero in the high-energy limit.

\subsection{Vector DM coupled to a Dirac mediator}
The coupling for this scenario can arise by constructing a tower of Kaluza-Klein neutrino and photon excited states and couple the first excited state of the antineutrino ($\bar{N}_{\rm{L}}$) to the first excited state of the photon (the DM candidate, $\chi^\mu$), and a standard $\nu_{\rm{L}}$ neutrino. We disregard the possibility of the excited state of the neutrino being the DM candidate, since this implies that it could interact with any SM fermion via its coupling to the SM Z boson. This thus, falls beyond the scenarios considered in this paper, where we assume annihilation to neutrinos to be the dominant channel. Furthermore, the constraints from direct detection in such scenarios are quite stringent and much stronger than the bounds that can be derived from collisional damping~\cite{Servant:2002hb, Profumo:2006bv}. For this scenario, the elastic scattering cross section is

\begin{widetext}
\begin{align}
\sigma_{\rm el}& = \frac{g^4}{96 \, \pi \, m^6_{\rm DM} \, y^2 \,  \left(m^2_{\rm N} - m^2_{\rm DM} \, (x + 1)\right)^2} \, \left(\frac{m_{\rm{DM}}^2 \, y^2 \, (4 \, m_{\rm{DM}}^{10} \, \left(y \, (y^3 + 5 \, y + 16) + 10\right)}{(y + 1)^2 \, \left(m_{\rm{DM}}^2 \, (y - 1) + m_{\rm{N}}^2\right) \, \left(m_{\rm{DM}}^2 - m_{\rm{N}}^2 \, (y + 1)\right)} \,  \right.  \nonumber \\ 
& + \, \left.  \frac{-4 \, m_{\rm DM}^8 \, m_{\rm N}^2 \, (y + 1) \, \left(y \, \left(y^3 + 10 \, y + 24\right) + 30\right) + m_{\rm DM}^6 \, m_{\rm N}^4 \, \left(y \, \left(y \, \left(y \, \left(y \, (8 \, y + 41) + 100\right) + 189\right) + 216\right) + 102\right)}{(y + 1)^2 \, \left(m_{\rm DM}^2 \, (y - 1) + m_{\rm N}^2\right) \, \left(m_{\rm DM}^2 - m_{\rm N}^2 \, (y + 1)\right)} \, \right.  \nonumber \\
& +\left. \frac{-m_{\rm{DM}}^4 \, m_{\rm{N}}^6 \, \left(y \, \left(y \, \left(y \, \left((17 - 3 \, y) \, y + 58\right) + 66\right) + 10\right) - 14\right)  - 6 \, m_{\rm{N}}^{10} \, (y - 3) \, (y + 1)^2)}{(y + 1)^2 \, (m_{\rm{DM}}^2 \, (y - 1) + m_{\rm{N}}^2) ( m_{\rm{DM}}^2 - m_{\rm{N}}^2 \, (y + 1))} \,  \right. \nonumber \\
& + \ \frac{m_{\rm{DM}}^2 \, m_{\rm{N}}^8 \, \left(y \, \left(y \, (5 \, y + 17) - 30\right) - 54\right) )}{(y + 1) \, (m_{\rm{DM}}^2 \, (y - 1) + m_{\rm{N}}^2) ( m_{\rm{DM}}^2 - m_{\rm{N}}^2 \, (y + 1))} \nonumber \\
& + \ \left. 2 \, \left(m_{\rm{DM}}^2 \, (y + 1) - m_{\rm{N}}^2) \, \left(4 \, m_{\rm{DM}}^6 \, ((y - 2) \, y + 5\right) - 4 \, m_{\rm{DM}}^4 \, m_{\rm{N}}^2 \, (y + 5) + m_{\rm{DM}}^2 \, m_{\rm{N}}^4 \, \left(y \, (y + 6) - 9\right) - 3 \, m_{\rm{N}}^6 \, (y - 3)\right) \right. \nonumber \\
& \times \ \left. \log\left[1 + \frac{m^2_{\rm{DM}}y^2}{m^2_{\rm{N}} \, (1 + y) -m^2_{\rm DM}}\right]\right) ~.
\end{align}
\end{widetext}

Consequently, in the low-energy limit, the elastic scattering cross section reads
\begin{equation}
\sigma_{\rm el} = \frac{g^4 \, y^2 \, m^2_{\rm DM}}{4 \, \pi \, (m^2_{\rm N} - m^2_{\rm DM})^2} + \mathcal{O}\left(y^3\right) \propto E^2_\nu ~,
\end{equation}
whereas for high neutrino energies,
\begin{equation}
\sigma_{\rm el} = \frac{g^4 \, m^4_{\rm N}}{32 \, \pi \, y \, m^6_{\rm DM}} + \mathcal{O}\left(\frac{1}{y^2}\right) \rightarrow 0 ~.
\end{equation}

In the region of the parameter space where the mediator and DM candidates are degenerate in mass, the elastic scattering cross section is given by
\begin{equation}
\sigma_{\rm el} = \frac{g^4}{96 \, \pi \, m_{\rm DM}^2 \, y^2 \, (y + 1)^2} \ \left(y \ \left(y \ (7 \ y + 29) + 18\right) \ + \right.  \left. 2 \ (5 \, y - 9) \ (y + 1)^2\right) \times \log(y + 1) ~, 
\end{equation}
so that for  low neutrino energies,
\begin{equation}
\sigma_{\rm el} = \frac{g^4}{8 \, \pi \, m^2_{\rm DM}} ~,
\end{equation}
while for high neutrino energies,
\begin{equation}
\sigma_{\rm el} = \frac{g^4}{96 \, \pi \, m^2_{\rm DM} \, y} \, \left(7 + 10 \, \log(y)\right) \, \rightarrow \, 0 ~.
\end{equation}

\subsection{Scalar DM coupled to a vector mediator}

In this case, the elastic scattering cross section is given by
\begin{equation}
\sigma_{\rm el} = \frac{g_\chi^2 \, g_\nu^2}{4 \, \pi \, m_{\rm{Z}'}^2} \left(1 - z \, \log\left[1 + \frac{1}{z}\right] \right) ~,
\label{el}
\end{equation}
with  
\begin{equation}
z = \frac{m_{\rm{Z}'}^2 \, (1 + y)}{m_{\rm DM}^2 \, y^2} ~.
\end{equation}

In the low-energy limit, the cross section is temperature-dependent,
\begin{equation}
\sigma_{\rm el} = \frac{g_\chi^2 \, g_\nu^2 \, m^2_{\rm DM} \, y^2}{8 \, \pi \, m^4_{\rm{Z}'}}\left(1 - y\right) + \mathcal{O}\left(y^4\right) \propto E_{\nu}^2 ~, 
\end{equation}
whereas at high energies, 
\begin{equation}
\sigma_{\rm el} = \frac{g_\chi^2 \, g_\nu^2}{4 \, \pi} \, \left(\frac{1}{ m_{\rm{Z}'}} + \frac{\log\left(\frac{m^2_{\rm{Z}'}}{m_{\rm DM}^2 \, y}\right)}{m^2_{\rm DM} \, y}\right) + \mathcal{O}\left(\frac{1}{y^2}\right) \propto \text{constant} ~.
\end{equation}

\subsection{Fermion DM coupled to a vector mediator}
This process occurs via the $t$- channel and the elastic scattering cross section and for Dirac DM is given by
\begin{equation}
\sigma_{\rm el} = \frac{g_\chi^2 \, g^2_\nu}{8 \, \pi \, m_{\rm DM}^2} \, \left(\frac{m_{\rm{Z}'}^2}{m_{\rm DM}^2 \, y^2 + m_{\rm{Z}'}^2 \, (y + 1)} + \frac{2 \, m_{\rm DM}^2}{m_{\rm{Z}'}^2} \, + \right.  \left. \frac{1}{y + 1} - 2 \, \frac{m^2_{\rm DM} \, (y + 1) + m_{\rm{Z}'}^2}{m^2_{\rm DM} \, y^2} \, \log\left[1 + \frac{m^2_{\rm DM} \, y^2}{m^2_{\rm{Z}'} \, (1 + y)}\right] \right) ~, \nonumber \\ 
\end{equation}
whereas for Majorana DM, 
\begin{align}
\sigma_{\rm el}& = \frac{g_\chi^2 \, g^2_\nu}{32 \, \pi \, m_{\rm DM}^2}\left(\frac{2 \, m_{\rm DM}^4 \, y^2}{m_{Z'}^2 \, \left(m_{\rm DM}^2 \, y^2 + m_{Z'}^2 \, (y + 1)\right)} \, + \, \right. \frac{m_{\rm DM}^2 \, (3 \, y^2 - 2) + 2 \, m_{Z'}^2 \, (y + 1)}{(y + 1) \, \left(m_{\rm DM}^2 \, y^2 + m_{Z'}^2 \, (y + 1)\right)}\left. \right. \,  \nonumber \\
& - \, \left. \frac{2 \, (m^2_{\rm DM} \, (y - 1) + m^2_{Z'})}{m^2_{\rm DM} \, y^2} \times \log\left[\frac{m_{\rm DM}^2 \, y^2}{m^2_{Z'} \, (y + 1)} + 1\right] \right) ~. \nonumber \\ 
\label{ElFerDMVecM}
\end{align}

In the low-energy limit, the cross section reduces to
\begin{equation}
\sigma_{\rm el} = \frac{g_\chi^2 \, g_\nu^2 \, m^2_{\rm DM} \, y^2}{8 \, \pi \, m^4_{\rm{Z}'}} \, \left(1 - y\right) + \mathcal{O}\left(y^4\right) \propto E_{\nu}^2 ~, 
\end{equation}
for Dirac DM and to
\begin{equation}
\sigma_{\rm el} = \frac{3 \, g_\chi^2 \, g_\nu^2 \, m^2_{\rm DM} \, y^2}{16 \, \pi \, m^4_{\rm{Z}'}} \, \left(1 - \frac{5 \, y}{3}\right) + \mathcal{O}\left(y^4\right) \propto E_{\nu}^2 ~, 
\end{equation}
for Majorana DM. 

In the high-$E_\nu$ limit, the cross section for Dirac DM is approximated as
\begin{eqnarray}
\sigma_{\rm el} & = & \frac{g_\chi^2 \, g_\nu^2}{4 \, \pi}\left(\frac{1}{ m_{\rm{Z}'}} + \frac{\left(\log\left[\frac{m^2_{\rm{Z}'}}{m_{\rm{DM}}^2 \, y}\right] + 1\right)}{m^2_{\rm{DM}} \, y}\right) + \mathcal{O}\left(\frac{1}{y^2}\right) \propto \text{constant} ~, \nonumber 
\end{eqnarray}
while for Majorana DM,
\begin{eqnarray}
\sigma_{\rm el} & = & \frac{g_\chi^2 \, g_\nu^2}{16 \, \pi}\left(\frac{1}{ m_{\rm{Z}'}} + \frac{\left(\log\left[\frac{m^2_{\rm{Z}'}}{m_{\rm{DM}}^2y}\right] + \frac{1}{2}\right)}{m^2_{\rm DM} \, y}\right) + \mathcal{O}\left(\frac{1}{y^2}\right)\propto \text{constant} ~. \nonumber \\ 
\end{eqnarray}

\subsection{Vector DM coupled to a vector mediator}
For this last scenario, the elastic scattering cross section is given by
\begin{widetext}
\begin{align}
\sigma_{\rm el}& = \frac{g^2_\chi \, g^2_\nu}{96 \, \pi \,  m_{\rm DM}^4}\left(\frac{24 \, m_{\rm DM}^6 \, y^2 \, (y + 1)^2 + m_{\rm DM}^4 \, m_{Z'}^2 \, \left(y \, \left(y \, \left(y \, \left(y \, (4 \, y - 35) - 48\right) + 40\right) + 72\right) + 24\right)}{m_{Z'}^2 \, (y + 1)^2 \, (m_{\rm DM}^2 \, y^2 + m_{\rm DM}^2 \, (y + 1))} \right. \nonumber \\
& + \, \left. \frac{m_{\rm DM}^2 \, m_{Z'}^4 \, (y + 1) \, \left(y \, \left(y \, (28 \, y - 15) - 64\right) - 24\right) + 6 \, m_{Z'}^6 \, (4 \, y + 3) \, (y + 1)^2}{m_{Z'}^2 \, (y + 1)^2 \, (m_{\rm DM}^2 \, y^2 + m_{\rm DM}^2 \, (y + 1))} \, \right. \nonumber \\
& + \ \left.\frac{2 \, (2 \, m_{\rm DM}^4 \, \left(y \, (5 \, y - 6)  - 6\right) + 4 \, m_{\rm DM}^2 \, m_{Z'}^2 \, \left((5 - 2 \, y) \, y + 3\right) - 3 \, m_{Z'}^4 \, (4 \, y + 3))}{m^2_{\rm DM} \, y^2} \times \log\left[\frac{m_{\rm DM}^2 \, y^2}{m^2_{Z'} \, (y + 1)} + 1\right]\right) ~,
\end{align}
\end{widetext}
so that for small energies, the cross section can be written as
\begin{equation}
\sigma_{\rm el} = \frac{g^2_\chi \, g^2_\nu \, m_{\rm DM}^2 \, y^2}{8 \, \pi \, m^4_{Z'}} \, \left(1 - y\right) + \mathcal{O}\left(y^4\right) \propto E^2_\nu ~.
\end{equation}
Finally, in the high-energy limit,
\begin{equation}
\sigma_{\rm el} = \frac{g^2_\chi \, g^2_\nu}{96 \, \pi \, m^2_{\rm DM}} \, \left(y + 4 \, (6 \, m^2_{\rm DM} - m^2_{Z'}) \times \log\left[\frac{m^2_{\rm DM} \, y}{m^2_{Z'}}\right] \, 
+ \, \frac{3 \, (9 \, m^4_{\rm DM} - 15 \, m^2_{\rm DM} \, m^2_{Z'} + 2 \, m^4_{Z'})}{m^2_{Z'}} \right) + \mathcal{O}\left(\frac{1}{y}\right) \propto E_\nu ~.
\end{equation}
Thus, this is the only scenario with the cross section proportional to the neutrino energy in the high-energy limit.

\section{Results for all scenarios}
\label{FullResults}

\subsection{Scalar DM and Dirac mediator}

\begin{figure}[H]
  \centering
    \subfloat{{\includegraphics[width=0.45\linewidth, height=7cm]{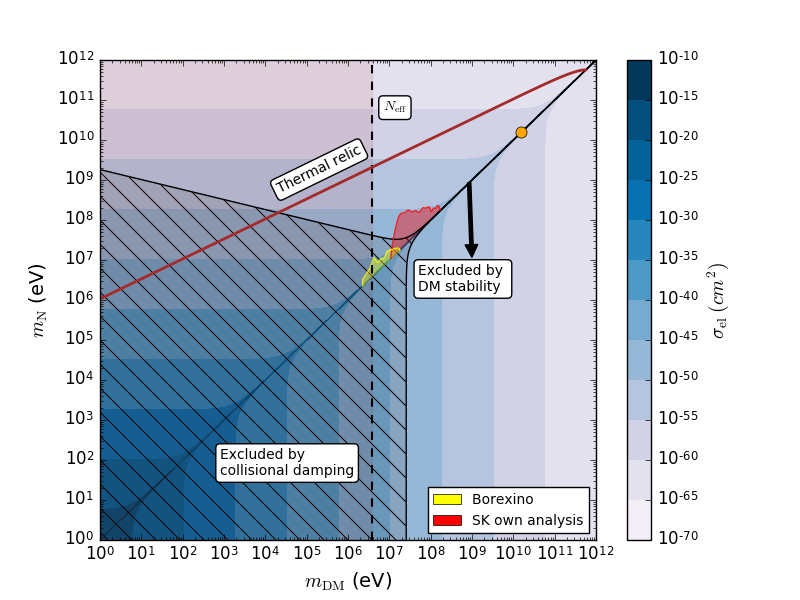} }}%
    \qquad
    \subfloat{{\includegraphics[width=0.45\linewidth, height=7cm]{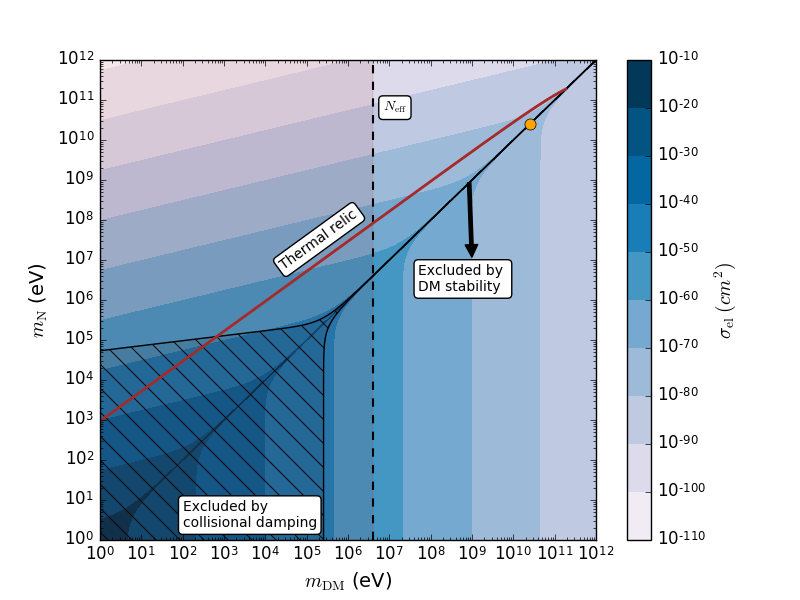} }}%
\caption{Same as Fig.~\ref{Fig2Paper2} but for complex DM (left) and real DM (right) with a Dirac mediator.}    
\end{figure}

\subsection{Real DM and Majorana mediator}

\begin{figure}[H]
\centering
\includegraphics[width=0.45\linewidth, height=7cm]{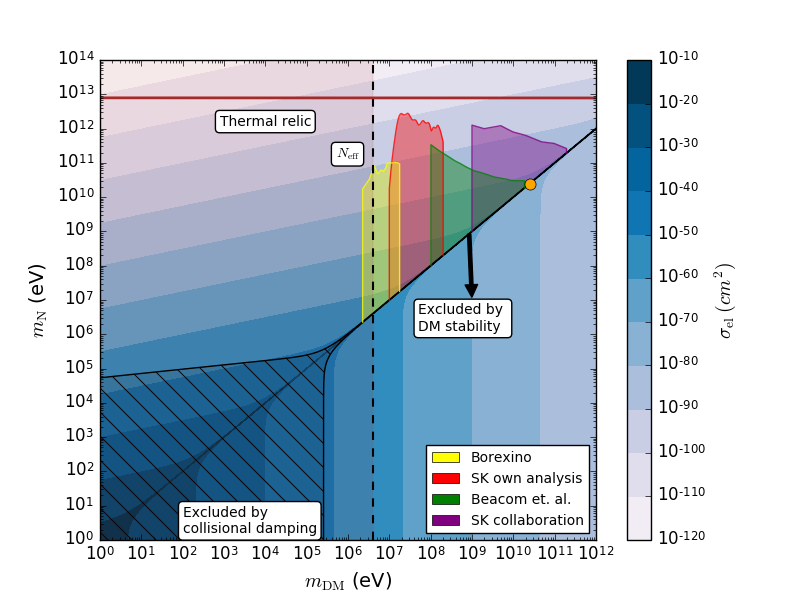}
\caption{Same as Fig.~\ref{Fig2Paper2}, but for real DM with a Majorana mediator.}
\end{figure}

\subsection{Vector DM and fermion mediator}

\begin{figure}[H]
  \centering
    \subfloat{{\includegraphics[width=0.45\linewidth, height=7cm]{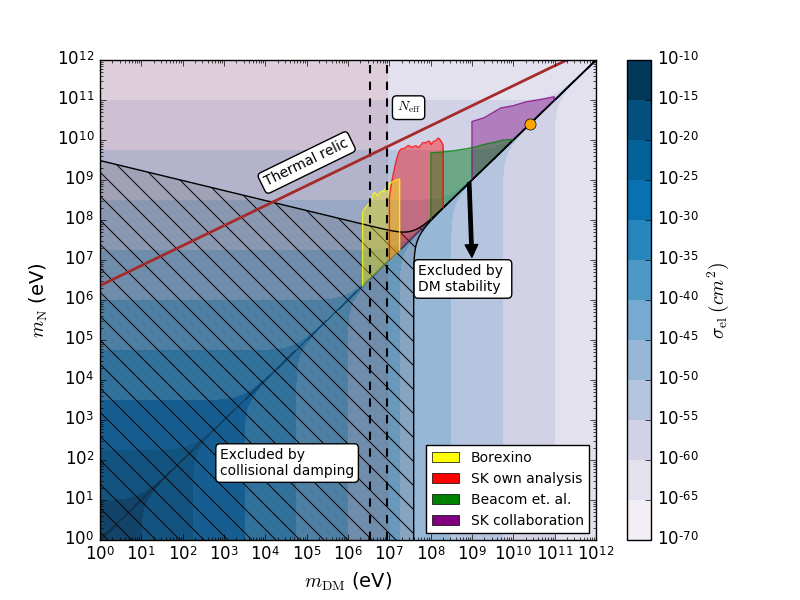} }}%
    \qquad
    \subfloat{{\includegraphics[width=0.45\linewidth, height=7cm]{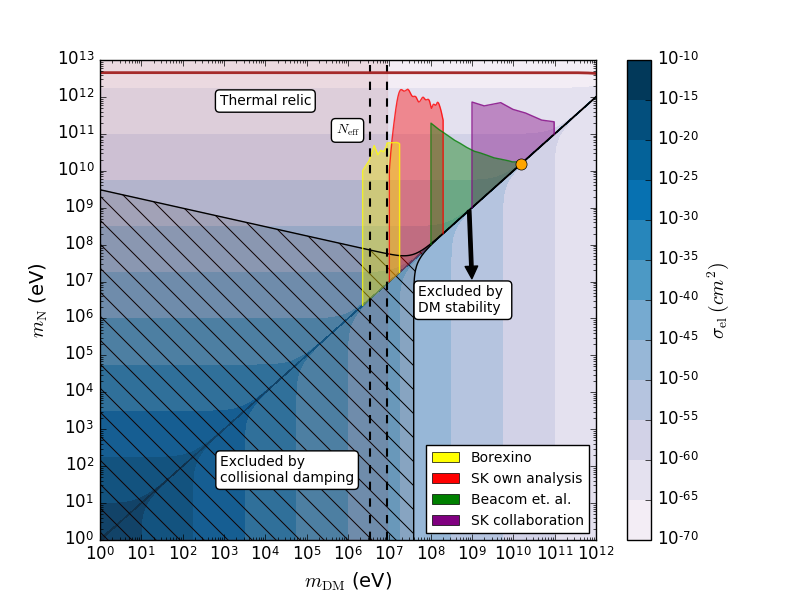} }}%
\caption{Same as Fig. \ref{Fig2Paper2} but for vector DM with a Dirac mediator (left) and a Majorana mediator (right). The dashed band for the $N_{\rm{eff}}$ bound is due to the fact that there is no precise bound for vector DM but it is expected to lie within the 1--10 MeV region.}    
\end{figure}

\subsection{Scalar DM and vector mediator}

\begin{figure}[H]
\centering
\includegraphics[width=0.45\linewidth, height=7cm]{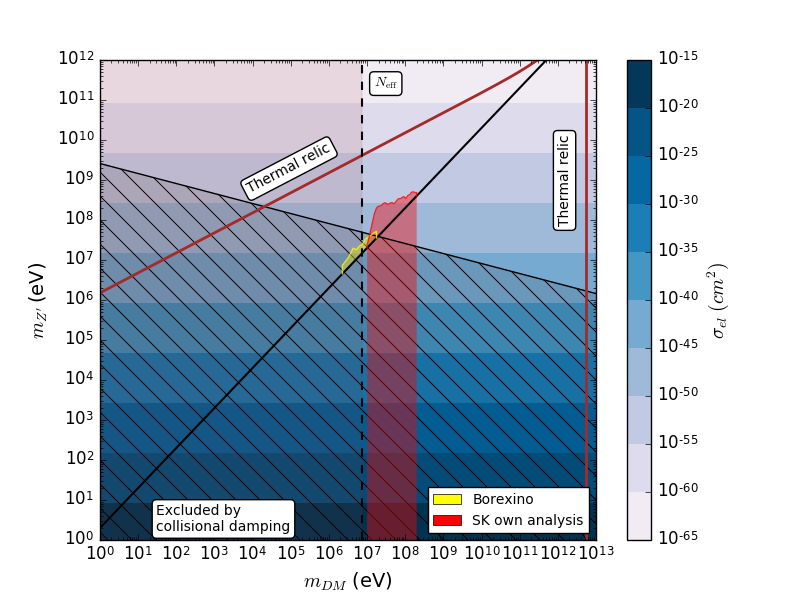}
\caption{Same as Fig.~\ref{Fig2Paper2}, but for scalar DM with a vector mediator. The vertical line represents the resonant region where $2 \, m_{Z'} \simeq m_{\rm DM}$.}
\end{figure}

\subsection{Majorana DM and vector mediator}

\begin{figure}[H]
\centering
\includegraphics[width=0.45\linewidth, height=7cm]{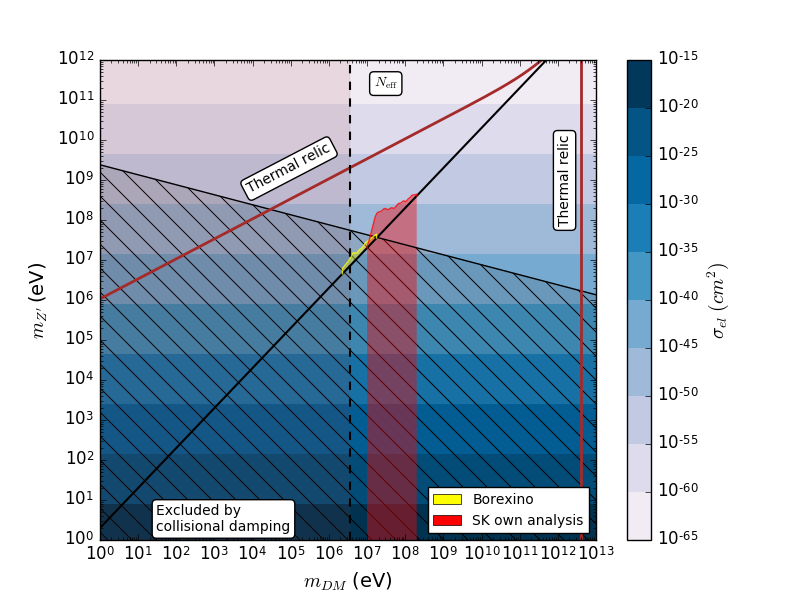}
\caption{Same as Fig.~\ref{Fig2Paper2}, but for Majorana DM with a vector mediator. The vertical line represents the resonant region where $2 \, m_{Z'} \simeq m_{\rm DM}$.}
\end{figure}

\subsection{Vector DM and vector mediator}

\begin{figure}[H]
\centering
\includegraphics[width=0.45\linewidth, height=7cm]{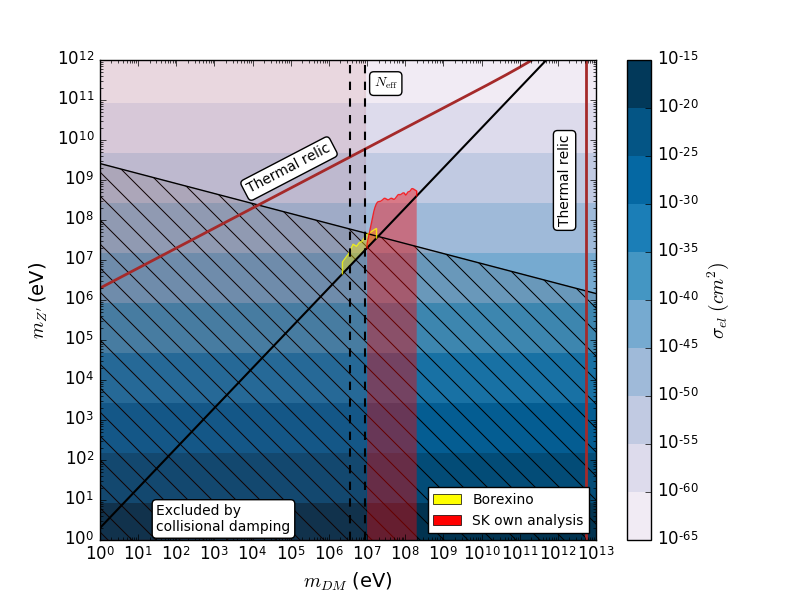}
\caption{Same as Fig.~\ref{Fig2Paper2} but for vector DM with a vector mediator. The vertical line represents the resonant region where $2 \, m_{Z'} \simeq m_{\rm DM}$. The dashed band for the $N_{\rm eff}$ bound is due to the fact that there is no precise bound for vector DM, although it is expected to lie within the 1--10 MeV region.}
\end{figure}

\end{document}